\documentclass[acmsmall]{acmart}
\usepackage[linesnumbered, ruled, vlined]{algorithm2e}
\usepackage{graphicx}
\usepackage{balance}
\usepackage{caption}
\usepackage{moresize}
\usepackage{pbox}
\usepackage{mathtools}
\usepackage[TABBOTCAP]{subfigure}
\usepackage{paralist}
\usepackage{tabularx}
\usepackage{hyperref}
\usepackage{balance}
\usepackage{multirow}
\usepackage{multicol}
\usepackage{amsmath}
\usepackage{listings}
\usepackage{setspace}
\usepackage{verbatim}
\usepackage{makecell}
\usepackage{float}
\usepackage[normalem]{ulem}
\usepackage{xspace}
\usepackage{wrapfig}
\usepackage{ifthen}
\usepackage{pifont}
\usepackage{tikz}
\usepackage{pict2e,picture}
\usepackage[most]{tcolorbox}
\definecolor{MidnightBlue}{HTML}{006895}

\include{macro}

\newcommand{\mybox}[4]{
    \begin{figure}[h!]
        \centering
    \begin{tikzpicture}
        \node[anchor=text,text width=\columnwidth-1.2cm, draw, rounded corners, line width=1pt, fill=#3, inner sep=5mm] (big) {\\#4};
        \node[draw, rounded corners, line width=.5pt, fill=#2, anchor=west, xshift=5mm] (small) at (big.north west) {#1};
    \end{tikzpicture}
    \end{figure}
}

\AtBeginDocument{%
  \providecommand\BibTeX{{%
    \normalfont B\kern-0.5em{\scshape i\kern-0.25em b}\kern-0.8em\TeX}}}

\setcopyright{acmlicensed}
\copyrightyear{2021}
\acmYear{2021}
\acmDOI{\#\#.\#\#\#\#/\#\#\#\#\#\#\#.\#\#\#\#\#\#\#}

\acmJournal{TOSEM}
\acmVolume{\#\#}
\acmNumber{\#}
\acmArticle{\#\#\#}
\acmMonth{\#}

\begin{document}

\title{A Systematic Literature Review on the Use of Deep Learning in Software Engineering Research}

\author{Cody Watson}
\authornote{Authors have contributed equally}
\email{cwatson@wlu.edu}
\affiliation{%
  \institution{Washington \& Lee University}
  \streetaddress{204 W Washington St.}
  \city{Lexington}
  \state{Virginia}
  \postcode{24450}
}

\author{Nathan Cooper}
\authornotemark[1]
\email{nacooper01@email.wm.edu}
\affiliation{%
  \institution{William \& Mary}
  \streetaddress{251 Jamestown Rd.}
  \city{Williamsburg}
  \state{Virginia}
  \postcode{23185}
}

\author{David Nader Palacio}
\authornotemark[1]
\email{danaderpalacio@email.wm.edu}
\affiliation{%
  \institution{William \& Mary}
  \streetaddress{251 Jamestown Rd.}
  \city{Williamsburg}
  \state{Virginia}
  \postcode{23185}
}

\author{Kevin Moran}
\email{kpmoran@cs.wm.edu}
\affiliation{%
  \institution{George Mason University}
  \streetaddress{4400 University Drive}
  \city{Fairfax}
  \state{Virginia}
  \postcode{22030}
}

\author{Denys Poshyvanyk}
\email{denys@cs.wm.edu}
\affiliation{%
  \institution{William \& Mary}
  \streetaddress{251 Jamestown Rd.}
  \city{Williamsburg}
  \state{Virginia}
  \postcode{23185}
}

\renewcommand{\shortauthors}{Watson, Cooper, Palacio, Moran \& Poshyvanyk}

\begin{abstract}
An increasingly popular set of techniques adopted by software engineering (SE) researchers to automate development tasks are those rooted in the concept of Deep Learning (DL). The popularity of such techniques largely stems from their automated feature engineering capabilities, which aid in modeling software artifacts. However, due to the rapid pace at which DL techniques have been adopted, it is difficult to distill the current successes, failures, and opportunities of the current research landscape. In an effort to bring clarity to this cross-cutting area of work, from its modern inception to the present, this paper presents a systematic literature review of research at the intersection of  SE \& DL. The review canvases work appearing in the most prominent SE and DL conferences and journals and spans \revision{\includedpapers papers across \includedsetasks ~unique SE tasks}. We center our analysis around the \textit{components of learning}, a set of principles that govern the application of machine learning techniques (ML) to a given problem domain, discussing several aspects of the surveyed work at a granular level. The end result of our analysis is a \textit{research roadmap} that both delineates the foundations of DL techniques applied to SE research, and highlights likely areas of fertile exploration for the future.
\end{abstract}



\maketitle

\section{Introduction}
\label{sec:introduction}

Software engineering (SE) research investigates questions pertaining to the design, development, maintenance, testing, and evolution of software systems. As software continues to pervade a wide range of industries, both open- and closed-source code repositories have grown to become unprecedentedly large and complex. This has resulted in an increase of unstructured, unlabeled, yet important data including requirements, design documents, source code files, test cases, and defect reports. Previously, the software engineering community has applied canonical machine learning (ML) techniques to identify patterns and unique relationships within this data to automate or enhance many tasks typically performed manually by developers. Unfortunately, the process of implementing ML techniques can be a tedious exercise in careful feature engineering, wherein researchers experiment with identifying salient attributes of data that can be leveraged to help solve a given problem or automate a given task.

However, with recent improvements in computational power and the amount of memory available in modern computer architectures, an advancement to traditional ML approaches has arisen called Deep Learning (DL). Deep learning represents a fundamental shift in the manner by which machines learn patterns from data by \textit{automatically} extracting salient features for a given computational task as opposed to relying upon human intuition. Deep Learning approaches are characterized by architectures comprised of several layers that perform mathematical transformations on data passing through them. These transformations are controlled by sets of learnable parameters that are adjusted using a variety of learning and optimization algorithms. These computational layers and parameters form models that can be trained for specific tasks by updating the parameters according to a model's performance on a set of training data. Given the immense amount of structured and unstructured data in software repositories that are likely to contain hidden patterns, DL techniques have ushered in advancements across a range of tasks in software engineering research including automatic program repair~\cite{Tufano2018}, code suggestion~\cite{Gu2018}, defect prediction~\cite{Wang2016}, malware detection \cite{Li2018}, feature location~\cite{Corley2015}, among many others~\cite{Ma2018, Wan2018, Liu2018, White2016, Xu2016, Guo2017, Tian2018a, Liu2017}. A recent report from the 2019 NSF Workshop on Deep Leaning \& Software Engineering has referred to this area of work as Deep Learning for Software Engineering (DL4SE)~\cite{dlse19-report}.

The applications of DL to improve and automate SE tasks points to a clear synergy between ongoing research in SE and DL. However, in order to effectively chart the most impactful path forward for research at the intersection of these two fields, researchers need a clear map of what has been done, what has been successful, and what can be improved. %
In an effort to map and guide research at the intersection of DL and SE, we conducted a systematic literature review (SLR) to identify and systematically enumerate the synergies between the two research fields. As a result of the analysis performed in our SLR, we synthesize a detailed \textit{research roadmap} of past work on DL techniques applied to SE tasks\footnote{It should be noted that another area, known as Software Engineering for Deep Learning (SE4DL), which explores improvements to engineering processes for DL-based systems, was also identified at the 2019 NSF workshop. However, the number of papers we identified on this topic was small, and mostly centered around emerging testing techniques for DL models. Therefore, we reserve a survey on this line of research for future work.} (\ie DL4SE), complete with identified open challenges and best practices for applying DL techniques to SE-related tasks and data. Additionally, we analyzed the impacts of these DL-based approaches and discuss some observed concerns related to the potential reproducibility and replicability of our studied body of literature. %

We organize our work around five major Research Questions (RQs) that are fundamentally centered upon the \textit{components of learning}. That is, we used the various components of the machine learning process as enumerated by Abu-Mostafa~\cite{abu-mastafa}, to aid in grounding the creation of our research roadmap and exploration of the DL4SE topic. Our overarching interest is to identify best practices and promising research directions for applying DL frameworks to SE contexts. %
Clarity in these respective areas will provide researchers with the tools necessary to effectively apply DL models to SE tasks. %
To answer our RQs, we created a taxonomy of our selected research papers that highlights important concepts and methodologies characterized by the types of software artifacts analyzed, the learning models implemented, and the evaluation of these approaches. We discovered that while DL in SE has been successfully applied to many SE tasks, there are common pitfalls and details that are critical to the components of learning that are often omitted. Therefore, in addition to our taxonomy that describes how the components of learning have been addressed, we provide insight into components that are often omitted, alongside strategies for avoiding such omissions. As a result, this paper provides the SE community with important guidelines for applying DL models that address issues such as sampling bias, data snooping, and over- and under-fitting of models. Finally, we provide an online appendix with all of our data and results to facilitate reproducability and encourage contributions from the community to continue to taxonomize DL4SE research\footnote{\url{http://wm-semeru.github.io/dl4se/}}~\cite{watson_palacio_cooper_moran_poshyvanyk, cody_watson_2021_4768587}.

\section{Research Question Synthesis}
\label{sec:rq_synthesis}

\begin{wrapfigure}{l}{0.5\textwidth} 
	\centering
	\includegraphics[width=0.5\columnwidth]{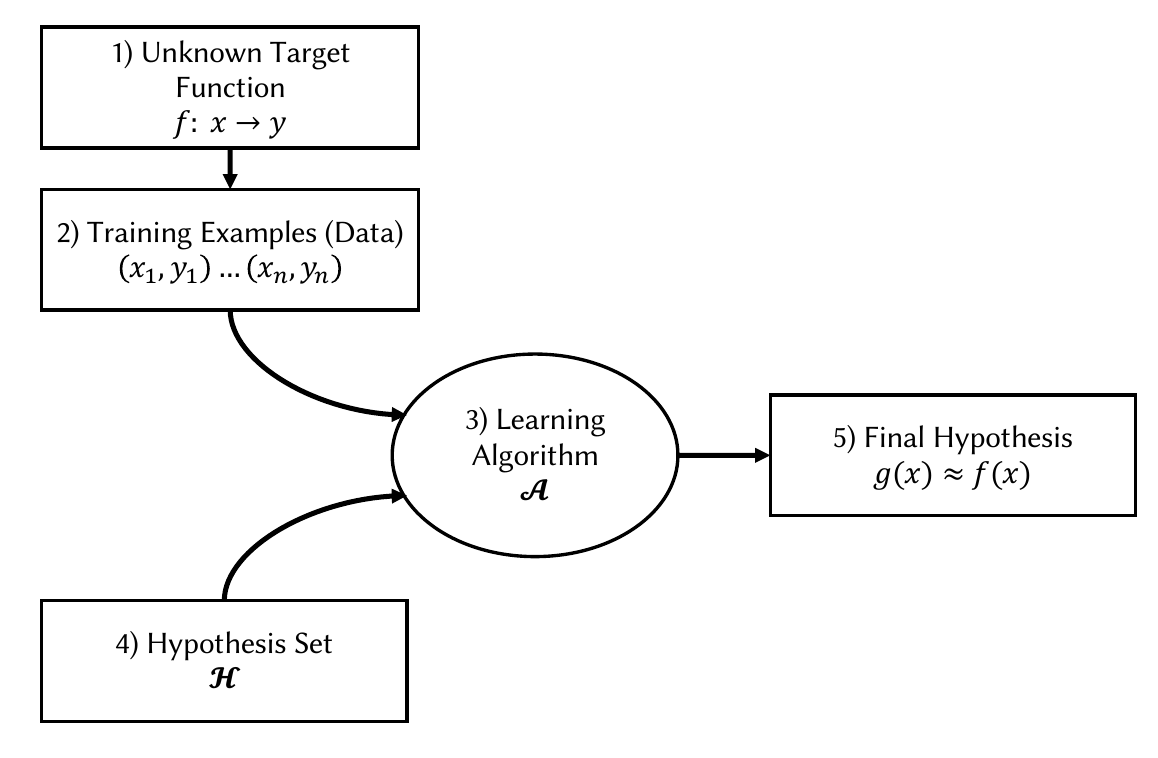}
	\vspace{-0.4cm}
	\caption{The Components of Learning}	
	\label{fig:components}
\end{wrapfigure}

The synthesis and generation of research questions (RQs) is an essential step to any systematic literature review (SLR). In order to study the intersection of DL \& SE, our intention was to formulate RQs that would naturally result in the derivation of a taxonomy of the surveyed research, establish coherent guidelines for applying DL to SE tasks, and address common pitfalls when implementing these complex models. Therefore, in order to properly accomplish these tasks and frame our review, we centered the synthesis of our RQs on the \textit{components of learning}~\cite{abu-mastafa}, which are illustrated in Figure~\ref{fig:components}. The components of learning are a formalization introduced by Abu-Mostafa~\cite{abu-mastafa} in an effort to enumerate the conditions for computational learning. By framing our top-level research questions according to these components, we can ensure that that analysis component of our literature review effectively captures the essential elements that any research project applying a \textit{deep} learning-based solution should discuss, allowing for a thorough taxonomic inspection. Given that these components represent essential elements that should be described in any application of computational learning, framing our research questions in this manner allows for the extrapolation of observed trends related to those elements that are commonly included or omitted from the surveyed literature. This, in turn, allows us to make informed recommendations to the community related to the reproducibility of our surveyed work. In the remainder of this section, we detail how each of our top-level research questions were derived from the elements of learning. Note that, in order to perform our analysis to a sufficient level of detail, in addition to our top-level RQs, we also define several Sub-RQs that allow for a deeper analysis of some of the more complex elements of learning. We provide the full list of all the research questions at the end of this section. %

\subsection{The First Element of Learning: The Target Function}

The first component of learning is an unknown \textit{target function} ($f:x\rightarrow y$), which represents the relationship between two observed phenomenon $x$ and $y$. The target function is typically tightly coupled to the task to which a learning algorithm is applied. By analyzing the target function to be learned, one can infer the input and output of the model, the type of learning, hypothetical features to be extracted from the data and potential applicable architectures. To capture the essence of this component of learning we formulated the following research question:

\begin{tcolorbox}[enhanced,skin=enhancedmiddle,borderline={1mm}{0mm}{MidnightBlue}]
	\textit{\textbf{RQ$_1$:} What types of SE tasks have been addressed by DL-based approaches?}
\end{tcolorbox}  

In understanding what SE tasks have been analyzed, we are able to naturally present a taxonomy of what tasks have yet to be explored using a DL-based approach. We were also able to infer why certain SE tasks may present unique challenges for DL models as well as the target users of these DL approaches, given the SE task they address.

\subsection{The Second Element of Learning: The (Training) Data}

The second component of learning is defined by the \textit{data} that is presented to a given learning algorithm in order to learn this unknown target function. Here, we primarily focused on studying the input and output training examples and the techniques used in DL approaches to prepare the data for the model. An understanding of the training examples presents greater insight into the target function while also providing further intuition about the potential features and applicable DL architectures that can be used to extract those features. Thus, in capturing this component of learning, we aimed to derive a taxonomy of the data used, how it was extracted and preprocessed, and how these relate to different DL architectures and SE tasks. This taxonomy captures relationships between data and the other elements of learning, illustrating effective (and ineffective) combinations for various SE-related applications. Our intention is that this information can inform researchers of effective combinations and potentially unexplored combinations of data/models/tasks to guide future work. Thus, our second RQ is formulated as follows:

\begin{tcolorbox}[enhanced,skin=enhancedmiddle,borderline={1mm}{0mm}{MidnightBlue}]
	\textit{\textbf{RQ$_2$:} How are software artifacts being extracted, prepared, and used in DL-based approaches for SE tasks?}
\end{tcolorbox}  

\noindent Given the multi-faceted nature of selecting, creating, and preprocessing data, we specifically examine three sub-research questions that explore the use of SE data in DL approaches in depth: 

\begin{itemize}\setlength\itemsep{0.4em}
	\item{\textit{\textbf{RQ$_{2a}$:} What types of SE data are being used?}}
	\item{\textit{\textbf{RQ$_{2b}$:} How is this data being extracted and pre-processed into formats that are consumable by DL approaches?}}
	\item{\textit{\textbf{RQ$_{2c}$:} What type of exploratory data analysis is conducted to help inform model design and training?}}
\end{itemize}

\noindent \textit{RQ$_{2a}$} explores the different types of data that are being used in DL-based approaches. Given the plethora of different software artifacts currently stored in online repositories, it is important to know which of those artifacts are being analyzed and modeled. \textit{RQ$_{2b}$} examines how data is being extracted and pre-processed into a format that a DL model can appropriately consume. The results of this RQ will enumerate the potential tools and techniques to mine different types of data for various DL applications within SE. Additionally, the representation of data is often dependent on the DL architecture and its ability to extract features from that representation, which lends importance to the discussion of the relationship between DL architectures and the data they process. \textit{RQ$_{2c}$} investigates what type of exploratory data analysis is conducted to help inform model design and training. In order to perform effectively, DL models typically require large-scale datasets, and the quality of the learned hypothesis is a product of the quality of data from which the model learns. Therefore, since the quality of a given DL model is often directly associated with its data, we examined how research performed (or didn't perform) various analyses to avoid common data- related pitfalls recognized by the ML/DL community, including sampling bias and data snooping.

\subsection{\hspace{-0.25cm}The Third \& Fourth Elements of Learning: The Learning Algorithm \& Hypothesis Set}

Next, we jointly examine both the third and fourth components of learning, the \textit{learning algorithm} and \textit{hypothesis set}, in a single research question due to their highly interconnected nature. The learning algorithm is a mechanism that navigates the hypothesis set in order to best fit a given model to the data. The learning algorithm typically consists of a numeric process that uses a probability distribution over the input data to appropriately approximate the optimal hypothesis from the hypothesis set. The hypothesis set is a set of all hypotheses, based on the learning algorithm, to which the input can be mapped. This set changes because it is a function of the possible outputs given the input space, and is dependent on the learning algorithm's ability to model those possible outputs. Taken together the learning algorithm and the hypothesis set are referred to as the learning model, thus, our third RQ is formulated as follows:

\begin{tcolorbox}[enhanced,skin=enhancedmiddle,borderline={1mm}{0mm}{MidnightBlue}]
	\textit{\textbf{RQ$_3$:} What deep learning models are used to support SE tasks?}
\end{tcolorbox} 

\noindent Given the various types of DL model architectures and optimization techniques that may be applicable to various SE tasks, we examine \textit{RQ$_3$} through the lens of three sub-RQs, which address the aforementioned attributes of the learning model individually. 

\begin{itemize}\setlength\itemsep{0.4em}
	\item{\textit{\textbf{RQ$_{3a}$:} What types of model architectures are used to perform automated feature engineering of the data related to various SE tasks?}}
	\item{\textit{\textbf{RQ$_{3b}$:} What learning algorithms and training processes are used in order to optimize the models?}}
	\item{\textit{\textbf{RQ$_{3c}$:} What methods are employed to combat over- and under-fitting of the models?}}
\end{itemize}

\noindent Firstly, \textit{RQ$_{3a}$} explores the different types of model architectures that are used to perform automated feature engineering of different SE artifacts for various SE tasks. As part of the analysis of this RQ we also examine how the type of architecture chosen to model a particular target function relates to the types of features that are being extracted from the data. Secondly, \textit{RQ$_{3b}$} examines the different learning algorithms and training processes that are used to optimize various DL models. As part of this analysis, we explore a variety of different learning algorithms whose responsibility is to properly capture the hypothesis set for the given input space. The different optimization algorithms and training processes used to tune the weights of the model are an important step for finding the target hypothesis that best represents the data. Lastly, \textit{RQ$_{3c}$} analyses the methods used to combat over- and under-fitting. Our intention with this RQ is to understand the specific methods (or lack thereof) used in SE research to combat over- or under-fitting, and the successes and shortcomings of such techniques.

\subsection{The Fifth Element of Learning: The Final Hypothesis}

Our fourth RQ addresses the component of learning known as the \textit{final hypothesis}, which is the target function learned by the model that is used to predict aspects of previously unseen data points. In essence, in order to investigate this component of learning in the context of SE applications, we examine the \textit{effectiveness} of the learned hypothesis as reported according to a variety of metrics across different SE tasks. Our intention with this analysis is to provide an indication of the advantages of certain data selection and processing pipelines, DL architectures, and training processes that have been successful for certain SE tasks in the past. Thus, our fourth RQ is formulated as follows:

\begin{tcolorbox}[enhanced,skin=enhancedmiddle,borderline={1mm}{0mm}{MidnightBlue}]
	\textit{\textbf{RQ$_4$:} How well do DL tasks perform in supporting various SE tasks?}
\end{tcolorbox}

\noindent Analyzing the effectiveness of DL applied to a wide range of SE tasks can be a difficult undertaking due to the variety of different metrics and evolving evaluation settings used in different contexts. Thus we examined two primary aspects of the literature as sub-RQs in order to provide a holistic illustration of DL effectiveness in SE research:

\begin{itemize}\setlength\itemsep{0.4em}
	\item{\textit{\textbf{RQ$_{4a}$:} What ``baseline'' techniques are used to evaluate DL models and what benchmarks are used for these comparisons?}}	
	\item{\textit{\textbf{RQ$_{4b}$:} How is the impact or automatization of DL approaches measured and in what way do these models promote generalizability?}}

\end{itemize}

\noindent Understanding the metrics used to quantify the comparison between DL approaches is important for informing future work regarding methods for best measuring the efficacy of newly proposed techniques. Thus, \textit{RQ$_{4a}$} explores trade-offs related to model complexity and accuracy. In essence, we examine applications of DL architectures through the lens of the \textit{Occam's Razor Principal}, which states that ``the least complex model that is able to learn the target function is the one that should be implemented''~\cite{NIPS2000_1925}.
We attempted to answer this overarching RQ by first delineating the baseline techniques that are used to evaluate new DL models and identifying what metrics are used in those comparisons. An evaluation that contains a comparison with a baseline approach, or even non-learning based solution, is important for determining the increased effectiveness of applying a new DL framework. \textit{RQ$_{4b}$} examines how DL-based approaches are impacting the automatization of SE tasks through measures of their effectiveness and in what ways these models generalize to practical scenarios, as generalizability of DL approaches in SE is vital for their usability. For instance, if a state-of-the-art DL approach is only applicable within a narrowly defined set of circumstances, then there may still be room for improvement.

\subsection{Analyzing Trends Across RQs}

Our last RQ encompasses all of the components of learning by examining the extent to which our analyzed body of literature properly accounts for and describes each element of learning. In essence, such an analysis explores the potential \textit{reproducibility} \& \textit{replicability} (or lack thereof) of DL applied to solve or automate various SE tasks. Therefore, our final RQ is formulated as follows:

\begin{tcolorbox}[enhanced,skin=enhancedmiddle,borderline={1mm}{0mm}{MidnightBlue}]
	\textit{\textbf{RQ$_5$:} What common factors contribute to the difficulty when reproducing or replicating DL4SE studies?}
\end{tcolorbox}

Our goal with this RQ is to identify common DL components which may be absent or underdescribed in our surveyed literature. In particular, we examined both the \textit{reproducibility} and \textit{replicability} of our primary studies as they relate to the sufficient presence or absence of descriptions of the elements of computational learning. Reproducibility is defined as the ability to take the exact same model with the exact same dataset from a primary study and produce the same results~\cite{acm-artifact-review}. Conversely, replicability is defined as the process of following the methodology described in the primary study such that a similar implementation can be generated and applied in the same or different contexts~\cite{acm-artifact-review}. The results of this RQ will assist the SE community in understanding what factors are being insufficiently described or omitted from approach descriptions, leading to difficulty in reproducing or replicating a given approach.

Lastly, given the analysis we perform as part of \textit{RQ$_5$} we derive a set of guidelines that both enumerate methods for properly applying DL techniques to SE tasks, and advocate for clear descriptions of the various different elements of learning. These guidelines start with the identification of the SE task to be studied and provide a step by step process through evaluating the new DL approach. Due to the high variability of DL approaches and the SE tasks they are applied to, we synthesized these steps to be flexible and generalizable. In addition, we provide checkpoints throughout this process that address common pitfalls or mistakes that future SE researchers can avoid when implementing these complex models. Our hope is that adhering to these guidelines will lead to future DL approaches in SE with an increased amount of clarity and replicability/reproducibility.

\subsection{Research Questions At-a-Glance}

We provide our full set of research questions below:
{\small
\begin{itemize}\setlength\itemsep{0.4em}
	
\item{\textit{\textbf{RQ$_1$: What types of SE tasks have been addressed by DL-based approaches?}}}

\item{\textit{\textbf{RQ$_2$: How are software artifacts being extracted, prepared, and used in DL-based approaches for SE tasks?}}}
	\begin{itemize}\setlength\itemsep{0.4em}
		\item{\textit{\textbf{RQ$_{2a}$:} What types of SE data are being used?}}
		\item{\textit{\textbf{RQ$_{2b}$:} How is this data being extracted and pre-processed into formats that are consumable by DL approaches?}}
		\item{\textit{\textbf{RQ$_{2c}$:} What type of exploratory data analysis is conducted to help inform model design and training?}}
	\end{itemize}

\item{\textit{\textbf{RQ$_3$: What deep learning models are used to support SE tasks?}}}
	\begin{itemize}\setlength\itemsep{0.4em}
		\item{\textit{\textbf{RQ$_{3a}$:} What types of model architectures are used to perform automated feature engineering of the data related to various SE tasks?}}
		\item{\textit{\textbf{RQ$_{3b}$:} What learning algorithms and training processes are used in order to optimize the models?}}
		\item{\textit{\textbf{RQ$_{3c}$:} What methods are employed to combat over- and under-fitting of the models?}}
	\end{itemize}

\item{\textit{\textbf{RQ$_4$: How well do DL tasks perform in supporting various SE tasks?}}}
	\begin{itemize}\setlength\itemsep{0.4em}
		\item{\textit{\textbf{RQ$_{4a}$:} What ``baseline'' techniques are used to evaluate DL models and what benchmarks are used for these comparisons?}}		
		\item{\textit{\textbf{RQ$_{4b}$:} How is the impact or automatization of DL approaches measured and in what way do these models promote generalizability?}}

	\end{itemize}

\item{\textit{\textbf{RQ$_5$: What common factors contribute to the difficulty when reproducing or replicating DL studies in SE?}}}

\end{itemize}}

\section{Methodology for Systematic Literature Review}
\label{sec:methodology}

We followed a systematic methodology to conduct our literature review in order to uphold the integrity of our analysis and provide a reproducible process. Within the field of SE, SLRs have become a standardized practice to communicate the past and present state of a particular research area. The most widely followed set of SLR standards were developed by Kitchenham \etal~\cite{Kitchenham2007}, thus we adopt these guidelines and procedures in our review process. As is described in Kitchenham's guidelines, we synthesized research questions (Sec. \ref{sec:rq_synthesis}) before beginning the search process. This aided in naturally guiding our SLR procedure and focused the search only on primary studies pertinent to these RQs. We then performed the following steps when conducting our review:

\begin{enumerate}
    \item Searching for primary studies;
    \item Filtering studies according to inclusion criteria;
    \item Probability Sampling;
    \item Non-Probability Sampling;
        \begin{enumerate}
            \item Snowballing and manual addition of studies;
            \item Applying exclusion criteria and performing alignment analysis;
        \end{enumerate}
    \item Data Extraction, Synthesis, and Taxonomy Derivation;
    \item Exploratory Data Analysis (or EDA)
\end{enumerate}

\begin{figure}[h]
	\centering
	\includegraphics[width=0.93\columnwidth]{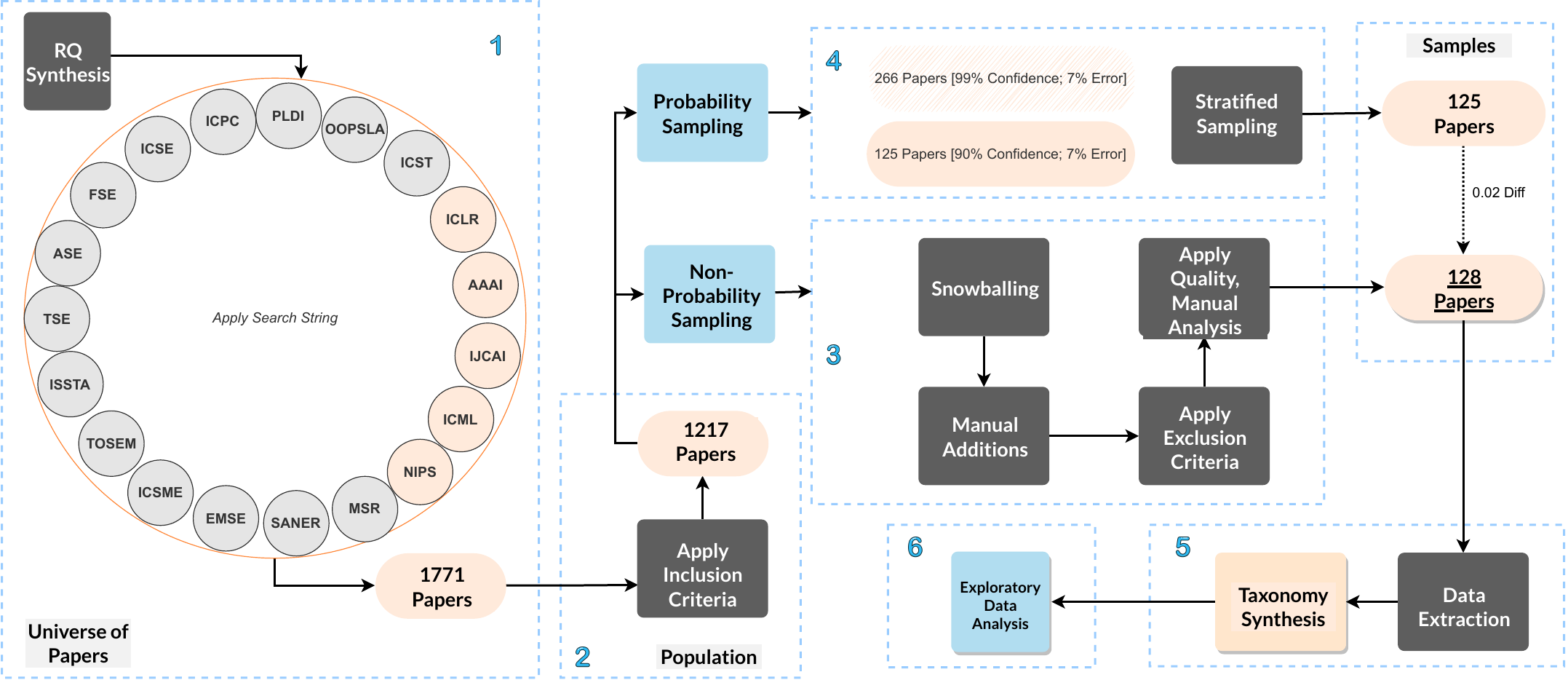}
	\vspace{-0.2cm}
	\caption{SLR Methodology. Numbers correspond to each review step.}	
	\label{fig:methodology}
	\vspace{0.2cm}
\end{figure}

A full enumeration of our methodology is illustrated in Figure~\ref{fig:methodology}.

\subsection{Searching for Primary Studies}

The first step in our methodology was to search for primary studies that would aid in our ability to properly address the synthesized RQs. We first began by focusing on the time period we would evaluate. We chose a 10 year period of January 1st, 2009 to June 1st, 2019, which corresponded roughly to when we started our search. We chose this 10 year period because we wanted to capture the time shortly before, during, and after the seminal AlexNet work by Krizhevsky \etal in 2012 \cite{NIPS2012_4824} that reinvigorated interest in neural networks. Next, we identified venues that would encompass our search. We selected the conference and journal venues which are generally recognized to be the top peer-reviewed and influential publication venues in the fields of SE and programming languages (PL), given in Table~\ref{tab:venues}. We opted to include PL venues as SE \& PL research often overlap, and we have observed DL approaches for code being published at PL venues. Additionally, we considered the top conferences dedicated to machine learning and deep learning in order to capture papers focused on the creation of a DL approach that also apply that approach to a SE task. These venues are also given in Table~\ref{tab:venues}. This selection of venues helps to ensure all relevant research was found and considered for the generation of our DL4SE taxonomy.

Once we established the venues to search, we developed a search string to query four electronic databases in order to identify appropriate studies from these venues. The databases that we considered were: IEEE Xplore, the ACM Digital Library, Springer Link, and DBLP. The development of the search string was primarily based upon key terms present in our formulated RQs. In order to ensure that our search string was robust enough to return as many relevant primary studies as possible, we empirically evaluated multiple search strings using a variety of derived terms. We compiled ten separate search strings with a variety of terms extracted from our synthesized RQs. Each string was evaluated for its ability to extract relevant studies for consideration. The results returned by each candidate string were cross-referenced with returned candidates from other search string combinations by one author to determine the most proficient search string (see Sec.~\ref{sec:threats} for more information). We eventually found that the string \noindent\textit{("Deep" OR "Learning" OR "Neural")} was best suited to our search for the following reasons: (i) this is the most general search string of our candidates and returned the widest variety of papers, and (ii) we believe that the generality of this term limited potential bias introduced by more focused terms. In summary, we decided to make a trade off by using a more general search term that was likely to have a higher rate of recall but resulted in more false positives for us to sort through.  %

The four major databases we considered provided a mechanism for advanced search that facilitates the accurate representation of our search string. Although there are nuances associated with each database, the method for gathering studies was consistent. The search string provided an initial filtering of the primary studies, then additional features of the advanced search allowed us to add discriminatory criteria. These additional features allowed us to limit our search results by year and venue. In addition to the four major databases, we also searched Google Scholar for papers through the \textit{Publish or Perish} software \cite{harzing}. To search our AI related venues, we augmented our search string with SE terms that would only return papers addressing a SE task (see our appendix for full term list \ref{appendix:A}). We gathered these terms from ICSE (the flagship academic conference in software engineering), as they represent a set of topics for technical papers generally agreed upon by the research community to represent topics of interest. We iterated through each term and appended it to our search string. The results of these searches were manually inspected for relevant papers to SE. \revision{After searching the databases with the specified search string, our initial results yielded \stringtotal potentially relevant studies.}

\subsection{Filtering Studies According to Inclusion Criteria}

In this step of the SLR, we defined the inclusion criteria that determined which primary studies would be included in our taxonomy. To an extent, part of our inclusion criteria was already used to extract the primary studies. The year and the venue were used in advanced searches to filter out extraneous research that our search string returned. However, after the search phase concluded, we filtered our results based on a more strict set of inclusion criteria. This ensured that the primary studies would be a useful addition to the taxonomy and would contain the necessary information to answer the RQs (\eg removing papers with search terms that appear only in reference lists). Our full set of inclusion and exclusion criteria considered is listed in our online appendix~\cite{watson_palacio_cooper_moran_poshyvanyk, cody_watson_2021_4768587}.

The primary studies gathered through the use of the search string, the snowballing method, or through manual addition (as enumerated in the following steps of the search) were subjected to the same criteria for inclusion. Three of the authors divided the works and labeled them for inclusion based upon a careful reading of the abstract and approach/methodology of the paper. A fourth author then manually reviewed each classification for potential errors. If the classifications between one of the three authors and the fourth were in disagreement, then all authors reviewed and discussed the work in question until a unanimous consensus was achieved. At the end of this phase of the SLR, we were left with 1,145 studies to be further considered for inclusion in our study, narrowed down from the 1,699 studies returned by our chosen search string. %

\subsubsection{Snowballing and Manual Addition of Studies}

Our next step was snowballing and manual inclusion of known studies. After extracting primary studies from the specified venues and subjecting those studies to our inclusion criteria, we performed snowballing on the resulting studies. Snowballing helped to ensure we were gathering other related works that were not included in our main search process. Our snowballing process looked at every reference within the studies passing our inclusion criteria and determined if any of those references also passed our inclusion criteria. Lastly, using our previous expertise in this research area, we manually added any related works that may have been missed according our knowledge of the field. We performed this manual addition for completeness of the survey and made note of which studies were manually added in our taxonomy. Only one paper which was published in arXiv in 2019 was manually added as we believe it is relevant to the DL4SE field and has been cited by a number of impactful, peer reviewed publications in well regarded SE venues \cite{Ahmed2021LearningLP,Karampatsis2020BigC,Tarlow2020LearningTF,Pradel2020TypeWriterNT}  %

\subsubsection{Exclusion Criteria and Alignment Analysis}

\begin{wrapfigure}{l}{0.50\textwidth} 
	\centering
	\includegraphics[width=0.50\textwidth]{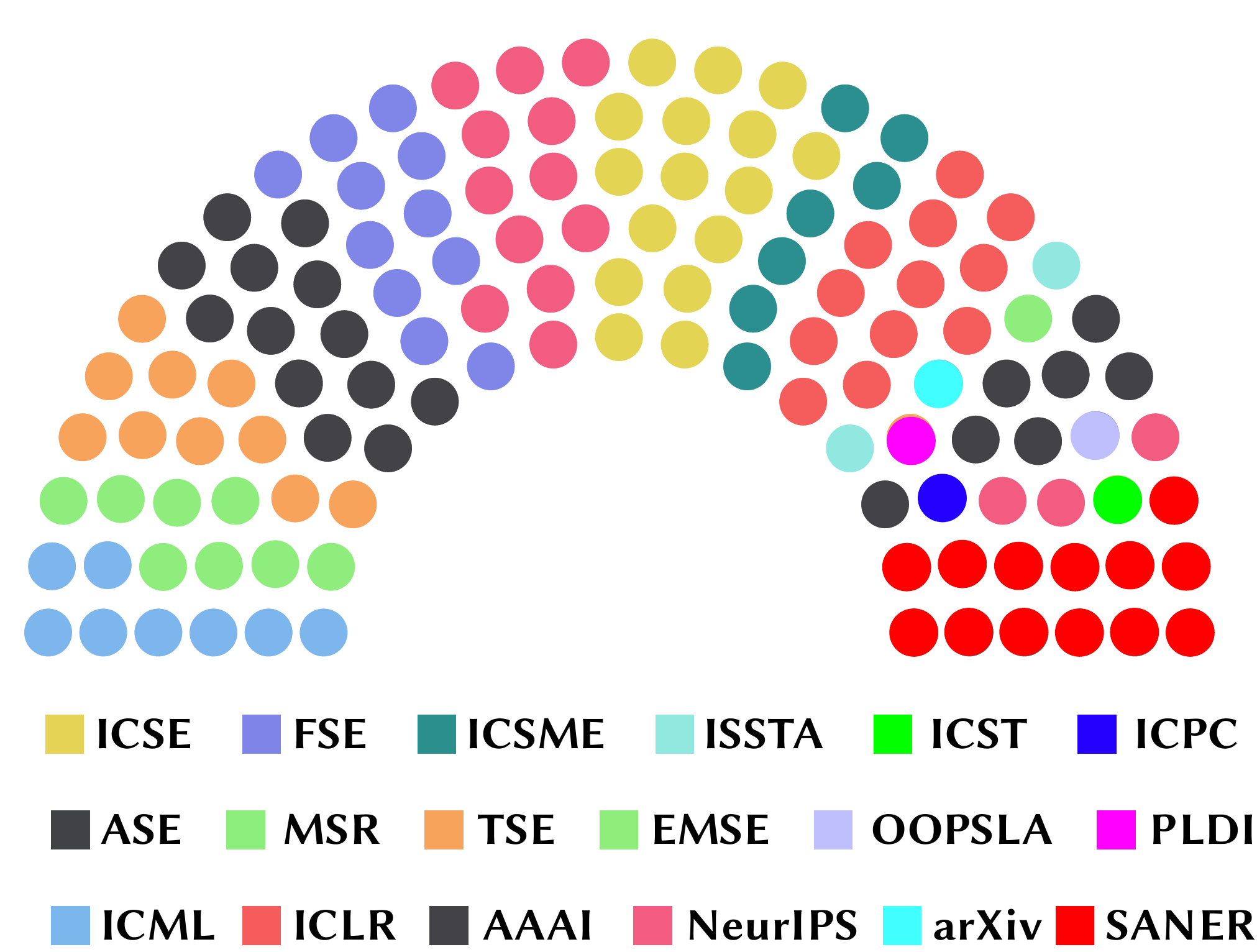}
	\vspace{-0.3cm}
	\caption{Venue distribution of DL4SE}	
	\label{fig:venues}
\end{wrapfigure}

Next, we applied our exclusion criteria to determine if there were any primary studies which were misaligned with our SLR and needed to be filtered. This involved a significant manual analysis to closely analyze how each study incorporated the use of DL when analyzing a SE task. Specifically, we excluded papers that fell outside of our SLR's timeline, did not solve some sort of software engineering task, were outside the scope of software engineering, were not published in one of our selected venues, did not implement a deep learning model, or only used a deep learning model as a baseline. In particular, we found many studies that only used a DL-based model as a baseline to compare against. We also found instances where DL was discussed as an idea or part of the future work of a study. We therefore excluded these works to ensure that every paper we analyzed both implemented and evaluated a DL approach to address a SE task. \revision{This process resulted in \includedpapers studies, all of which were included in the data extraction and taxonomy synthesis phases. Of these \includedpapers papers, 110 were captured by our initial search string, 17 were added as a result of snowballing, and one was added manually. In Figure~\ref{fig:venues} we illustrate the venues from which these \includedpapers studies originated.} Next, we provide empirical evidence that our filtering methodology according to our inclusion and exclusion criteria was able to capture a representative group of DL4SE papers.

\subsection{Comparison Between Probability Sampling of Papers and Non-Probability (Expert) Sampling Based on Exclusion Criteria}
\label{subsec:prob-sampling}

\begin{wraptable}{t}{3.8in}
\centering
\vspace{-0.5cm}
\caption{Venues Included After SLR Filtering}
\label{tab:venues}
\vspace{-0.3cm}
\scalebox{0.73}{%
\revision{
\begin{tabular}{llllllc}
\hline
\multicolumn{1}{c}{\multirow{2}{*}{\textbf{Topic}}} &
  \multicolumn{1}{c}{\multirow{2}{*}{\textbf{Venue}}} &
  \multicolumn{1}{c}{\multirow{2}{*}{\textbf{\begin{tabular}[c]{@{}c@{}}Search \\ String\end{tabular}}}} &
  \multicolumn{1}{c}{\multirow{2}{*}{\textbf{\begin{tabular}[c]{@{}c@{}}Inclusion\\ Criteria\\ Filtering\end{tabular}}}} &
  \multicolumn{2}{c}{\textbf{SLR Sampling}} &
  \multirow{2}{*}{\textbf{\begin{tabular}[c]{@{}c@{}}Relative\\ Difference\\ $\Delta_{.90}$\end{tabular}}} \\ \cline{5-6}
\multicolumn{1}{c}{} &
  \multicolumn{1}{c}{} &
  \multicolumn{1}{c}{} &
  \multicolumn{1}{c}{} &
  \multicolumn{1}{c}{\textbf{\begin{tabular}[c]{@{}c@{}}Random\\ Stratified\end{tabular}}} &
  \multicolumn{1}{c}{\textbf{\begin{tabular}[c]{@{}c@{}}Non-\\ Random\end{tabular}}} &
   \\ \hline
\multirow{12}{*}{\begin{tabular}[c]{@{}l@{}} SE \& PL \end{tabular}} &
  \textbf{ICSE} &
  455 &
  90 &
  10 &
  \multicolumn{1}{l|}{16} &
  0.39 \\
 &
  \textbf{FSE} &
  142 &
  130 &
  14 &
  \multicolumn{1}{l|}{11} &
  0.22 \\
 &
  \textbf{ASE} &
  162 &
  142 &
  15 &
  \multicolumn{1}{l|}{13} &
  0.15 \\
 &
  \textbf{OOPSLA} &
  36 &
  32 &
  3 &
  \multicolumn{1}{l|}{4} &
  0.13 \\
 &
  \textbf{ISSTA} &
  38 &
  25 &
  3 &
  \multicolumn{1}{l|}{2} &
  0.26 \\
 &
  \textbf{EMSE} &
  144 &
  59 &
  6 &
  \multicolumn{1}{l|}{1} &
  0.84 \\
 &
  \textbf{ICSME} &
  105 &
  91 &
  10 &
  \multicolumn{1}{l|}{7} &
  0.29 \\
 &
  \textbf{MSR} &
  52 &
  27 &
  3 &
  \multicolumn{1}{l|}{8} &
  0.27 \\
 &
  \textbf{TSE} &
  97 &
  97 &
  11 &
  \multicolumn{1}{l|}{10} &
  0.14 \\
 &
  \textbf{ICPC} &
  22 &
  15 &
  2 &
  \multicolumn{1}{l|}{2} &
  0.19 \\
 &
  \textbf{PLDI} &
  76 &
  76 &
  8 &
  \multicolumn{1}{l|}{1} &
  0.88 \\
 &
  \textbf{TOSEM} &
  24 &
  24 &
  3 &
  \multicolumn{1}{l|}{0} &
  1.00 \\ 
  &
  \textbf{SANER} &
  72 &
  72 &
  7 &
  \multicolumn{1}{l|}{13} &
  0.43 \\ 
  &
  \textbf{ICST} &
  43 &
  42 &
  5 &
  \multicolumn{1}{l|}{1} &
  0.78 \\ \hline
\multirow{6}{*}{\begin{tabular}[c]{@{}l@{}}AI/ML/DL \end{tabular}} &
  \textbf{AAAI} &
  103 &
  98 &
  11 &
  \multicolumn{1}{l|}{7} &
  0.34 \\
 &
  \textbf{ICLR} &
  45 &
  44 &
  5 &
  \multicolumn{1}{l|}{12} &
  0.60 \\
 &
  \textbf{ICML} &
  56 &
  55 &
  6 &
  \multicolumn{1}{l|}{8} &
  0.15 \\
 &
  \textbf{NIPS} &
  30 &
  29 &
  3 &
  \multicolumn{1}{l|}{12} &
  0.74 \\
 &
  \textbf{IJCAI} &
  69 &
  69 &
  7 &
  \multicolumn{1}{l|}{0} &
  1.00 \\ \hline
\multicolumn{1}{c}{\multirow{3}{*}{Statistics}} &
  \textbf{Total} &
  \stringtotal &
  \inclusiontotal &
  \textbf{\samplingtotal} &
  \multicolumn{1}{l|}{\textbf{\nonsamplingtotal}} &
  \diffpopulation \\
\multicolumn{1}{c}{} &
  \multicolumn{1}{c}{\textbf{$Avg \pm Std$}} &
  $93.2 \pm 97.3$ &
  $64.1 \pm 36.9$ &
  $6.6 \pm 3.8$ &
  \multicolumn{1}{l|}{$6.74 \pm 5.2$} &
  \textbf{\diffstratified} \\
\multicolumn{1}{c}{} &
  \textbf{Median} &
  69 &
  59 &
  6 &
  \multicolumn{1}{l|}{7} &
  \diffmedian \\ \hline
\end{tabular}
}
}\vspace{-0.3cm}
\end{wraptable} %

The goal of a given sampling strategy is to select a representative number of papers according to a given goal. Our goal in our SLR is to use non-probabilistic (also called expert) sampling methodology by applying our exclusion criteria in order to sample papers only relevant to our goals. Our exclusion criteria removed a significant number of papers from our initial list of papers that passed our inclusion criteria. However, in this subsection, we aim to empirically illustrate that our methodology is comparable to stratified sampling techniques that could be used to derive a random statistically significant sample from our set of \revision{\inclusiontotal papers}. As such we argue that our expert filtering methodology returned a \textit{representative} set of DL4SE papers.%

 We employed \textit{Stratified Sampling} to compute the required number of samples by conference to obtain a representative set of papers. This sampling was performed for a total universe of 18 venues: 12 venues in Software Engineering and 6 venues in AI. We used confidence intervals at 90\% and 99\% with a margin error of $\%7$. We then calculated the relative difference between the non-probability and the stratified sampling in order to quantify how distant our ``expert criteria'' filtering procedure (\ie applying our inclusion and exclusion criteria) was from a statistically significant sampling strategy. The relative difference is defined as $\Delta_{ci} = (r - s)/f(r-s)$ where $r$ is the number of samples obtained by stratification, $s$ is the number of samples obtained by expert criteria, and $ci$ is the Confidence Interval. We utilized  $f(r,s) = max(r,s)$ as the relative function.

The required conventional sampling at 90\% confidence is \samplingtotal papers and at 99\% is 266 papers for a population of \inclusiontotal papers (after applying inclusion criteria) from 18 venues. On the other hand, our expert criteria filtering reduced the population to a total of \nonsamplingtotal papers. Consequently, the relative stratified difference is $\widehat{\Delta_{.99}} =$ \diffstratified  and the relative population difference is $\widehat{\Delta_{.90}} = $ \diffpopulation  between expert sampling and probability sampling. In summary, the sampling strategy suggests that the target set of \nonsamplingtotal papers is statistically representative for a confidence interval at 90\% with a relative median difference  $\Tilde{x}_{.90} = $ \diffmedian. This sampling strategy remains true under the given preconditions of time period and conferences' quality.

\subsection{Data Extraction, Synthesis, and Taxonomy Derivation}

\subsubsection{Data Extraction}

\begin{wrapfigure}{R}{0.40\textwidth} %
    \centering
    \vspace{-2em}
    \includegraphics[width=0.40\textwidth]{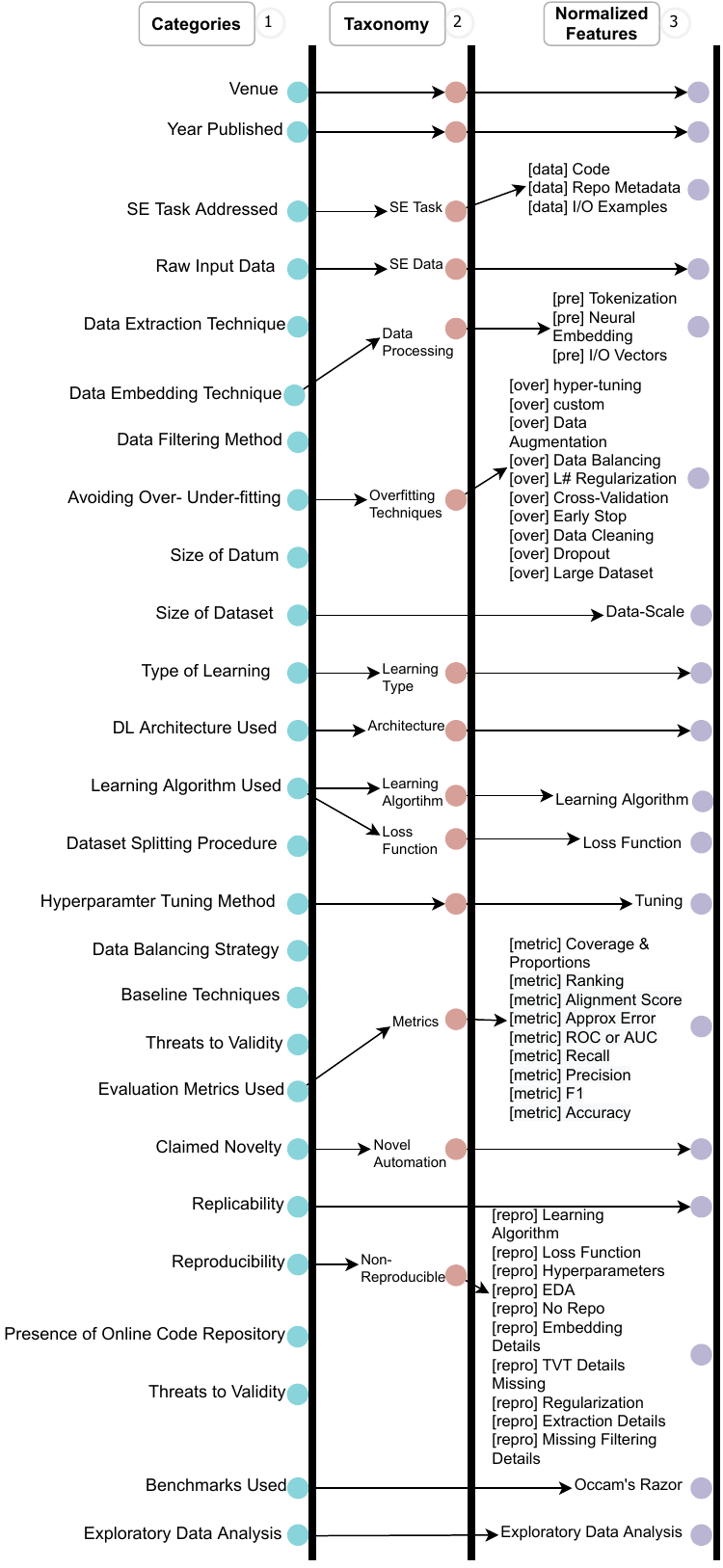}
    \caption{SLR Concept Map: 1. Categories for Data Extraction, 2. Taxonomy Derivation, and 3. Normalized Features from KDD}
    \label{fig:data-extraction}
\end{wrapfigure}

Our next step in the SLR methodology was to extract the necessary data from the primary studies. This step involved the development of an extraction form for identifying attributes of papers that correspond to our research questions. %
Each author was tasked with extracting data from a subset of the \includedpapers primary studies selected using the methodology outlined above. The results of the extraction phase were confirmed by a separate author to ensure that all important details were gathered and that papers were properly represented within the taxonomy. Each category of our extraction form is listed in the first column of Figure \ref{fig:data-extraction} and we provide our completed data extraction forms as part of our online appendix~\cite{watson_palacio_cooper_moran_poshyvanyk, cody_watson_2021_4768587}.

\subsubsection{Data Synthesis and Taxonomy Derivation}

In order to build our taxonomy on the use of DL approaches in SE research, we followed several complementary methodologies. The first of these was an open coding methodology consistent with constructivist grounded theory~\cite{Charmaz:groundedtheory}.  
Following the advice of recent work within the SE community \cite{Stol:ICSE16}, we stipulate our specific implementation of this type of grounded theory while discussing our deviations from the methods in the literature. We derived our implementation from the material discussed in \cite{Charmaz:groundedtheory} involving the following steps: (i) establishing a research problem and questions, (ii) data-collection and initial coding, and (iii) focused coding.  We excluded other steps described in \cite{Charmaz:groundedtheory}, such as memoing because we were focused on the construction of a taxonomy. The first two steps of this process were largely encompassed by the previously discussed methodological steps of the SLR, whereas the focused coding was used to derive the final category labels of our taxonomy that aided in answering each of our RQs. The initial coding was performed by one of the authors and then refined during the focused coding process by two others until an agreement was made among all three. The results of these coding steps formed our taxonomy. Finally, we normalized each topic in our taxonomy into detailed classes to perform a data analysis on them. This process is visualized in Figure \ref{fig:data-extraction}.

After taxonomy construction, we make use of a combination of \textit{descriptive statistics} and our \textit{exploratory data analysis} in order to synthesize and describe our results to each RQ. Our use of descriptive statistics enumerates various trends we identify in our taxonomy, whereas our exploratory data analysis investigates statistical relationships between various aspects of our taxonomy treated as ``features''. We further describe this analysis methodology in the following subsection.
The answers to our RQs naturally define a holistic taxonomy that researchers can use to determine what types of SE tasks can be better studied using certain DL-based approaches, as well as look for future applications of DL to model complex software artifacts. We oriented our discussion of the results of our RQs to provide an understanding about the process of applying a DL-based approach in SE.

\subsection{Exploratory Data Analysis}

In order to gain a more holistic view of our taxonomy constructed based on the extracted data, we performed an analysis to determine the types of relationships between the various taxonomy categories, which we use to form a set of \textit{features} used to conduct an exploratory data analysis (EDA). %
The mined associations between different paper attributes allowed us to provide more complete and holistic answers to several research questions, thus supporting the conclusions of our SLR. We utilized a data mining pipeline and set of statistical processes in order to uncover hidden relationships among the different extracted attributes of our primary studies.

Our data mining and analysis process was inspired by classical Knowledge Discovery in Databases, or KDD~\cite{Fayyad_Piatetsky-Shapiro_Smyth_1996}. The KDD process extracts knowledge from the data gathered during our extraction process that was then converted into a database format, and involved five stages:

\textbf{1. Selection}. This stage was encompassed by the data extraction process explained in the beginning of this section. After collecting all the papers and extracting the relevant paper attributes, we organized the data into 55 explicit features or attributes extracted from the original data taken from the primary studies. A complete list of these features is provided in our online repository~\cite{watson_palacio_cooper_moran_poshyvanyk, cody_watson_2021_4768587}. %

\textbf{2. Preprocessing}. We applied a preprocessing technique that transformed and normalized categories into nominal features as depicted by the concept map in Figure \ref{fig:data-extraction}. For instance, the category ``Avoiding Over-UnderFitting'' was transformed into ``Over-fitting Techniques'' in the taxonomy; then, such taxonomy was normalized into 10 different features (e.g., ``Tokenization'', ``Neural Embedding'', ``I/O Vectors'', ``Hyper-Tuning'', and so on). Similarly, this preprocessing pipeline was applied to other categories such as ``SE Task Addressed'' and ``Evaluation Metrics Used''. This normalization into more detailed classes contributes to a more holistic understanding of the papers via data mining methods.

\textbf{3. Data Mining.} In this stage, we employed two well-established data mining tasks to better understand our data: Correlation Discovery and Association Rule Learning. We oriented our KDD process to uncover hidden relationships on the normalized features. We provide further details related to this stage in Section~\ref{correlation-discovery} and \ref{association-rule-learning}.

\textbf{4. Interpretation/Evaluation} We used the knowledge discovered to automatically find patterns in our papers that resemble actionable knowledge. This actionable knowledge was generated by applying a reasoning process on the data mining outcomes. Although this reasoning process produced a support analysis formally presented in our online appendix \cite{watson_palacio_cooper_moran_poshyvanyk, cody_watson_2021_4768587}, each summary of the outcome of a RQ contains a brief description of the results of the exploratory data analysis. 
We used RapidMiner \cite{rapidminer_2020} to conduct the data analysis, and the procedures and pipelines for using this tool are published in our companion online repository~\cite{watson_palacio_cooper_moran_poshyvanyk, cody_watson_2021_4768587}. Before carrying out any mining tasks, we decided to address a basic descriptive statistics procedure for each feature. These statistics exhibit basic relative frequencies, counts, and quality metrics. We then utilized four quality metrics: ID-ness, Stability, Missing, and Text-ness. ID-ness measures to what extent a feature resembles an ID, where as the title is the only feature with a high ID-ness value. Stability measures to what extent a feature is constant (or has the same value). Missing refers to the number of missing values. Finally, Text-ness measures to what extent a feature contains free-text. The results for each feature were calculated using the RapidMiner interface.

\begin{wrapfigure}{r}{0.50\textwidth} %
    \centering
    \includegraphics[width=0.50\textwidth]{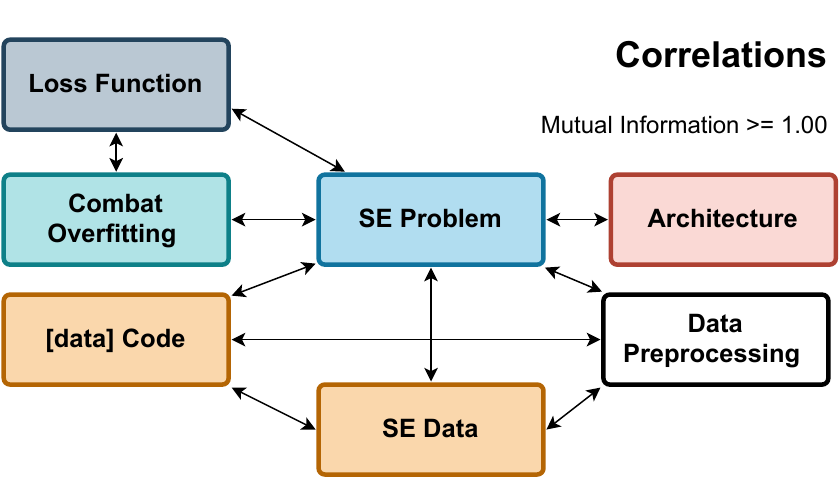}
    \vspace{-2em}
    \caption{SLR Feature Correlations}
    \label{fig:correlations}
\end{wrapfigure}

\subsubsection{Correlation Discovery} \label{correlation-discovery} Due to the nominal nature of the SLR features, we were unable to infer classic statistical correlations for our data (\ie Pearson's correlation). However, we adapted an operator based on attributes information dependencies. This operator is known as mutual information \cite{10.5555/971143}, which is measured in bits $B$. Similar to covariance or Pearson's correlation, we were able to represent the outcomes of the operator in a confusion matrix. Figure~\ref{fig:correlations} depicts feature correlations greater than $1.0B$ from the confusion matrix.

The mutual information measures to what extent one feature "knows" about another. High mutual information values represent less uncertainty; therefore, we built arguments such as whether the deep learning architecture used on a paper is more predictable given a particular SE task or the reported architectures within the papers are mutually dependent upon the SE task. The difference between correlation and association rules depends on the granularity level of the data (\eg paper, feature, or class). The correlation procedure was performed at the feature level, while the association rule learning was performed at the class or category level.

\subsubsection{Association Rule Learning}\label{association-rule-learning} \begin{wrapfigure}{l}{0.50\textwidth} %
    \centering
    \vspace{0.0em}
    \includegraphics[width=0.50\textwidth]{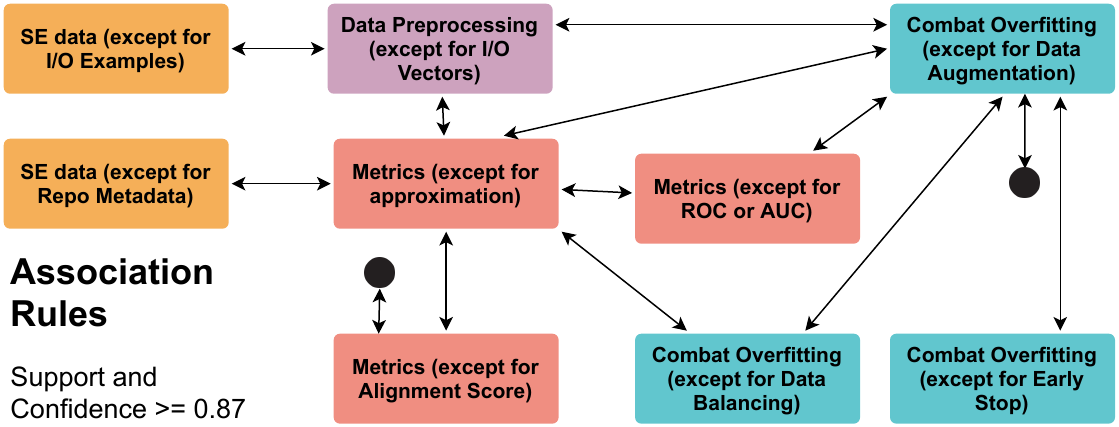}
    \vspace{-1.5em}
    \caption{SLR Association Rules}
    \label{fig:rules}
\end{wrapfigure}
For generating the association rules, we employed the classic Frequent Pattern (FP)-Growth algorithm in order to observe associations among variables. 
FP-Growth computes frequently co-occurring items in our transactional database with a minimum support of 0.95 and a minimum number of itemsets of 100 (Table ~\ref{tab:components} depicts the parameter values employed for FP-Grow and Association Rule Mining components).
These items comprise each class per feature. For instance, the feature Loss Function exhibited a set of classes (or items) such as NCE, Max Log Likelihood, Cross-Entropy, MSE, Hinge Loss, N/A-Loss, and so on. The main premise of the FP-Growth can be summarized as: if an item set is frequent (\ie ~{MSE, RNN}),  then all of its item subsets are also frequent (\ie ~{MSE} and {RNN}).

Once the FP-Growth generated the item sets (\eg ~{MSE, Hinge Loss}), the algorithm analyzed the item sets support in continuous scans of the database (DB). Such support measures the occurrences of the item set in the database. The FP-Growth scans the database, which brings about a FP-tree data structure. We recursively mined the FP-tree data structure to extract frequent item sets \cite{Han00miningfrequent}. Nonetheless, the association rule learning requires more than the FP-tree extraction.

\begin{wraptable}{r}{2.2in}
\centering
\vspace{-1em}
\caption{RapidMiner Parameters}
\vspace{-1em}
\label{tab:components}
\scalebox{0.73}{%

\begin{tabular}{cll}
\hline
\textbf{Component}                                       & \multicolumn{1}{c}{\textbf{Parameter}} & \multicolumn{1}{c}{\textbf{Value}} \\ \hline
\multirow{5}{*}{FP-Grow} & min support           & 0.95   \\
                         & min items per itemset & 1      \\
                         & max items per itemset & 0      \\
                         & max \# of itemsets    & $10^6$ \\
                         & min \# of itemsets    & 100    \\ \hline
\begin{tabular}[c]{@{}c@{}}Association\\ Rules\end{tabular} & min confidence                         & 0.8                                \\ \hline
\end{tabular}
}\vspace{-1em}
\end{wraptable}
An association rule serves as an if-then (or premise-conclusion) statement based on frequent item set patterns. Let's observe the following rule mined from our dataset: \textit{Software Engineering data, excluding Input/Output Vectors, determine the type of data preprocessing performed by the authors with a support of 0.88 and a confidence of 0.99}. We observe the association rule has an antecedent (\ie the feature {SE data, but for I/O vectors}) and a consequent (\ie the feature {Data Preprocessing}). These relationships are mined from item sets that usually have a high support and confidence. The confidence is a measure of the number of times that premise-conclusion statement is found to be true. We kept association rules that have both confidence and support greater than 0.8.

It is possible that Association Rule Mining can lead to spurious correlations. In order to avoid this, we organized the rules into an interconnected net of premises/conclusions based upon our formulated RQs in order to find explanations around techniques and methodologies reported on the papers. Figure~\ref{fig:rules} depicts such an interconnected net with the highest support and confidence.   Any non-logical rule was disregarded as well as rules that possessed a lower support. Non-logical rules are association rules where the premise and conclusion are easy to falsify. Generally, if we decrease the minimum support parameter, non-logical association rules might arise.

We discuss the results of this data analysis process where appropriate as they relate to our RQs in the following sections of this SLR. The full results of our exploratory data analysis and source code of our entire EDA process can be found in our supplemental material~\cite{watson_palacio_cooper_moran_poshyvanyk, cody_watson_2021_4768587}.

\section{RQ$_1$: What types of SE tasks have been addressed by DL-based approaches?}
\label{sec:rq1}

\begin{figure*}[t]
	\centering
	\includegraphics[width=1\textwidth]{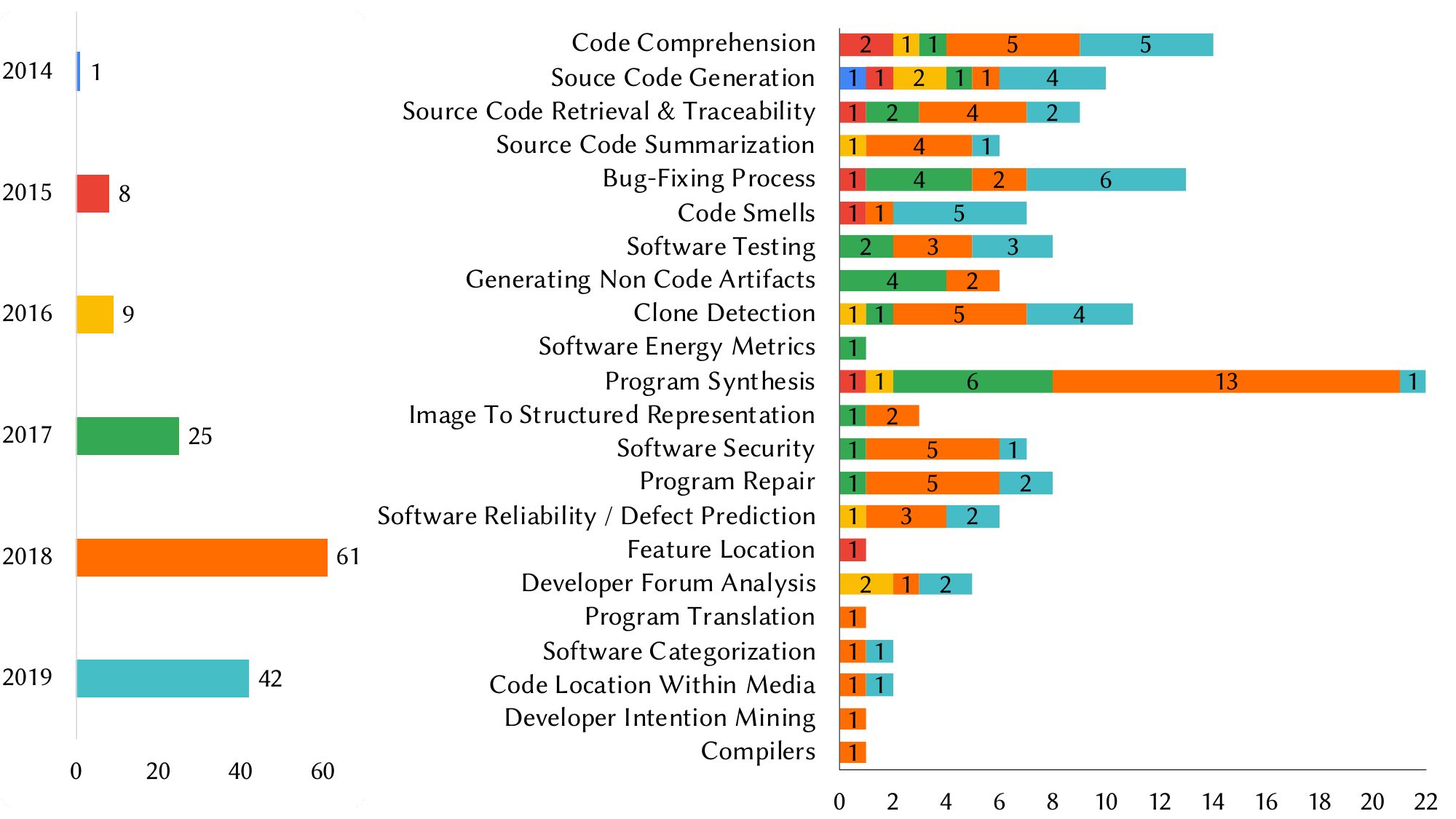}
	\vspace{-0.4cm}
	\caption{Papers published per year according to SE task. Note that a single paper can be associated with multiple SE Tasks.}	
	\label{fig:sebreakdown}
\end{figure*}

This RQ explores and quantifies the different applications of DL approaches to help improve or automate various SE tasks. Out of the \includedpapers papers we analyzed for this SLR, we identified \includedsetasks ~separate SE tasks where a DL-based approach had been applied. Figure \ref{fig:sebreakdown} provides a visual breakdown of how many SE tasks we found within these \includedpapers primary studies across a 10 year period. %
Unsurprisingly, there was very little work done between the years of 2009 and 2014. However, even after the popularization of DL techniques brought about by results achieved by approaches such as  AlexNet~\cite{NIPS2012_4824}, it took the SE community nearly $\approx3$ years to begin exploring the use of DL techniques for SE tasks. This also coincides with the offering and popularization of DL frameworks such as PyTorch and TensorFlow. The first SE tasks to use a DL technique were those of  \textit{Source Code Generation}, \textit{Code Comprehension}, \textit{Source Code Retrieval \& Traceability}, \textit{Bug-Fixing Processes}, and \textit{Feature Location}. Each of these tasks uses source code as their primary form of data. Source code served as a natural starting point for applications of DL techniques given the interest in large scale mining of open source software repositories in the research community, and relative availability of large-scale code datasets to researchers. Access to a large amount of data and a well-defined task is important for DL4SE, since in order for DL to have an effective application two main components are needed: i) a large-scale dataset of data to support the training of multi-parameter models capable of extracting complex representations and ii) a task that can be addressed with some type of predictable target. One of the major benefits of DL implementations is the ability for automatic feature extraction. However, this requires data associated with the predicted target variable.

It was not until 2017 that DL was used extensively in solving SE tasks as shown in Figure \ref{fig:sebreakdown}, with a large increase in the number of papers, more than doubling over the previous year from 9 to 25. During this period, the set of target SE tasks also grew to become more diverse, including tasks such as \textit{Code Smell Detection}, \textit{Software Security}, and \textit{Software Categorization}. However, there are three main SE tasks that have remained the most active across the years: \textit{Code Comprehension}, \textit{Source Code Retrieval \& Traceability}, and \textit{Program Synthesis}. %
The most popular of the three being Program Synthesis, composing a total of 22 papers out of the \includedpapers we collected. We suspect that a variety of reasons contribute to the multiple applications of DL in program synthesis. First and foremost, is that the accessibility to data is more prevalent. Program synthesis is trained using a set of input-output examples. This makes for accessible, high quality training data, since one can train the DL model to generate a program, given a set of existing or curated specifications. The second largest reason is the clear mapping between training examples and a target programs. Given that it has proven difficult to engineer effective features that are capable to predict or infer programs, DL techniques are able to take advantage of the structured nature of this problem and extracting effective hierarchical representations. We display the full taxonomy in Table \ref{tab:setask}, which associates the cited primary study paired with its respective SE task.

\begin{table}[t]
\centering
\caption{SE Task Taxonomy}
\vspace{-0.3cm}
\label{tab:setask}
\footnotesize
\begin{tabular}{|c|c|}
\hline
\multicolumn{1}{|c|}{\textbf{SE Task}}                      	& \multicolumn{1}{c|}{\textbf{Papers}} \\ \hline
\multicolumn{1}{|c|}{Code Comprehension}                    	& \multicolumn{1}{c|}{\cite{Levy2017, BenNun2018, Piech2015, Hellendoorn2018, Le2018a, allamanis2018learning, Allamanis2015, codereviewlearn, 10.5555/3015812.3016002, DBLP:conf/iclr/YinNABG19, 10.1109/ICSE.2019.00084, 8812062, 10.1145/3196321.3196334, guo:saner19}} \\ \hline
\multicolumn{1}{|c|}{Souce Code Generation}                 	& \multicolumn{1}{c|}{\cite{Karampatsis2019, Hellendoorn2017, White2015a, Gu2016, Chen2018, sun2019, 10.1145/2594291.2594321, gao:saner19, nguyen:saner19, cvitkovic:icml19}} \\ \hline
\multicolumn{1}{|c|}{Source Code Retrieval \& Traceability} 	& \multicolumn{1}{c|}{\cite{Chen2018a, Guo2017, Gu2018, Deshmukh2017, Lam2015, Chen2019, allamanis2018learning, 10.1145/3196398.3196408, xie:saner19}} \\ \hline
\multicolumn{1}{|c|}{Source Code Summarization}					& \multicolumn{1}{c|}{\cite{Chen2018a, Wan2018, Allamanis2016, 10.1109/ICSE.2019.00087, 10.1145/3196398.3196408, 10.1145/3196321.3196334}} \\ \hline
\multicolumn{1}{|c|}{Bug-Fixing Process} 						& \multicolumn{1}{c|}{\cite{Lee2017, Murali2017, Deshmukh2017, Lam2015, gupta2019, 10.5555/3298239.3298436, DBLP:conf/iclr/YinNABG19, Liu2018d, 10.1109/ICSE.2019.00021, 10.1145/3276517, 10.1145/3360588, huo:tse19, zhang:saner19}} \\ \hline
\multicolumn{1}{|c|}{Code Smells} 								& \multicolumn{1}{c|}{\cite{Liu2018, Allamanis2015, 8812134, 10.1109/ICSE.2019.00021, 8811893, thaller:saner19, fakhoury:saner19}} \\ \hline
\multicolumn{1}{|c|}{Software Testing} 							& \multicolumn{1}{c|}{\cite{Liu2017, Godefroid2017, Cummins2018, Si2018, DBLP:conf/aaai/LiuLPW19, Zhang2018a, 8730177}} \\ \hline
\multicolumn{1}{|c|}{Non Code Related Software Artifacts} 		& \multicolumn{1}{c|}{\cite{Jiang2017, Schroeder2017, Choetkiertikul2018, Choetkiertikul2017, Choetkiertikul2019, Huang2018, 10.1145/3213846.3213876}} \\ \hline
\multicolumn{1}{|c|}{Clone Detection} 							& \multicolumn{1}{c|}{\cite{Li2017, White2016, Saini2018, Gao2018, Liu2018c, Tufano2018a, 10.1145/3236024.3236068, 10.1109/ICPC.2019.00021, 8812062, perez:msr19, buch:saner19}} \\ \hline
\multicolumn{1}{|c|}{Software Energy Metrics} 					& \multicolumn{1}{c|}{\cite{Romansky2017}} \\ \hline
\multicolumn{1}{|c|}{Program Synthesis} 						& \multicolumn{1}{p{5.5cm}|}{\cite{Gaunt2017, Cai2017, Zhang2018, Devlin2017, Sun2018, DBLP:conf/iclr/MuraliQCJ18, DBLP:journals/corr/ReedF15, Liu2016, Balog2016, DBLP:conf/icml/DevlinUBSMK17, Bunel2018, Vijayakumar2018, DBLP:conf/iclr/ChenLS18, Hellendoorn2018a, Arabshahi2018, Zohar2018, Shin2018, Ellis2018, Ellis2018a, Liang2018, DBLP:conf/iclr/ParisottoMS0ZK17, 10.1145/3360594}} \\ \hline
\multicolumn{1}{|c|}{Image To Structured Representation} 		& \multicolumn{1}{c|}{\cite{10.5555/3305381.3305483, Chen2018, Moran2018}} \\ \hline
\multicolumn{1}{|c|}{Software Security} 						& \multicolumn{1}{c|}{\cite{Han2017, Zhao2018, Chen2018d, Gao2018, Harer2018, Dam2018, 8812083}} \\ \hline
\multicolumn{1}{|c|}{Program Repair} 							& \multicolumn{1}{c|}{\cite{Bhatia2018, Wang2017, Harer2018, Tufano2018b, Liu2018d, gupta2019, 10.5555/3298239.3298436, white:saner19}} \\ \hline
\multicolumn{1}{|c|}{Software Reliability / Defect Prediction} 	& \multicolumn{1}{c|}{\cite{Wang2016, Wen2018, 8502853, dam:msr19, hoang:msr19, liu:saner19}} \\ \hline
\multicolumn{1}{|c|}{Feature Location} 							& \multicolumn{1}{c|}{\cite{Corley2015}} \\ \hline
\multicolumn{1}{|c|}{Developer Forum Analysis} 					& \multicolumn{1}{c|}{\cite{Chen2016, Xu2016, Lin2018, wang:msr19, guo:saner19}} \\ \hline
\multicolumn{1}{|c|}{Program Translation} 						& \multicolumn{1}{c|}{\cite{Chen2018e}} \\ \hline
\multicolumn{1}{|c|}{Software Categorization} 					& \multicolumn{1}{c|}{\cite{DBLP:conf/aaai/BuiJY18, bui:saner19}} \\ \hline
\multicolumn{1}{|c|}{Compilers} 				            	& \multicolumn{1}{c|}{\cite{katz:saner19}} \\ \hline
\multicolumn{1}{|c|}{Code Location Within Media} 				& \multicolumn{1}{c|}{\cite{Ott2018, 8811922}} \\ \hline
\multicolumn{1}{|c|}{Developer Intention Mining} 				& \multicolumn{1}{c|}{\cite{Huang2018}} \\ \hline
\multicolumn{1}{|c|}{Software Resource Control} 				& \multicolumn{1}{c|}{\cite{8811988, 10.1145/3213846.3213876}} \\ \hline
\end{tabular}%
\end{table}

\subsection{Results of Exploratory Data Analysis}

In performing our exploratory data analysis, we derived two primary findings. \revision{First, it is clear that SE researchers apply DL techniques to a diverse set of tasks, as 70\% of our derived SE task distribution was comprised of distinct topics that were evenly distributed ($\approx$ 3-5\%). Our second finding is that the SE task was the most informative feature we extracted ($\approx 4.04B$), meaning that it provides the highest level of discriminatory power in predicting the other features (\eg elements of learning) related to a given study. In particular, we found that SE tasks had strong correlations to data (1.51B), the loss function used (1.14B) and the architecture employed (1.11B). This suggests that there are DL framework components that are better suited to address specific SE tasks, as authors clearly chose to implement certain combinations of DL techniques associated with different SE tasks.} %
For example, we found that SE tasks such as program synthesis, source code generation and program repair were highly correlated with the preprocessing technique of tokenization. Additionally, we discovered that the SE tasks of source code retrieval and source code traceability were highly correlated with the preprocessing technique of neural embeddings. When we analyzed the type of architecture employed, we found that code comprehension, prediction of software repository metadata, and program repair were highly correlated with both recurrent neural networks and encoder-decoder models. When discussing some of the less popular architectures we found that clone detection was highly correlated with siamese deep learning models and security related tasks were highly correlated with deep reinforcement learning models. %
Throughout the remaining RQs, we look to expand upon the associations we find to better assist software engineers in choosing the most appropriate DL components to build their approach. 

\subsection{Opportunities for Future Work}

Although the applications of DL-based approaches to SE related tasks is apparent, there are many research areas of interest in the SE community as shown in ICSE'20's topics of interest\footnote{https://conf.researchr.org/track/icse-2020/icse-2020-papers\#Call-for-Papers} that DL has not been used for. Many of these topics have no \textit{readily apparent} applicability for a DL-based solution. Still, some potentially interesting topics that seem well suited or positioned to benefit from DL-based techniques have yet to be explored by the research community or are underrepresented. Topics of this unexplored nature include software performance, program analysis, cloud computing, human aspects of SE, parallel programming,  feature location, defect prediction and many others. Some possible reasons certain SE tasks have yet to gain traction in DL-related research is likely due to the following:

\begin{itemize}
    \item There is a lack of available, ``clean'' data in order to support such DL techniques;
    \item The problem itself is not well-defined, such that a DL model would struggle to be effectively trained and optimized;
    \item No current architectures are adequately fit to be used on the available data.
\end{itemize}

We believe that one possible research interest could be in the application of new DL models toward commonly explored SE tasks. For example, a DL model that is gaining popularity is the use of transformers, such as BERT, to represent sequential data \cite{DBLP:journals/corr/abs-1810-04805}. It is possible that models such as this could be applied to topics related to clone detection and program repair. There is sufficient exploration in the use of DL within these topics to determine if these new architectures would be able to create a more meaningful representation of the data when compared to their predecessors.

\mybox{\textbf{Summary of Results for RQ$_{1}$}:}{gray!60}{gray!20}{Researchers have applied DL techniques to a diverse set of tasks, wherein \textit{program synthesis}, \textit{code comprehension}, and \textit{source code generation} are the most prevalent. The SE task targeted by a given study is typically a strong indicator of the other details regarding the other components of learning, suggesting that certain SE tasks are better suited to certain combinations of these components. Our associative rule learning analysis showed a strong correlation amongst SE task, data type, preprocessing techniques, loss function used and DL architecture implemented, indicating that the SE task is a strong signifier of what other details about the approach are present. While there has been a recent wealth of work on DL4SE, there are still underrepresented topics that should be considered by the research community, including different topics in software testing and program analysis.}

\vspace{-1em}
\section{RQ$_2$: How are software artifacts being extracted, prepared, and used in DL-based approaches for SE tasks?}
\label{sec:rq2}

In this research question, we analyze the type of SE data that is modeled by DL approaches applied to various SE tasks. Our aim with this analysis is to understand the various types of SE data used, how the data is extracted or preprocessed, the scale of the data being modeled, and the type of learning applied given the data. All four of these points are crucial to effectively understanding how a given approach is able to model specific software artifacts. These points also ground our discussion regarding future research and potential improvements of data filtering and preprocessing in order to improve the effectiveness of DL approaches; as in many cases, data must be cleaned or filtered in such a manner that can limit the ability of a DL technique to model a desired relationship. %

\subsection{RQ$_{2A}$: What types of SE data are being used?}
\label{rq2a}

\begin{figure*}[t]
	\centering
	\includegraphics[width=1\textwidth]{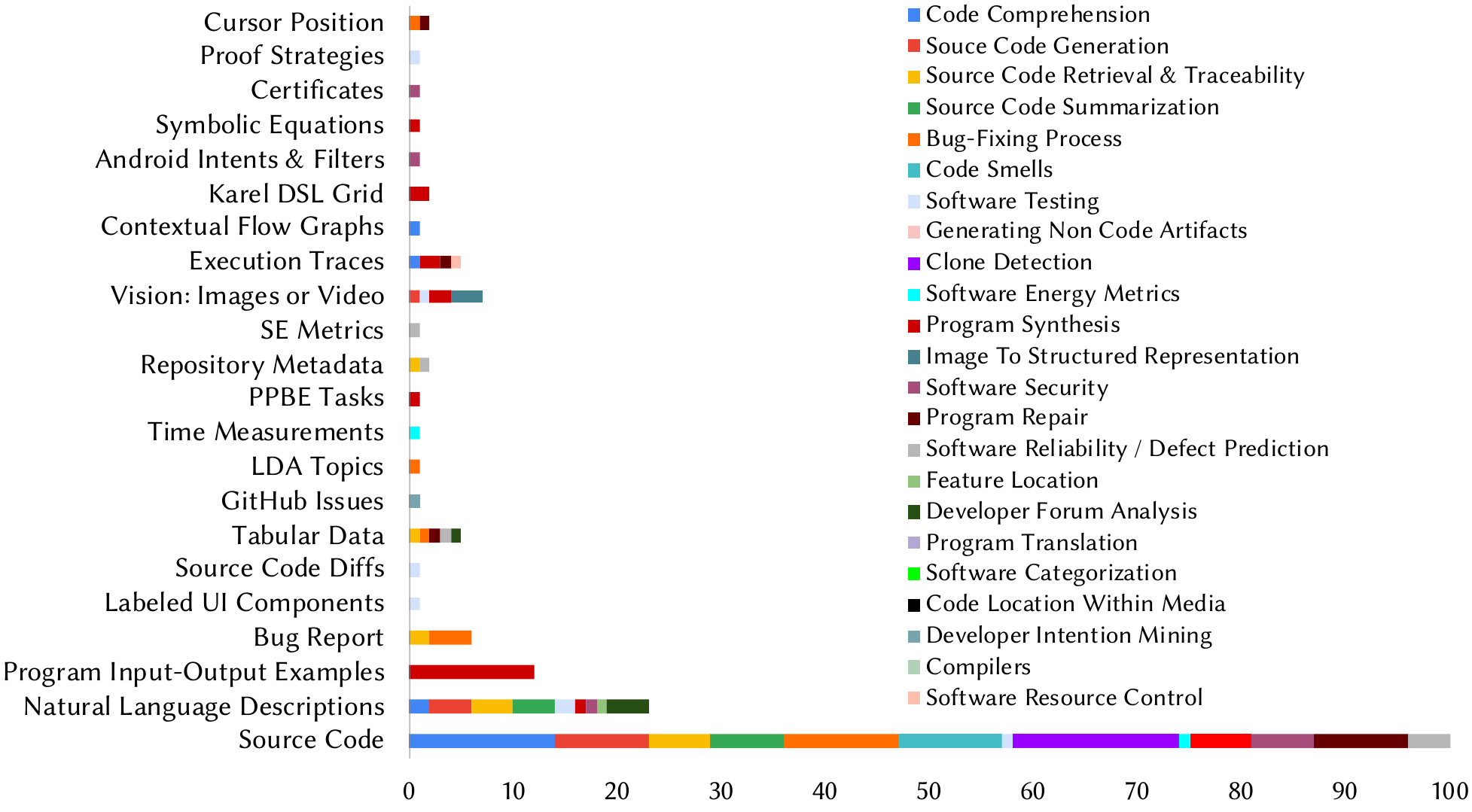}
	\vspace{-2em}
	\caption{Data Types With SE Tasks}	
	\label{fig:datatypes}
	\vspace{0.1cm}
\end{figure*}

To analyze the types of data being used in DL-based approaches, we provide a high level classification, along with descriptive statistics, as to why some types of data were used for particular tasks. Figure~\ref{fig:datatypes} provides an illustration of different data types used in conjunction with different SE tasks. We found that overwhelmingly the most common type of data being used is \textit{source code}, albeit at a range of different granularities. In our identified studies, we found that source code is used at the binary level, code snippet level, method level, class level and project level. Source code is a popular data construct for DL-approaches for a number of reasons. First, source code is plentiful and can be found in a number of different online repositories. This availability helps to appease the "data-hungry" nature of DL techniques in order to learn an effective, representative target function. Second, a majority of SE tasks revolve around the generation, understanding, analysis, and documentation of source code, making it a popular artifact in the (deep) learning process.

In total we identified 152 uses of data in our DL4SE papers, 86 of them are attributed to source code, wherein certain studies utilized more than one type of data for their approach (\eg source code \& natural language). Although source code is the primary type of data that DL models attempt to learn from, we found that the type of data used is heavily dependent on the SE task. Thus, the SE tasks that focus on the comprehension, generation, summarization, and testing of source code will frequently use source code at various granularities as part of their dataset. However, there are many SE tasks that address problems where source code may not be the most representative type of data from which to learn. As an example, the SE task of program synthesis primarily uses 12/27 input and output examples to comprise their dataset. This type of data incorporates attributes, which more strongly correlates to desired relationship for the model to learn. This pattern continues for SE tasks that can learn from textual data, software artifacts, and repository metadata. 

Although we identified a variety of different data types used, the data must be accessible in large quantities in order to extract and learn the relevant target hypothesis. With an increase in the number of online source code repositories, opportunities for mining them have increased. This partially explains the dramatic increase in DL papers addressing SE tasks, which was a trend discussed in Section \ref{sec:rq1}. Interestingly, the growth of online repositories does not only increase the amount of accessible source code but also other types of software artifacts. For example, analyzing SE tasks such as software security, bug fixing, etc. have only recently been addressed, in part due to the increasing amount of accessible online repository data/metadata. We anticipate that this trend continues as new datasets are gathered and processed for use in DL techniques.

When exploring online repositories, one attribute of the data that can be problematic for the use of some DL approaches and should be considered is that the data is frequently ``unlabeled'', meaning that there is no inherent target to associate with this data. For unsupervised learning techniques this is not problematic, however, for supervised learning techniques this type of data is not usable without first establishing a target label for the data. The use of code mined from online software repositories as unlabeled data was explored early in 2015 by White \etal~\cite{White2015a}. Our findings illustrate that source code is used as an unlabeled, yet useful set of data across several SE tasks including: the localization of buggy files through bug reports \cite{Lam2015}, specification mining \cite{Le2018a}, mining fix patterns for bug violations \cite{Liu2018d}, identification of source code clones \cite{Tufano2018a} and repairing vulnerable software \cite{Harer2018}. %
Additionally, we also found that researchers mined unlabeled data and manually labeled it in order to apply a particular DL architecture. We found examples of this when generating accurate method and class names \cite{Allamanis2015}, learning proper API sequences \cite{Gu2016} and source code summarization \cite{Allamanis2016}. %
With the increase in the number of online repositories and the variety of unlabeled data within those repositories, we expected the number of DL-based approaches analyzing software artifacts to increase.

\subsubsection{Results of Exploratory Data Analysis}

Our exploratory data analysis also highlighted these points. Our analysis demonstrated that the main data employed in DL4SE papers are input-output (I/O) examples, source code, natural language, repository metadata and visual data. \revision{In fact, these categories of data types comprised $\approx78.3\%$ of the distribution we found in this SLR. In conjunction with this finding, execution traces and bug reports represent $\approx5.8\%$ of the distribution of data. The remaining $\approx15\%$ is comprised of a variety of different data types.}

\subsubsection{Opportunities for Future Work} Throughout our study, it was clear that certain types of SE artifacts have not yet been analyzed using DL-based approaches. Specifically, we identified that software requirements, software dependencies, and software licenses have not been mined or studied using DL architectures. This suggests that these and other underrepresented data types not included in Fig.~\ref{fig:datatypes} could be ripe for future DL applications. In addition to analyzing some of the underrepresented data types, we believe there is an opportunity to combine multiple data types for a more complete representation of the SE data. The ability to consider and combine multiple representations should provide a more meaningful representation of the SE task and could lead to better deep learning solutions. Additionally, there are a number of different data types that could potentially be applied to different SE tasks. For example, contextual flow graphs (CFGs) have only been considered when looking at code comprehension task but information contained in CFGs could prove to be useful for tasks such as code summarization and program repair among others.
\vspace{-1em}
\mybox{\textbf{Summary of Results for RQ$_{2A}$}:}{gray!60}{gray!20}{Our analysis found that researchers have explored a variety of different types of SE data in conjunction with DL techniques, with the main types of data utilized being \revision{\textit{source code} ($\approx59.78\%$), \textit{I/O examples} ($\approx7.89\%$), \textit{natural language} ($\approx12.16\%$), \textit{repository metadata} ($\approx7.24\%$), and \textit{visual data} ($\approx5.26\%$)}. These data types were often tightly associated with a given SE task.}

\vspace{-1em}
\subsection{\textit{RQ$_{2B}$}: How is this data being extracted and pre-processed into formats that are consumable by DL approaches?}
\label{rq2b}

In Section \ref{rq2a}, we analyzed the types of data that were being used to model complex relationships between software artifacts and the target output function as it relates to a specific SE task. In RQ$_{2b}$, we examine the mechanism behind the extraction and preprocessing of this data. %
Typically, DL models are not amenable to raw data mined from the online repositories. Rather, the data is subjected to a preparation and formatting process before given to the DL model. For example, image data is downsampled and scaled, and text is preprocessed in order to preserve relevant words. This is an extremely important step for those applying DL to SE, since the process can dramatically affect the performance of the model and the resulting target hypothesis. For example, some primary studies represent source code in an abstracted format. This abstraction will inherently transform the features of the code and can affect the applicability of these features to different tasks. Such transformations could also address issues often associated with certain types of DL architectures, such as the ``open vocabulary'' problem in DL-based language models wherein the vocabulary size of a model is untenable due to source code's unrestricted vocabulary. Conversely, a limitation of these types of transformation processes is that it removes some complex, hierarchical features from the dataset, which can limit the model's ability to accurately predict the target in some cases.

\begin{figure*}[t]
	\centering
	\vspace{-0.5cm}
	\includegraphics[width=\textwidth]{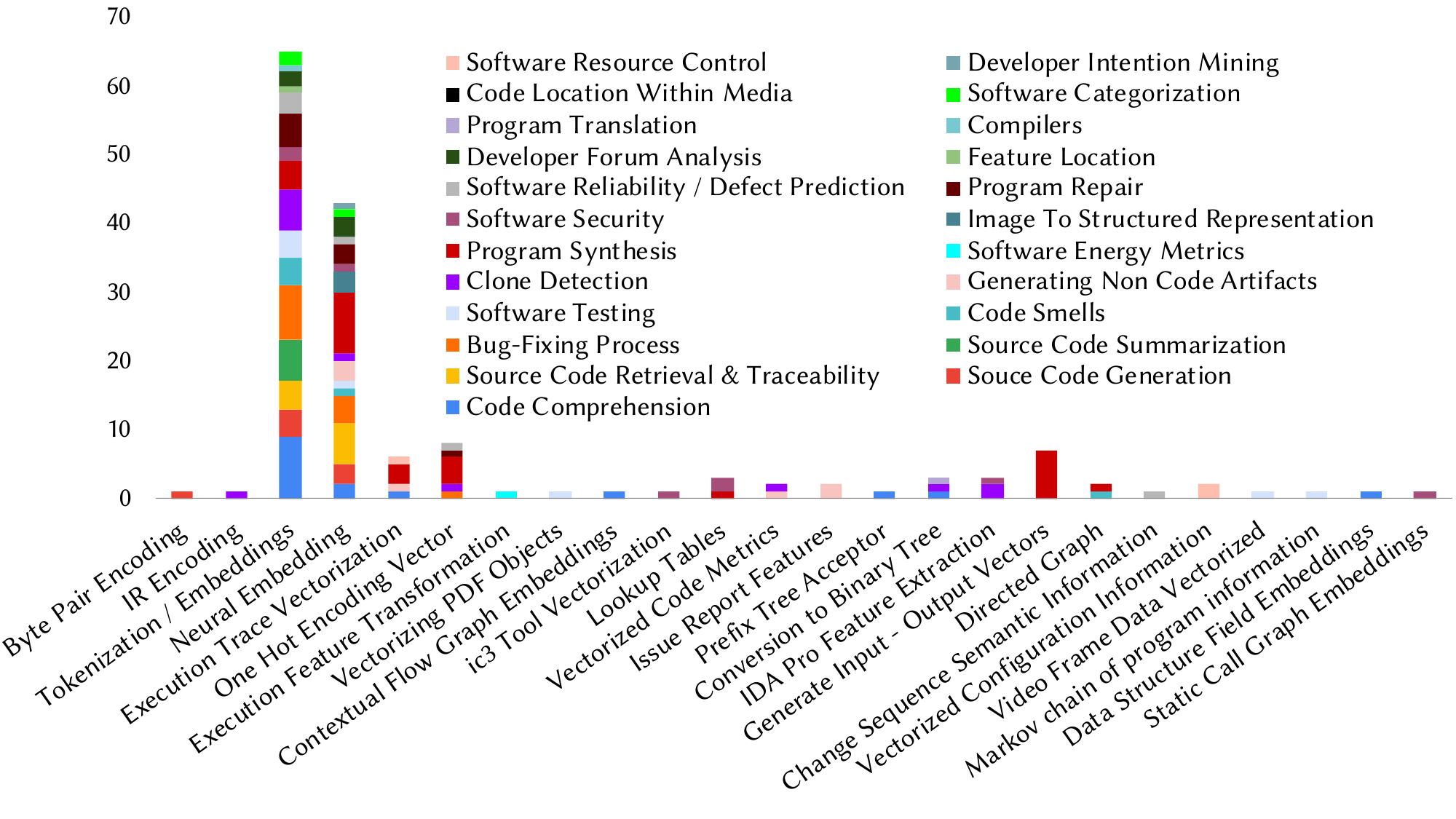}
	\vspace{-0.9cm}
	\caption{Preprocessing Techniques by SE Task}	
	\label{fig:preprocessing}
	\vspace{0.1cm}
\end{figure*}

In light of the importance of preprocessing in DL-based approaches, we synthesized a taxonomy of preprocessing and formatting techniques. We also looked to analyze the relationships between the SE tasks, types of data and the preprocessing steps taken. %
A breakdown of the preprocessing and formatting techniques used in the primary studies we analyzed can be found in Figure \ref{fig:preprocessing}. It is important to understand that Figure \ref{fig:preprocessing} shows the \textit{general} preprocessing and formatting techniques according to SE tasks. However, within the same preprocessing technique, there exist nuances across different studies.  There is rarely a standard method for the ability to preprocess the data in such a way that will fit the desired DL model. Therefore, we suggest interested readers refer to the primary studies for more details pertaining to a specific preprocessing or formatting technique of interest. However, there are some dominant techniques researchers use in order to prepare their data to be analyzed by a DL model. Specifically, the use of tokenization and neural embeddings are popular for a variety of different tasks. This was expected, given the prevalence of source code as a primary type of data used. Tokenization of that source code is an effective process for preparing different granularities of code to be fed into a DL architecture.

Even more popular than the use of tokenization was the use of neural embeddings ($23.85\%$ of the distribution). This technique uses canonical machine learning techniques or other DL architectures to process the data, meaning that the output from these other models are then used as input to an additional DL architecture. We found that Word2Vec and recurrent neural networks (RNNs) were the most popular types of models for the preprocessing of data. %
Source code, natural language, and other repository metadata often have a sequential nature, which can be captured by these techniques. In both cases, the outputs of these models are a series of vectorized numerical values, which capture features about the data they represent. These vectors do not require much additional manipulation before they can be used again as input data to another DL model.

Although tokenization and neural embeddings are the most popular type of preprocessing techniques, there are many more that are required for a different types of data or SE tasks. The ability to accurately represent the data to the DL model is what provides training data for the algorithm to learn from. Any alteration to this process can result in the learning algorithm focusing on different features, leading to an entirely different final hypothesis.

\subsubsection{Results of Exploratory Data Analysis} \revision{Our exploratory data analysis discovered that the steps taken in preprocessing were strongly dependent on the type of data employed ($1.19B$, according to the mutual information measure). %
This is intuitive, as the data and preprocessing techniques are inherently linked.} We also found that the SE task and venue had high discriminatory power in determining the type of data preprocessing (given larger self-information values from the correlation analysis). This indicates that certain SE Tasks and preprocessing combinations are more likely for different SE venues. %

\subsubsection{Opportunities for Future Work} In the midst of our analysis it became clear that preprocessing techniques were often uniquely specialized to particular DL architectures. However, there are many commonalities amongst these preprocessing \textit{pipelines} that could be standardized dependent upon the type of data being represented. For example, tokenization and neural embeddings were used extensively in a variety of tasks. It would be advantageous to standardize the data processing pipelines for these approaches related to source code, input/output examples, or textual data found in repository metadata. Additionally, exploring less popular preprocessing techniques such as directed graphs, lookup tables, and execution trace vectors for different SE tasks could lead to results which are orthogonally beneficial to those found using a more common techniques. %

We also noticed many preprocessing techniques have not been formalized based on the SE task and DL architecture used. Additionally, it is not just the use of certain techniques, but the details used in applying those techniques that may lead to an increase or decrease in performance for DL models. For example, tokenization is a broad technique that takes sequential data and separates them into tokens. However, details such as token filtering, removal of stop words and vocabulary size can have a large affect on the process. We believe a more in-depth look at how preprocessing techniques affect the quality of DL solutions for SE tasks would be beneficial for future applications.

\mybox{\textbf{Summary of Results for RQ$_{2B}$}:}{gray!60}{gray!20}{\revision{Our analysis found that, while a number of different data preprocessing techniques have been utilized, \textit{tokenization} (\tokcount) and \textit{neural embeddings} (\neuralcount) are by far the two most prevalent.} We also found that data-preprocessing is tightly coupled to the DL model utilized, and that the SE task and publication venue were often strongly associated with specific types of preprocessing techniques. %
This coupling of paper attributes is likely due to the often tight coupling between preprocessing and allowable inputs of various DL architectures.} %

\subsection{\textit{RQ$_{2C}$}: What type of exploratory data analysis is conducted to help inform model design and training?}
\label{rq2c}

For RQ$_{2C}$, we analyzed our primary studies to determine if precautions were taken to limit the number of confounding factors that may exist should researchers have not properly analyzed their datasets prior to training a DL model.  Primarily, we were interested in whether \textit{sampling bias} or \textit{data snooping} are present in studies that apply DL models to SE tasks. We found that when DL is applied to an SE task, there were many instances %
where no methods were discussed %
to protect against these confounding variables. This can severely limit the conclusions that can be drawn from the learned target hypothesis. Through analyzing the various strategies used to combat sampling bias and data snooping we found that many SE tasks working with source code (\ie source code generation, source code completion and bug fixing tasks) did not check for \textit{duplicated examples} within their training and testing sets %
We also found examples of sampling bias in studies pulling code from multiple projects on GitHub. This included SE tasks that analyzed source code clones and software test methods. Extracting data from projects that are only of a particular size, developer, or rating could all lead to biased results that affect study claims. These findings corroborate the findings presented by Allamanis et al. \cite{81206469}, which described the adverse affects of duplicated examples within the training and testing sets of machine learning approaches. This is not only a problem for the SE task of clone detection, but it may also impact the findings of previous DL techniques trained on large-scale code datasets. %

\textit{Sampling Bias} occurs when data is collected from a dataset in such a way that it does not accurately represent the data's distribution due to a non-random selection process~\cite{sampling}. The repercussions of sampling bias normally result in a target hypothesis that is only useful for a small subset of the actual distribution of the data. When sampling bias is not appropriately considered authors can unknowingly make claims about their predicted target function that overestimate the actual capabilities of their model. One effective mechanism for mitigating sampling bias is to limit or carefully consider the types of filtering criteria applied to a dataset. %

\textit{Data Snooping} occurs when data is reused for the purpose of training and testing the model~\cite{sage}. This means that the data the model was trained on is also incorporated into the testing and evaluation of the model's success. Since the model has already seen the data before, it is unclear whether the model has actually learned the target hypothesis or simply tuned itself to effectively reproduce the results for previously seen data points (overfitting). The overall result of data snooping is an overinflated accuracy reported for a given model. Therefore, it is important that software engineers apply methods to reduce data snooping in order to protect the integrity of their results. This can be done through a simple exploratory data analysis or the removal of duplicate data values within the training and testing sets.

In order to evaluate the extent to which methods that mitigate sampling bias and data snooping were used in our primary studies, we noted instances where the authors conducted some form of exploratory data analysis before training. Ideally, we hoped to see that authors performed exploratory data analysis in order to identify if the dataset that has been extracted is a good representation of the target distribution they are attempting to model.  Exploratory data analysis can also provide insight into whether data snooping occurred.

There was a noticeable number of primary studies, \revision{$\approx90\%$ and $\approx82\%$} ,that did not mention or implement any methods to mitigate sampling bias or data snooping respectively. However, it should be noted that we focused our analysis on primary studies that included any sign of exploratory analysis on their dataset. This exploratory analysis ranged from basic statistical analyses to an in-depth study into the distribution of the dataset in order to mitigate these confounding factors. A majority of the methods we discovered that addressed sampling bias included putting the fewest number of limitations on the filtering of source data as possible. For example, when mining GitHub for potential project candidates, studies would normally restrict the results based on necessary attributes of the data. %
Additionally, some studies included evaluations on entire projects that were not considered during training. %
A smaller number of studies %
ensured that data gathered was balanced by class. The studies that did attempt to combat data snooping \revision{($\approx16\%$)} did so by ensuring the removal of duplicates within their dataset. This means that every input-output pairing used for training was unique in both the training and testing sets. Lastly, we found that 41 of the primary studies explicitly mention the use of a validation set to find optimal hyperparameter configurations. %
The use of this validation set helps to ensure exclusivity in the model evaluation on the test set, bolstering potential generalizability. %

\subsubsection{Results of Exploratory Data Analysis} \revision{Our data analysis showed that nearly $\approx52\%$ of the papers report exploratory techniques for analyzing their datasets. However, even among those papers that did perform some sort of exploratory analysis, most of the techniques utilized were relatively simple (e.g., checking for duplicate data entries). We posit that SE researchers could benefit dramatically from an expanded repertoire of EDA techniques. Furthermore, we detected that approximately $2\%$ of the studies \textit{directly} addressed sampling bias and data snooping. On the one hand, nearly $8\%$ of the inspected approaches are explicitly susceptible to sampling bias. On the other hand, $16\%$ of the data employed are explicitly susceptible to data snooping. In conclusion, we are unable to determine in $90\%$ of the studies whether data snooping or sampling bias were controlled for. Similarly, in $82\%$ of the studies, we were unable to determine the potential for data snooping due to a limited amount of associated descriptions} %

\subsubsection{Opportunities for Future Work} Our primary objective within this RQ was to bring attention to the oversight of confounding factors that affect the treatment and preparation of data for DL models. DL models are heavily dependent on data to properly model a target hypothesis, thus we hope to encourage future work that carefully considers the makeup of their dataset and their methods for extraction. In particular, there are research opportunities related to standardizing the process for preventing sampling bias and data snooping. To our knowledge there has been no attempt at the generation of guidelines related to how one might prevent sampling bias when considering SE data. Likewise, there is potential for the creation of analytical tools to help evaluate the likelihood of data snooping in DL based approaches, or automated tools for combating bias by automatically separating data into a training, validation and testing set that removes duplicate examples between each set. It is also important to note that as time has progressed, the primary studies we analyzed generally included more details about their data exploration process. We hope to see this trend continue as DL is used to address SE tasks. %

\mybox{\textbf{Summary of Results for RQ$_{2C}$}:}{gray!60}{gray!20}{\revision{Our analysis found that as many as 1/2 of our analyzed studies do not perform any type of exploratory data analysis in order to combat confounding factors such as data snooping or bias.} Some of the most popular mitigation techniques employed were a detailed analysis of the duplicates found within the training and testing set, the use of a validation set to prevent tuning the parameters of the model to best perform on a test set, and the removal of restrictive data filtering in order to extract datasets that are as diverse as possible.} %

\section{RQ$_3$: What Deep Learning Models are Used to Support SE Tasks?}
\label{sec:rq3}

In Section \ref{sec:rq2} we investigated how different types of SE data were used, preprocessed, and analyzed for use in DL techniques. In this section, we shift our focus to the two key components of DL models: the \textit{architecture} and the \textit{learning algorithm}. The type of architecture selected for use in a DL application reveals key aspects of the types of features that researchers hope to model for a given SE task. Thus, we aim to empirically determine if certain architectures pair with specific SE tasks. Additionally, we aim to explore the diversity of the types of architectures used across different SE tasks and whether or not idiosyncrasies between architectures might be important when considering the specific SE task at hand. We also examined how various architectures are used in conjunction with different learning or optimization algorithms. Specifically, we aimed to create a taxonomy of different learning algorithms and determine if there was a correlation between the DL architectures, the learning algorithms and the SE tasks. %

\subsection{\textit{RQ$_{3A}$}: What types of model architectures are used to perform automated feature engineering of the data related to various SE tasks?} 
\label{rq3a}

 \begin{figure*}[t]
	\centering
	\vspace{0.2cm}
	\includegraphics[width=\columnwidth]{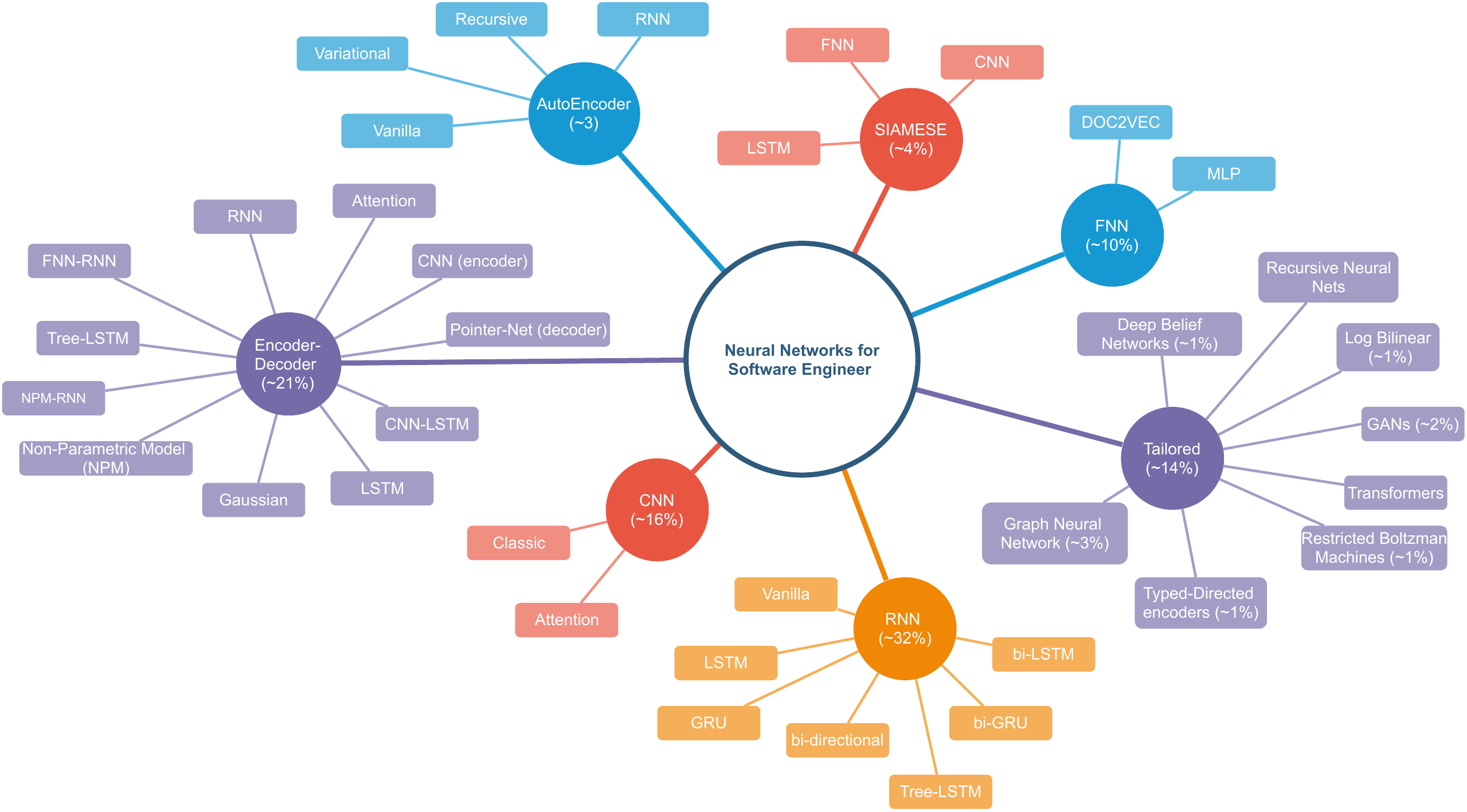}
	\vspace{-0.4cm}
	\caption{DL Model Taxonomy \& Type Distribution} %
	\label{fig:dlmodel}
	\vspace{0.1cm}
\end{figure*} 

In this section, we discuss the different types of DL models software engineers are using to address SE tasks. \revision{Figure~\ref{fig:dlmodel} illustrates the various different DL architecture types that we extracted from our selected studies. We observe seven major architecture types: \textit{Recurrent Neural Networks (RNNs)} (\rnncount), \textit{Encoder-Decoder Models} (\encdeccount), \textit{Convolutional Neural Networks (CNNs)} (\cnncount), \textit{Feed-Forward Neural Networks (FNNs)} (\fnncount), \textit{AutoEncoders} (\autocount), \textit{Siamese Neural Networks} (\siamcount), as well as a subset of other custom, highly tailored architectures.} We observe an additional level of diversity within each of these different types of architectures with Encoder-Decoder models illustrating the most diversity, followed by RNNs and the tailored techniques. The diversity of Encoder-Decoder models is expected, as this type of model is, in essence, a combination of two distinct model types, and is therefore extensible to a range of different combinations and hence architectural variations. The variance in RNNs is also logical. RNNs excel in modeling sequential data since the architecture is formulated such that a weight matrix is responsible for representing the features between the sequence of inputs \cite{Goodfellow2016}, making them suitable to source code. Given that one of the most popular SE data types is source code which is inherently sequential data, the varied application of RNNS is expected. We also observe a number of architectures, such as Graph Neural Networks, that are specifically tailored for given SE tasks. For instances, graph-based neural networks have been adapted to better model the complex \textit{structural} relationships between code entities.

\begin{figure*}[t]
	\centering
	\vspace{0.2cm}
	\includegraphics[width=\columnwidth]{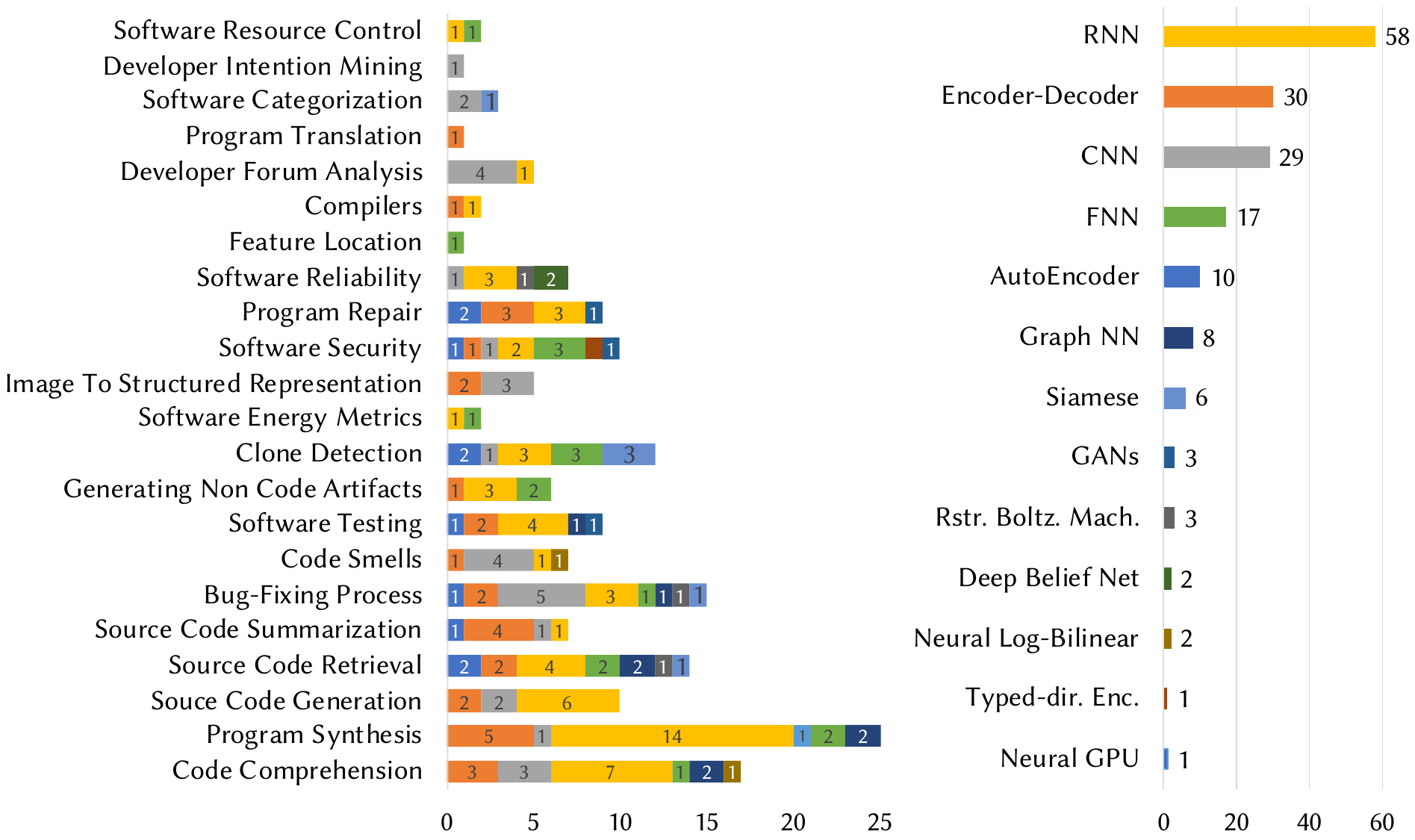}
	\vspace{-0.4cm}
	\caption{DL Architectures by the Task}	
	\label{fig:architecture}
	\vspace{0.1cm}
\end{figure*}

Figure~\ref{fig:architecture} delineates the prevalence of various different types of architectures according to the SE tasks to which they are applied. The popularity of our identified techniques closely mirrors their diversity. Examining this data, we find that RNNs are the most prevalent architectures, followed by Encoder-Decoder models, CNNs, and FNNs. The prevalence of RNNs is not surprising given the prevalence of source code as a utilized data type, as discussed above.  The flexibility of Encoder-Decoder models is also expected as they excel at understanding and ``translating'' between parallel sets of sequential data, which is a common scenario in SE data (\eg code and natural language). %
The encoder's responsibility is to translate the raw data into a latent representation that the decoder is capable of understanding and decoding into the target. %
Therefore, since neural embeddings were such a popular preprocessing technique for data formatting and preparation, it aligns with the high prevalence of the encoder-decoder DL architecture. CNNs serve as the most popular architectures for processing visual data, such as images, and hence are popular for visual SE data.

In addition to the prevalence, we observed certain trends between the DL architecture utilized and the corresponding SE task, as illustrated in Figure~\ref{fig:architecture}. %
As expected, most of the SE tasks having to do with source code generation, analysis, synthesis, traceability, and repair make use of RNNs and encoder-decoder models. %
Likewise, SE tasks involving the use of images or media data have CNNs commonly applied to them. 

We also observed some pertinent trends related to some of the less popular types of DL architectures, including: siamese networks, deep belief networks, GNNs and auto-encoders. While these architectures have only been applied to a few tasks it is important to note that they have only recently gained prominence and become accessible outside of ML/DL research communities. %
It is possible that such architectures can highlight orthogonal features of SE data that other architectures may struggle to observe. For example, the use of GNNs may better capture the structure or control flow of code or possibly the transition to different mobile screens within a mobile application. There may also be an opportunity for the use of Siamese networks in software categorization, as they have been shown to classify data into unique classes accurately based only on a few examples \cite{Saini2018}.
One notable absence from our identified architecture types is \textit{deep reinforcement learning}, signaling its relative lack of adoption within the SE community. Deep reinforcement learning excels at modeling decision-making tasks. One could argue that deep reinforcement learning is highly applicable to a range of SE tasks that can be modeled as decisions frequently made by developers. This is a fairly open area of DL in SE that has not been sufficiently explored. The only type of SE task that had an application of Reinforcement Learning was related to program verification. In this paper the authors propose an approach that constructs the structural external memory representation of a program. They then train the approach to make multi-step decisions with an autoregressive model, querying the external memory using an attention mechanism. Then, the decision at each step generates subparts of the loop invariant~\cite{Si2018}. %

In addition to the discussion around the DL architectures and their relations to particular SE tasks, it is also important to understand trends related to the \textit{explicit} and \textit{implicit} features extracted from these different architectures. As we discussed in Section \ref{rq2b} (RQ$_{2B}$), it is common for data to be fed into DL models only after being subjected to certain preprocessing steps. However, in supervised learning, once that data has been preprocessed, the DL model automatically extracts implicit features from the preprocessed data in order to associate those features with a label or classification. In unsupervised learning, the model extracts implicit features from the preprocessed data and groups similar datum together as a form of classification. We refer to the preprocessing steps as highlighting certain explicit features, since these steps frequently perform dimensionality reduction while maintaining important features. In our analysis we found the most common techniques for highlighting explicit features to be tokenization, abstraction, neural embeddings and vectorizing latent representations. These techniques attempt to highlight explicit features that are uniquely tailored to the data being analyzed. Once the data is fed into the model itself, the model is responsible for extracting implicit features to learn a relationship between the input data and target function. %
The extraction of explicit and implicit features dramatically impacts a DL model's ability to represent a target function, which can be used to predict unobserved data points. 

 \begin{figure*}[t]
	\centering
	\vspace{-0.2cm}
	\includegraphics[width=\columnwidth]{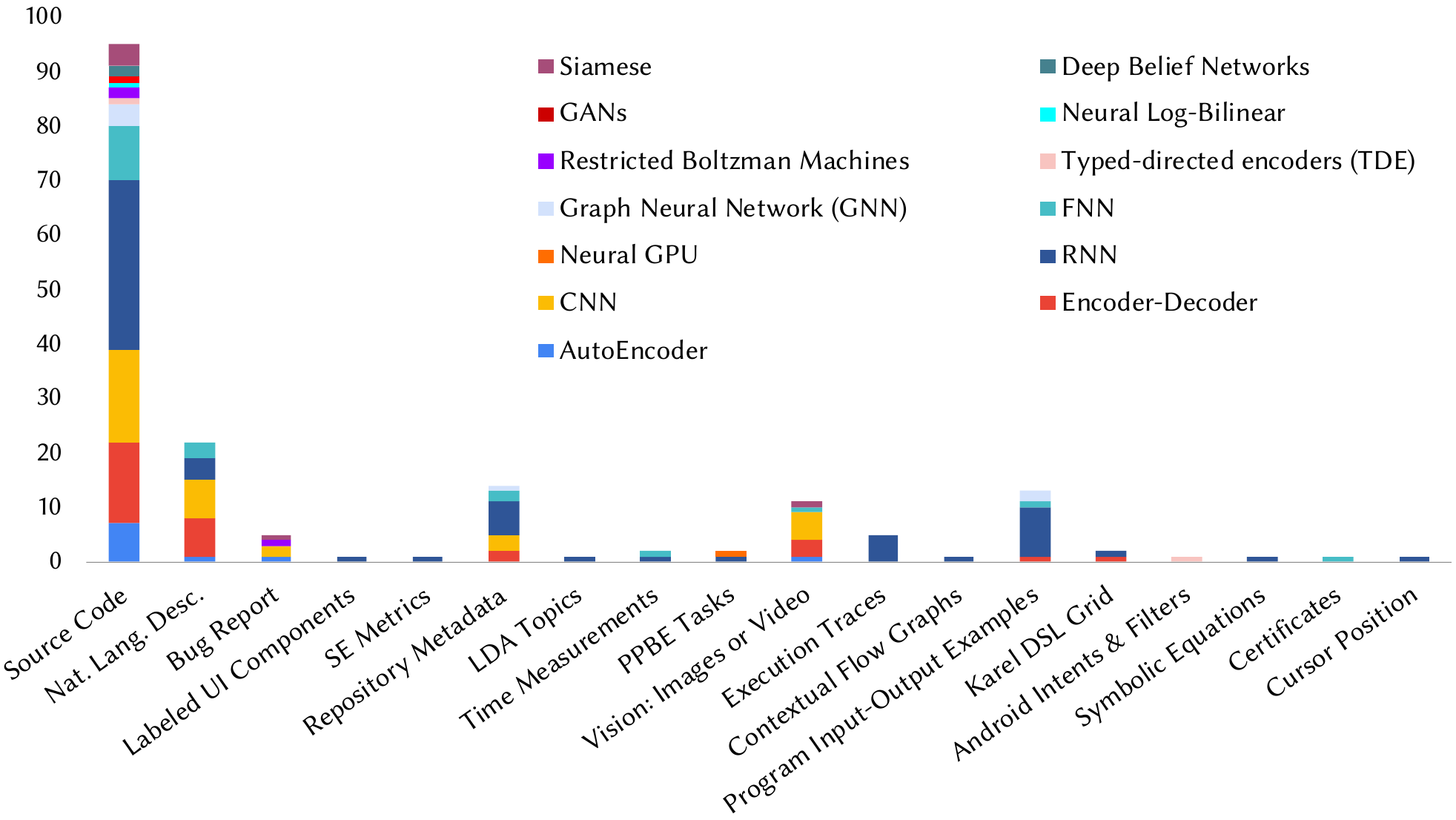}
	\vspace{-0.7cm}
	\caption{DL Architectures by Data Type}	
	\label{fig:archData}
\end{figure*}

Figure \ref{fig:archData} shows a breakdown of DL architectures by the type of data to which they are applied. This relationship between data and architecture is important since the architecture is partially responsible for the type of implicit features being extracted. For example, images and other visual data are commonly represented with a CNN. This is because CNNs are particularly proficient at modeling the spatial relationships of pixel-based data. We also discovered a strong correlation between RNNs and sequential data such as source code, natural language and program input-output examples. This correlation is expected due to RNNs capturing implicit features relating to the sequential nature of data. The models are able to capture temporal dependencies between text and source code tokens. Another correlation we observed was the use of CNNs for visual data or data which requires dimensionality reduction. This included the data types of images, videos, and even natural language and source code. CNNs have the ability to reduce features within long sequential data which makes them useful for tasks involving sentiment analysis or summarization. We also observed less popular combinations such as the use of deep belief networks (DBNs) for defect prediction~\cite{Wang2016}. Here, a DBN is used to learn semantic features of token vectors from a program's AST graph to better predict defects. A DBN can be a useful architecture in this situation due to its ability to extract the necessary semantic features from a tokenized vector of source code tokens. Those features are then used within their prediction models to drastically increase performance. %
\subsubsection{Results of Exploratory Data Analysis}

In our exploratory data analysis, we found that SE tasks greatly influence the architecture adopted in an approach. \revision{The mutual information value between the features of a SE task and a DL architecture is $1.11B$.} We also note that the SE research landscape has primarily focused on SE tasks that consist primarily of text-based data, including source code. \revision{This helps to explain why RNNs are used in \rnncount~of the papers analyzed in this SLR. The encoder-decoder architecture was also seen frequently (\encdeccount~of papers), which generally makes use of RNNs.}

\subsubsection{Opportunities for Future Work} We were able to correlate different DL architectures with particular SE tasks and data types, primarily due to the fact that a given architecture is typically suited for a specific type of implicit feature engineering. However, there exists a fundamental problem in the ability of current research to validate and quantify these implicit features the model is extracting. This leads to decreased transparency in DL models, which in turn, can impact their practical applicability and deployment for real problems. Thus, there exists an open research problem related to being able to explain how a given model was capable of predicting the target function, \textit{specifically} as it relates to SE data \cite{yuan2021explainability,yang2019enhancing,ANGELOV2020185,gilpin2019explaining,xie2020explainable}. While interpretability is a broader issue for the DL community, insights into implicit feature engineering specifically for SE data would be beneficial for DL4SE work. It is necessary for developers to understand what complex hierarchical features are used by the model for this prediction. This could demystify their ability to correctly predict the output for a given input datum.

The ability to increase the interpretability of DL4SE solution also contributes toward the novel field of SE4DL, where SE methods are applied to the creation, sustainability and maintenance of DL software. The ability to interpret DL based solutions could help to create more complete testing suites for DL based software. This paradigm becomes even more important as new and innovative DL architectures are being developed. The SE community could take inspiration from the recent success in the NLP community on developing benchmarks for explainability~\cite{ribeiro:2020}. Peeling back the "black box" nature of DL models should allow for an analysis on the integrity of the learning algorithms and an ability to better understand and build usable tools around their predictions. %

\mybox{\textbf{Summary of Results for RQ$_{3A}$}:}{gray!60}{gray!20}{\revision{Our analysis revealed seven major types of DL architectures that have been used in work on DL4SE including: \textit{Recurrent Neural Networks (RNNs)} (\rnncount), \textit{Encoder-Decoder Models} (\encdeccount), \textit{Convolutional Neural Networks (CNNs)} (\cnncount), \textit{Feed-Forward Neural Networks (FNNs)} (\fnncount), \textit{AutoEncoders} (\autocount), \textit{Siamese Neural Networks} (\siamcount), as well as a subset of other custom, highly tailored architectures.} RNNs and Encoder-Decoder models were both the most prevalent architecture used in our surveyed studies and the most diverse in terms of their varying configurations. We also discovered strong correlations between particular DL architectures to data types. For example, we found that architectures capable of capturing temporal differences within sequential data are used to study source code, natural language, repository metadata and program input-output examples. Likewise, architectures capable of capturing spatial and structural features from data have been used to study images, bug reports and program structures (ASTs, CFGs, etc.). }%

\subsection{\textit{RQ$_{3B}$}: What learning algorithms and training processes are used in order to optimize the models?}
\label{rq3b}

In addition to the variety of DL models that can be used within a DL-based approach, the way in which the model is trained can also vary. To answer RQ$_{3B}$ we aimed to analyze the learning algorithms used in three primary ways: according to (i) the manner in which the weights of the model are updated, (ii) the overall error calculation, and (iii) by the optimization algorithm, which governs the parameters of the learning algorithm as training progresses. Learning algorithms that have been defined in ML/DL research are typically used in an ``off-the-shelf'' manner, without any alteration or adjustment, in the context of SE research. This is likely a result of researchers in SE being primarily interested in DL applications, rather than the intricacies of learning algorithms. 

In terms of the process for adjusting weights, the most prevalent technique employed among our analyzed studies was the incorporation of the gradient descent algorithm. The breakdown of learning algorithms throughout our SLR are as follows: \revision{We found $\approx76\%$ of the primary studies used some version of gradient descent to train their DL model. The remaining studies used gradient ascent $\approx2\%$, or policy based learning $\approx2\%$. Other studies did not explicitly specify their learning algorithm in the paper $\approx18\%$. Our exploratory data analysis revealed that papers published in recent years (2018 and 2019) have begun to employ learning algorithms that differ from gradient descent, such as reward policies or gradient ascent.} %

Our analysis reveled that there are a variety of ways that DL-based implementations calculate error. \revision{However, we did find that a majority of the papers we analyzed used cross entropy as their loss function \entropycount, which was most commonly paired with gradient descent algorithms. Other common loss functions that were used with gradient descent algorithms were negative log likelihood (\neglogcount), maximum log likelihood (\maxlogcount), and cosine loss (\cosinelog). There were a number of papers which did not provide any indication about the loss function within their learning algorithm ($\approx42\%$).} We did find that when the primary study was not using gradient descent as a way to adjust the weights associated with the DL model, the error functions used became a lot more diverse. For example, the work done by Ellis \etal learned to infer graphical programs from deep learning hand-drawn images. They used gradient ascent rather than descent as their learning algorithm and also used surrogate likelihood function as a way to calculate the error of the model \cite{Ellis2018a}. We found that approaches that implement reinforcement algorithms are based on a developed policy, which calculates the error associated with the action taken by the model and adjusts the weights.

Lastly, we examined the use of optimization algorithms to determine if there were any relevant patterns. We discovered that the choice of optimization algorithm is somewhat agnostic to the model, the weight adjustment algorithm and the error function. In many cases, the optimization algorithm was not reported within the primary study ($\approx53\%$ of the time). %
\revision{However, we did analyze the papers that provided this information and identified four major optimization algorithms: Adagrad (\adagradcount) , AdaDelta (\adadeltacount), RMSprop (\rmspropcount), and Adam (\adamcount).} Below, we briefly address each optimization algorithm in order to point out potential situations in which they should be used.

\emph{Adagrad} is an algorithm that adapts the learning rate based on the impact that the parameters have on classification. When a particular parameter is frequently involved in classification across multiple inputs, the amount of adjustment to those parameters is lower. Likewise, when the parameter is only associated with infrequent features, then the adjustment to that parameter is relatively high~\cite{Duchi2011}. A benefit of AdaGrad is that it removes the need for manual adjustment of the learning rates. However, the technique that AdaGrad calculates the degree by which it should adjust the parameters is using an accumulation the sum of the squared gradients. This can lead to summations of the gradient that are too large, often requiring an extremely small learning rate. 

\emph{AdaDelta} was formulated out of AdaGrad in order to combat the gradient size problem. Rather than consider all the sums of the past squared gradients, AdaDelta only considers the sum of the past squared gradients limited to a fixed size. Additionally, this optimization algorithm does not require a default learning rate as it is defined by an exponentially decaying average of the calculated squared gradients up to a fixed size \cite{Zeiler2012}. 

\emph{RMSprop} is the next optimization algorithm, however, this algorithm has not been published or subjected to peer review. This algorithm was developed by Hinton \etal and follows the similar logic of AdaDelta. The way in which RMSprop battles the diminishing learning rates that AdaGrad generates is by dividing the learning rate by the recent average of the squared gradients. The only difference is that AdaDelta uses the root means squared error in the numerator as a factor that contributes to the adjustment of the learning rate where RMSprop does not.

\emph{Adam}, the last of our optimization algorithms discussed, also calculates and uses the exponentially decaying average of past squared gradients similar to AdaDelta and RMSprop. However, the optimization algorithm also calculates the exponentially decaying average of the past gradients. Keeping this average dependent on gradients rather than just the squared gradients allows Adam to introduce a term which mimics the momentum of how the learning rate is moving. It can increase the rate at which the learning rate is optimized \cite{Kingma2014}. %

\subsubsection{Results of Exploratory Data Analysis}

\revision{We found that the loss function is correlated to the chosen technique to combat overfitting with a mutual dependence of $1.00B$. However, the SE community omits reporting the loss function in $\approx33\%$ of the papers we analyzed. Additionally, the loss function is correlated to SE task with a mutual dependence of $1.14B$}

\subsubsection{Opportunities for Future Work} A consequential highlight of our analysis of employed learning algorithms was the lack of data available from the primary studies. However, we did find a strong correlation between certain loss functions paired to specific learning algorithms. One aspect we believe could provide vital insight into the DL process is an analysis regarding how learning algorithms affect the parameters of the model for certain types of data. It would not only be important to study the type of data that learning algorithms and loss functions are associated with, but also what preprocessing techniques influence the learning algorithms and loss functions chosen. It is possible that some loss functions and learning algorithms are more efficient when applied to data that has been subjected to a particular preprocessing technique. Finding the optimal pairing of loss function and learning algorithm for an architecture/data pair remains an open problem. %

\mybox{\textbf{Summary of Results for RQ$_{3B}$}:}{gray!60}{gray!20}{Our analysis revealed four different techniques for updating the weights of the DL models, with the large majority making use of gradient descent. \revision{We found four major techniques that were utilized for calculating error, including cross entropy \entropycount, negative log likelihood \neglogcount, maximum log likelihood \maxlogcount, and cosine loss \cosinelog -- with cross entropy being the most prevalent. Finally, we observed the use of four major optimization algorithms, including Adagrad (\adagradcount) , AdaDelta (\adadeltacount), RMSprop (\rmspropcount), and Adam (\adamcount).}}

\subsection{\textit{RQ$_{3C}$}: What methods are employed to combat over- and under-fitting?}
\label{rq3c}

Two potential problems associated with the use of any type of learning based approach, whether that be canonical machine learning or deep learning, are \textit{overfitting} and \textit{underfitting}. Both of these issues are related to the notion of generalization, \ie how well does a trained ML/DL model perform on unseen data. Overfitting is the process of a model learning to fit the training data extremely well, yet not being able to generalize to unseen data, and hence is a poor approximation of the actual target function to be learned~\cite{Tetko1995NeuralNS}. %
Underfitting is typically described as the scenario in which a given model incurs a high error rate on a training set. This can occur when the model lacks the necessary complexity, is overly constrained, or has not had the sufficient training iterations to appropriately approximate the target function.   For RQ$_{3C}$, we are primarily interested in the specific methods employed by researchers to combat these two problems in the context of SE tasks.

\begin{figure*}[t]
	\centering
	\vspace{-0.2cm}
	\includegraphics[width=0.75\columnwidth]{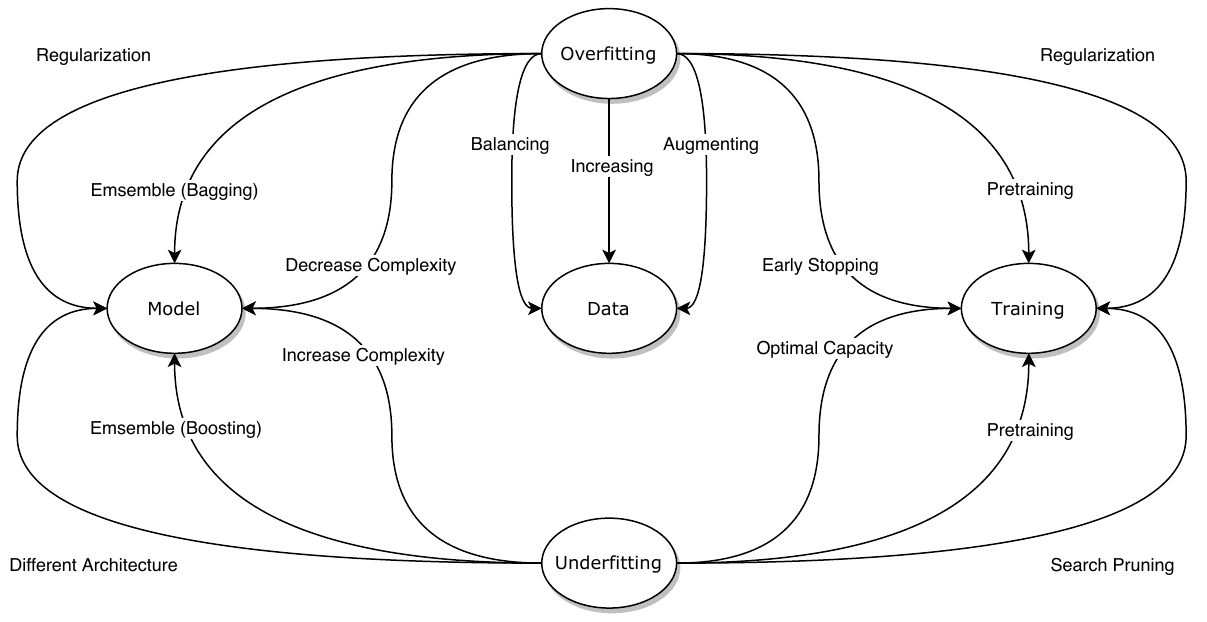}
	\vspace{-0.4cm}
	\caption{Overfitting and Underfitting Overview}	
	\label{fig:over_under_fit_overview}
\end{figure*}

Figure \ref{fig:over_under_fit_overview} provides an overview of some general methods used to combat overfitting and underfitting\footnote{Generated through an analysis of the following sources: \textit{\url{https://elitedatascience.com/overfitting-in-machine-learning}}, \textit{\url{https://hackernoon.com/memorizing-is-not-learning-6-tricks-to-prevent-overfitting-in-machine-learning-820b091dc42}}, \textit{\url{https://towardsdatascience.com/dont-overfit-how-to-prevent-overfitting-in-your-deep-learning-models-63274e552323}}, \textit{\url{https://elitedatascience.com/bias-variance-tradeoff}}}. The figure also addresses what parts of an ML/DL approach are affected by these techniques. As illustrated, there are three main types of regularization. The first regularizes the model, which includes things such as adding Dropout layers \cite{JMLR:v15:srivastava14a} or Batch Normalization \cite{DBLP:journals/corr/IoffeS15}. The second regularizes the data itself, either through adding more data or cleaning the data already extracted. The third type of regularization is applied to the training process, which modifies the loss function with L1 regularization, L2 regularization or incorporates early stop training.

As outlined in \cite{abu-mastafa}, the use of a validation set is a commonly used method for detecting if a model is overfitting or underfitting to the data, which is why it is very common to split data into training, validation and evaluation sets. The splitting of data helps to ensure that the model is capable of classifying unseen data points. This can be done in parallel with a training procedure, to ensure that overfitting is not occurring. We see cross-validation in $\approx11\%$ papers we analyzed. However, other potentially more effective techniques were seen less frequently. %

\begin{figure*}[t]
	\centering
	\vspace{-0.5cm}
	\includegraphics[width=0.90\textwidth]{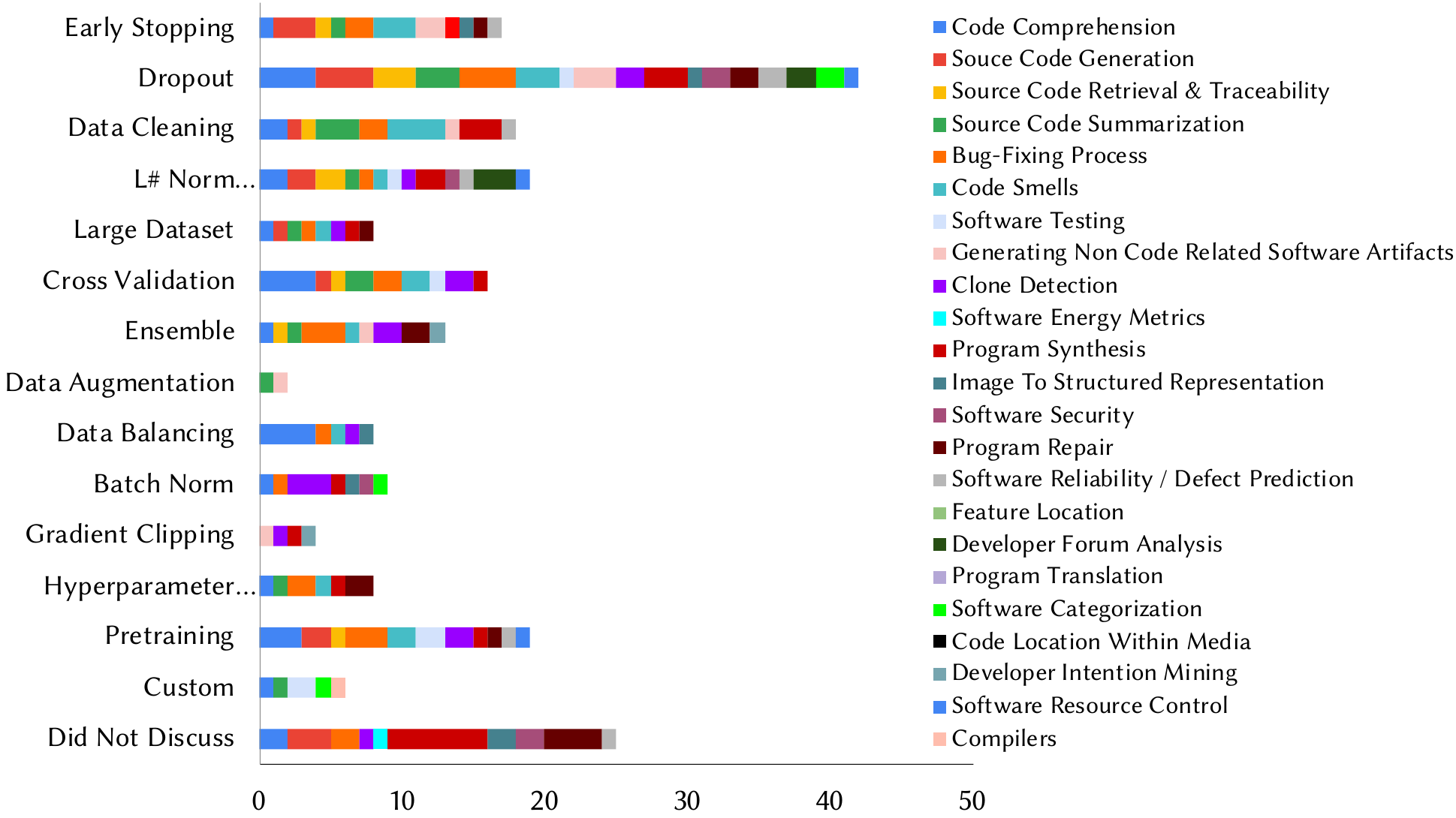}
	\vspace{-1em}
	\caption{Overfitting Techniques per Task Type}	
	\label{fig:overfit_task}
\end{figure*}

We aimed to determine if a given SE task had any relationship with the methods employed to prevent over/under-fitting. Figure \ref{fig:overfit_task} analyzes the relationship between DL approaches and the techniques that combat overfitting. This figure shows that there are some techniques that are much more commonly applied to SE tasks than others. \revision{For example, \textit{dropout} (\dropoutcount) was the most commonly used regularization technique and is used in a variety of DL approaches that address different SE tasks, followed by \textit{data cleaning} (\cleancount), \textit{L1/L2 regularization} (\regcount), and \textit{early stopping} (\earlycount).} %
Dropout is one of the most popular regularization techniques because of its effectiveness and ease of implementation. Dropout randomly blocks signals from a node within a layer of the neural network with a certain probability determined by the researcher. This ensures that a single node doesn't overwhelmingly determine the classification of a given data point. We also observed a number of custom methods that were employed. These methods are configured to address the specific neural network architecture or data type being used. For example, in Sun et al. \cite{Sun2018}, they encourage diversity in the behavior of generated programs by giving a higher sampling rate to the perception primitives that have higher entropy over $K$ different initial states. In Delvin et al. \cite{Devlin2017} they perform multiple techniques to combat overfitting which include the even sampling of the dataset during training and ensuring that each I/O grid of every example is unique. %
In addition to the aforementioned techniques, we found a subset of more unique approaches including the use of deep reinforcement learning instead of supervised learning \cite{Wan2018}, gradient clipping, lifelong learning \cite{Gaunt2017}, modification of the loss function \cite{Bunel2018}, pretraining \cite{Wan2018, Si2018}, and ensemble modeling~\cite{Jiang2017}.

We also analyzed the relationships between techniques to combat over/under-fitting, and the underlying data type that a given model operated upon. We observed similar patterns in that there are a variety of techniques to combat overfitting regardless of the data type. The only exception to this pattern was seen when analyzing natural language, where L1/L2 regularization was predominately used.  Figure \ref{fig:overfit_task} illustrates that the techniques used to combat overfitting do not have a strong association with the SE task. Therefore, we observe that a range of different techniques are applicable across many different contexts. %

\revision{One of the more concerning trends that we observed is the number of papers categorized into the \textit{Did Not Discuss} (\naoverfitcount) category.} Given the importance of combating overfitting when applying a DL approach, it is troublesome that so many primary studies did not mention these techniques. We hope that our observation of this trend signals the importance of recording such information.

Combating underfitting is a more complex process, as there aren't a well-defined set of standard techniques that are typically applied. %
One method that can be used to combat underfitting is searching for the optimal capacity of a given model. The optimal capacity is the inflection point where the model starts to overfit to the training data and performs worse on the unseen validation set. One technique for achieving this optimal capacity include maximizing training time while monitoring performance on validation data. %
Other techniques include the use of a more complex model or a model better suited for the target function, which can be determined by varying the number of neurons, varying the number of layers, using different DL architectures, pretraining the model, and pruning the search space. From our SLR, the most commonly used underfitting techniques applied were pruning the search space of the model \cite{Vijayakumar2018, DBLP:conf/iclr/ChenLS18}, curriculum training \cite{Zhang2018, DBLP:journals/corr/ReedF15, DBLP:conf/iclr/ChenLS18} and pretraining \cite{Wan2018, Si2018}. We found that only 6/\includedpapers primary studies explicitly stated the implementation of an underfitting technique. This is a stark contrast to the number of studies implementing an overfitting technique, 97/\includedpapers. %

Surprisingly, more than \naoverfitcount \hspace{.001cm} of our studied papers did not discuss any techniques used to combat overfitting or underfitting. Combating this issue is a delicate balancing act, as attempting to prevent one can begin to cause the other if the processes are not carefully considered. For example, having a heavily regularized learning model to prevent overfitting to a noisy dataset can lead to an inability to learn the target function, thus causing underfitting of the model. This is also possible while attempting to prevent underfitting. An increase in the number of parameters within the architecture to increase the complexity of the model can cause the model to learn a target function that is too specific to the noise of the training data. Therefore, the incorporation of techniques to address over- and under-fitting is crucial to the generalizability of the DL approach. %

\subsubsection{Opportunities for Future Research}

Given the relative lack of discussion of techniques to combat the over- and under-fitting observed in our studies, it is clear that additional work is needed in order to better understand different mitigation techniques in the context of SE tasks and datasets, culminating in a set of shared guidelines for the DL4SE community. In addition, more work needs to be done to analyze and understand specialized techniques for SE tasks, data types, and architectures. Similar to preprocessing data, the implementation of over- and underfitting techniques are subject to a set of variables or parameters that define how they work. An in-depth analysis on how these details and parameters change depending on the type of SE task, architecture or data, is beyond the scope of this review. However, it would be useful to the SE community to provide some intuition about what combination of over- and underfitting techniques to apply and what parameters inherent to those techniques will likely lead to beneficial results.%

\mybox{\textbf{Summary of Results for RQ$_{3c}$}:}{gray!60}{gray!20}{\revision{Our analysis shows that \textit{dropout} (\dropoutcount) was the most commonly used method to combat over/under-fitting, followed by \textit{data cleaning} (\cleancount), \textit{L1/L2 regularization} (\regcount), and \textit{early stopping} (\earlycount). Nearly 1/4 of papers did not discuss such techniques.}}

\section{\textbf{RQ4: How well do DL tasks perform in supporting various SE tasks?}}
\label{sec:rq4}

In this RQ, we aim to explore the impact that DL4SE research has had through an examination of the effectiveness of the techniques proposed in our selected studies. we primarily analyze metrics on a per task basis and summarize the current state of benchmarks and baselines in DL4SE research. %

\subsection{\textbf{\textit{RQ$_{4A}$}: What ``baseline'' techniques are used to evaluate DL models and what benchmarks are used for these comparisons?}}
\label{rq4a}

For RQ$_{4A}$, we examine the baseline techniques and evaluation metrics used for comparison in DL4SE work. %
In general, while we did observe the presence of some common benchmarks for specific SE tasks, we also found that a majority of papers self-generated their own benchmarks. %
We observed that baseline approaches are extremely individualized, even within the same SE task. Some DL4SE papers do not compare against any baseline approaches while others compare against 3-4 different models. Therefore, we included the listing of baselines that each paper compared against in our supplemental material~\cite{watson_palacio_cooper_moran_poshyvanyk, cody_watson_2021_4768587}. We found that many of the baseline approaches were canonical machine learning models or very simple neural networks. We suspect the reason for this is in part due to DL4SE being a relatively new field, meaning that there were not many available DL-based approaches to compare against. %
As the field of DL4SE begins to mature, we expect to see a transition to evaluations that include comparisons against previously implemented DL approaches.

One somewhat troubling pattern that we observed is that many model implementations do not include a publicly available implementation of a DL approach. This, in part, explains why there are so many highly individualized, baseline approaches. Since researchers do not have access to common baselines used for comparison, they are forced to implement their own version of a baseline. %
The robustness of the results of such papers may suffer from the fact that many papers did not include any information about the baselines themselves. Additionally, a unique implementation of the same baselines could lead to confounding results when attempting to examine purported improvements. While we expect that the set of existing, publicly available baselines will continue to improve over time, we also acknowledge the need for well-documented and publicly available baselines, and guidelines that dictate their proper dissemination. %

 \begin{figure*}[t]
	\centering
	\vspace{-0.3cm}
	\includegraphics[width=0.95\columnwidth]{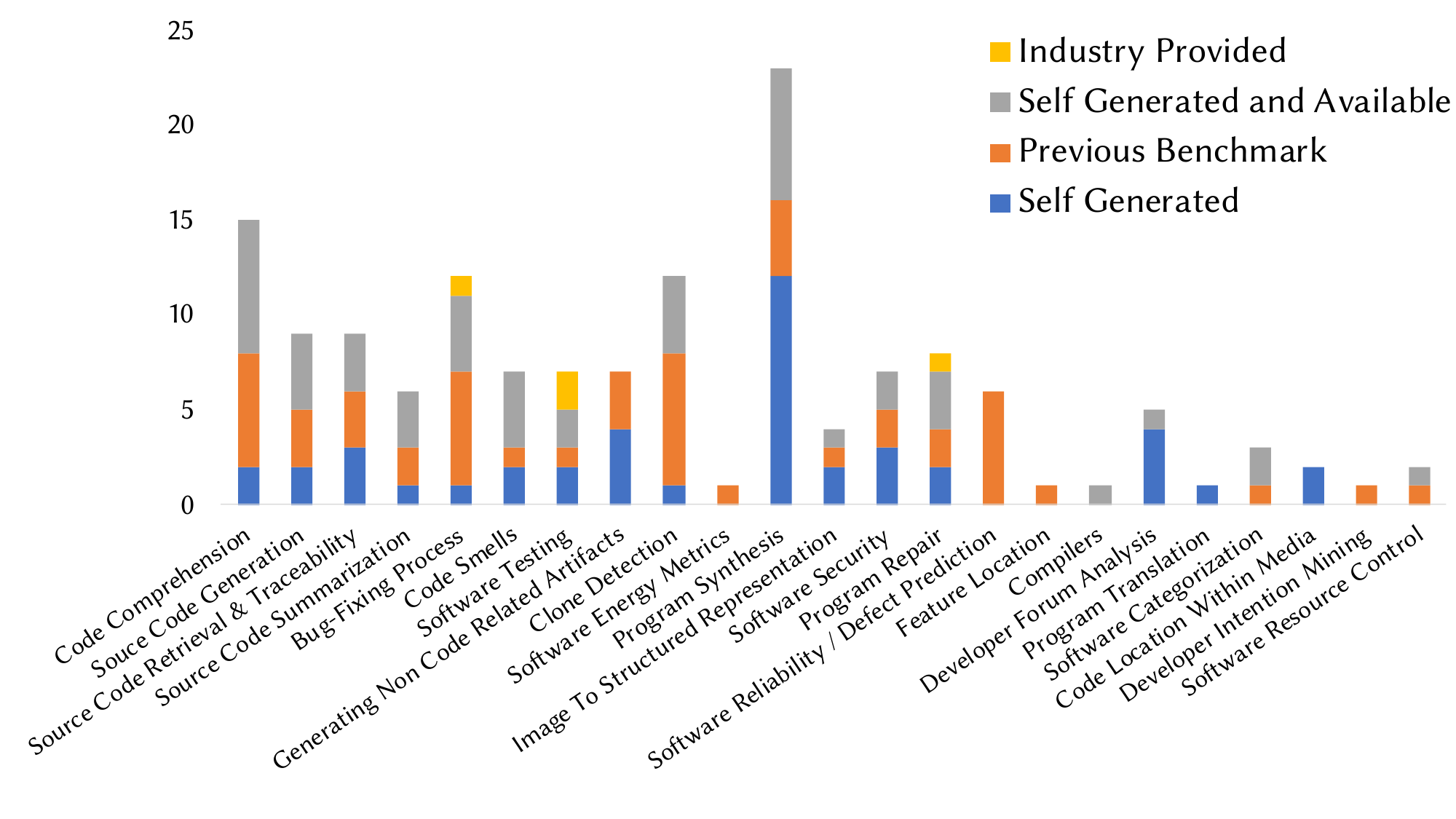}
	\vspace{-0.9cm}
	\caption{Benchmark Usage DL in SE}	
	\label{fig:benchmarks}
\end{figure*}

Our online appendix~\cite{watson_palacio_cooper_moran_poshyvanyk, cody_watson_2021_4768587} includes a list of all the benchmarks and baselines used for each paper within our SLR. The diversity and size of this list of benchmarks prohibited its inclusion to the text of this manuscript. %
However, we recorded the number of primary studies that used a previously curated benchmark as opposed to ones that curated their own benchmark. We noted that there is an overwhelming number of self-generated benchmarks. Additionally, we classified self-generated benchmarks into those that are publicly available and those that are not. Unfortunately, we found a majority of self-generated benchmarks may not be available for public use. The full breakdown of benchmarks used in the primary studies can be seen in Figure \ref{fig:benchmarks}. This trend within DL4SE is worrying as there are few instances where DL approaches can appropriately compare against one another with available benchmarks. We hope that our online repository aids researchers by providing them with an understanding about which benchmarks are available for an evaluation of their approach within a specific SE task. Additionally, we urge future researchers to make self-generated benchmarks publicly available, which will provide a much needed resource not only for comparisons between approaches, but also for available data applicable to DL techniques.

Although the use of previously established benchmarks was not common among our studies, we did observe a subset of benchmarks that were used multiple times within our primary studies. For the SE task of clone detection, we found that the dataset BigCloneBench~\cite{6976121} was used frequently to test the quality of the DL frameworks. Also, for the task of defect prediction, we saw uses of the PROMISE dataset~\cite{Sayyad-Shirabad+Menzies:2005} as a way to compare previous DL approaches that addressed defect prediction in software. %

\subsubsection{Opportunities for Future Research} The use of baselines and benchmarks in DL4SE studies, for the purpose of evaluation, is developing into a standard practice. However, there exists a need for replicable, standardized, baseline approaches that can be used for comparison when applying a new DL approach to a given SE task. The baseline implementations should be optimized for the benchmark used as data for a non-biased evaluation. This requires a thorough and detailed analysis of each published benchmark, within a specific SE task, for high quality data that does not suffer from sampling bias, class imbalance, etc. Many of the primary studies used a comparative approach for their evaluation, however, with a lack of standardized baselines the evaluations are dependent on how optimized the baseline is for a particular dataset. This can lead to confusing or conflicting  results across SE tasks. We have started to see recent progress in the derivation and sharing of large-scale datasets with efforts such as the CodeXGlue dataset from Microsoft~\cite{lu2021codexglue}. %

\mybox{\textbf{Summary of Results for RQ$_{4A}$}:}{gray!60}{gray!20}{Our analysis revealed a general lack of well-documented, reusable baselines or benchmarks for work on DL4SE. A majority of the baseline techniques utilized in the evaluations of our studied techniques were self-generated, and many are not publicly available or reusable. While a subset of benchmark datasets do exist for certain tasks, there is a need for well-documented and vetted benchmarks.}

\subsection{\textbf{\textit{RQ$_{4B}$}: How is the impact or automatization of DL approaches measured and in what way do these models promote generalizability?}}
\label{rq4b}

\revision{
\begin{table}[t]
\centering
\caption{Metrics Used for Evaluation}%
\vspace{-0.35cm}
\label{tab:metrics}
\resizebox{\columnwidth}{!}{%
{\begin{tabular}{lll}
\toprule
Measurement Type                            & Metrics                                           & Studies \\ \midrule
\hspace{3mm}Alignment Scores                
                                            & \hspace{3mm}Rouge-L                               & \hspace{3mm}\cite{Wan2018} \\ 
                                            & \hspace{3mm}BLEU Score                            & \hspace{3mm}\cite{Chen2018a, Jiang2017, Wan2018, Gu2016, Chen2018, Harer2018, 10.1145/3196321.3196334, sun2019, 10.1109/ICSE.2019.00087, gao:saner19} \\ 
                                            & \hspace{3mm}METEOR Score                          & \hspace{3mm}\cite{Chen2018a, Wan2018} \\ \midrule
\hspace{3mm}Classification Measures         
                                            & \hspace{3mm}Precision                             & \hspace{3mm}\makecell[l]{\cite{Tufano2018, Liu2018, Li2017, Han2017, Lam2015, Allamanis2015, White2016, Xu2016, Wang2016, Guo2017, Deshmukh2017, Saini2018, Hellendoorn2018, Choetkiertikul2018, Choetkiertikul2017, Arabshahi2018, Gu2018, 10.1145/3213846.3213876}\\\cite{Lin2018, Dam2018, Huang2018, Wen2018, Tufano2018a, Ott2018, Moran2018, 8812134, 8811893, 8811922, DBLP:conf/aaai/BuiJY18, 10.1145/3196398.3196408, 10.1145/3360588, 10.1145/3236024.3236068, 10.1109/ICPC.2019.00021, 8812062, 10.1109/ICSE.2019.00084, thaller:saner19, wang:msr19, perez:msr19, huo:tse19, xie:saner19, guo:saner19, fakhoury:saner19}} \\\\
                                            & \hspace{3mm}Recall                                & \hspace{3mm}\makecell[l]{\cite{Liu2018, Li2017, Han2017, Allamanis2015, Xu2016, Wang2016, Guo2017, Deshmukh2017, Saini2018, Hellendoorn2018, Choetkiertikul2018, Choetkiertikul2017, Arabshahi2018, 10.1145/3213846.3213876, Lin2018, Liu2018c, Dam2018, Huang2018, Wen2018}\\\cite{Chen2019, Tufano2018a, Ott2018, 8812134, 8811893, 8811922, DBLP:conf/aaai/BuiJY18, 10.1145/3196398.3196408, 10.1145/3276517, 10.1145/3360588, 10.1145/3236024.3236068, DBLP:conf/iclr/YinNABG19, 10.1109/ICPC.2019.00021, 8812062, 10.1109/ICSE.2019.00084, thaller:saner19, wang:msr19, perez:msr19, guo:saner19, xie:saner19, guo:saner19}} \\\\
                                            & \hspace{3mm}Confusion Matrix                      & \hspace{3mm}\cite{Moran2018, 10.1145/3276517} \\\\
                                            & \hspace{3mm}Accuracy                              & \hspace{3mm}\makecell[l]{\cite{Lee2017, Murali2017, Hellendoorn2017, Li2017, Gaunt2017, 10.5555/3305381.3305483, Levy2017, Han2017, Cai2017, BenNun2018, Zhang2018, Devlin2017, Sun2018, DBLP:conf/iclr/MuraliQCJ18, Bhatia2018, DBLP:journals/corr/ReedF15, Piech2015, Chen2016, Xu2016, Liu2016, DBLP:conf/icml/DevlinUBSMK17, Deshmukh2017, Bunel2018, Vijayakumar2018, DBLP:conf/iclr/ChenLS18}\\\cite{ Hellendoorn2018a, Saini2018, Hellendoorn2018, Arabshahi2018, Chen2018, 10.1145/3213846.3213876, Lin2018, Chen2018e, Gao2018, Wang2017, allamanis2018learning, Zohar2018, Shin2018, Ellis2018a, Liang2018, Choetkiertikul2019, Huang2018, Ott2018, Harer2018, 8811988, 10.1109/ICSE.2019.00021, 10.1145/3276517, sun2019, 10.5555/3015812.3016002, 10.5555/3298239.3298436, DBLP:conf/iclr/YinNABG19, bui:saner19, liu:saner19, white:saner19, nguyen:saner19, cvitkovic:icml19}} \\\\
                                            & \hspace{3mm}ROC/AUC                               & \hspace{3mm}\cite{Zhao2018, Saini2018, Wen2018, Choetkiertikul2018, Choetkiertikul2017, Gao2018, Dam2018, Wen2018, 10.1145/3196398.3196408, codereviewlearn, dam:msr19, hoang:msr19, liu:saner19, buch:saner19} \\
                                            & \hspace{3mm}F-Score                               & \hspace{3mm}\cite{Liu2018, Han2017, Zhao2018, Allamanis2015, Xu2016, Allamanis2016, Wang2016, Choetkiertikul2018, Choetkiertikul2017, Le2018a, 10.1145/3213846.3213876, Dam2018, Huang2018, Wen2018, Tufano2018a, 8812134, 8811893, 8811922, DBLP:conf/aaai/BuiJY18, 10.1145/3360588, codereviewlearn, 10.1145/3236024.3236068, 10.1109/ICPC.2019.00021, 8812062, thaller:saner19, dam:msr19, wang:msr19, perez:msr19, guo:saner19, xie:saner19, fakhoury:saner19} \\
                                            & \hspace{3mm}Matthews Correlation                  & \hspace{3mm}\cite{Choetkiertikul2018, thaller:saner19} \\
                                            & \hspace{3mm}Scott-Knott Test                      & \hspace{3mm}\cite{liu:saner19} \\
                                            & \hspace{3mm}Exam-Metric                           & \hspace{3mm}\cite{zhang:saner19} \\
                                            & \hspace{3mm}Clustering-Based                      & \hspace{3mm}\cite{8812083} \\ \midrule
\hspace{3mm}Coverage \& Proportions
                                            & \hspace{3mm}Rate or Percentages                   & \hspace{3mm}\cite{Gu2018, Cummins2018, Chen2018d, gupta2019, DBLP:conf/aaai/LiuLPW19, DBLP:conf/iclr/YinNABG19, 8730177, 8502853, fakhoury:saner19} \\
                                            & \hspace{3mm}Coverage-Based                        & \hspace{3mm}\cite{Liu2017, Godefroid2017, DBLP:conf/aaai/LiuLPW19, DBLP:conf/iclr/ParisottoMS0ZK17, Zhang2018a, wang:msr19} \\
                                            & \hspace{3mm}Solved Tasks                          & \hspace{3mm}\cite{Si2018, Ellis2018, Wen2018, 10.1145/3360594, gupta2019} \\
                                            & \hspace{3mm}Cost-Effectiveness                     & \hspace{3mm}\cite{liu:saner19, white:saner19} \\
                                            & \hspace{3mm}Total Energy or Memory Consumption    & \hspace{3mm}\cite{Romansky2017} \\ \midrule
\hspace{3mm}Distance Based
                                            & \hspace{3mm}CIDER                                 & \hspace{3mm}\cite{Wan2018, Zohar2018} \\
                                            & \hspace{3mm}Cross Entropy                         & \hspace{3mm}\cite{Hellendoorn2018a} \\
                                            & \hspace{3mm}Jaccard Distance                       & \hspace{3mm}\cite{DBLP:conf/iclr/MuraliQCJ18} \\
                                            & \hspace{3mm}Model Perplexity                      & \hspace{3mm}\cite{White2015a, katz:saner19, dam:msr19} \\
                                            & \hspace{3mm}Edit Distance                         & \hspace{3mm}\cite{katz:saner19, gao:saner19} \\
                                            & \hspace{3mm}Exact Match                           & \hspace{3mm}\cite{gao:saner19} \\
                                            & \hspace{3mm}Likert Scale                          & \hspace{3mm}\cite{Jiang2017} \\ \midrule
\hspace{3mm}Approximation Error
                                            & \hspace{3mm}Mean Absolute Error                   & \hspace{3mm}\cite{Choetkiertikul2018, Choetkiertikul2019} \\
                                            & \hspace{3mm}Minimum Absolute Difference           & \hspace{3mm}\cite{DBLP:conf/iclr/MuraliQCJ18} \\
                                            & \hspace{3mm}Macro-averaged Mean Absolute Error    & \hspace{3mm}\cite{Choetkiertikul2018, Choetkiertikul2017} \\
                                            & \hspace{3mm}Root Mean Squared Error               & \hspace{3mm}\cite{Schroeder2017} \\
                                            & \hspace{3mm}Median Absolute Error                 & \hspace{3mm}\cite{Choetkiertikul2019} \\
                                            & \hspace{3mm}Macro-averaged Mean Cost Error        & \hspace{3mm}\cite{Choetkiertikul2017} \\ \midrule
\hspace{3mm}Ranking
                                            & \hspace{3mm}F-Rank                                & \hspace{3mm}\cite{Gu2018} \\
                                            & \hspace{3mm}Top K - Based                         & \hspace{3mm}\cite{Shin2018, Moran2018, Liu2018d, 10.1145/2594291.2594321, huo:tse19} \\
                                            & \hspace{3mm}Spearmans Rank                        & \hspace{3mm}\cite{Tufano2018a} \\
                                            & \hspace{3mm}MRR                                   & \hspace{3mm}\cite{Karampatsis2019, Chen2018a, Hellendoorn2017, Lam2015, Corley2015, Gu2018, Chen2019, huo:tse19} \\
                                            & \hspace{3mm}Kruskal's $\gamma$                      & \hspace{3mm}\cite{Zhao2018} \\ \midrule
Timing                                      & \hspace{3mm}Time                                  & \hspace{3mm}\cite{White2016, Balog2016, Ellis2018, Ellis2018a, 8811988} \\ \midrule
\bottomrule
\end{tabular}}
}
\end{table}
}

Table \ref{tab:metrics} describes the distribution of metrics found in this SLR. In our analysis of utilized metrics within work on DL4SE, we observed that the metrics chosen are often related to the type of learning. \revision{Therefore, many of the supervised learning methods have metrics that analyze the resulting hypothesis, such as the accuracy (\accpercent),  precision (\precpercent), recall (\recallpercent), or F1 measure (\fpercent). In fact, classification metrics are reported in $\approx74\%$ of the papers.} These metrics are used to compare the supervised learning algorithms with the outputs representing the target hypothesis. Intuitively, the type of metric chosen to evaluate the DL-based approach is dependent upon the data type and architecture employed by the approach. The ``other'' category illustrated in Figure \ref{tab:metrics} is comprised of less popular metrics including: \textit{likert scale}, \textit{screen coverage}, \textit{total energy consumption}, \textit{coverage of edges}, \textit{ROUGE}, \textit{Jaccard similarity}, \textit{minimum absolute difference}, \textit{cross entropy}, \textit{F-rank}, \textit{top-k generalization}, \textit{top-k model-guided search accuracy}, \textit{Spearman's rank correlation coefficient}, and \textit{confusion matrices}. In addition to the use of these metrics, we found a limited number of statistical tests to support the comparison between two approaches. These statistical tests included: \textit{Kruskal's $\gamma$}, \textit{macro-averaged mean cost-error}, \textit{Matthew's correlation coefficient}, and \textit{median absolute error}. \revision{Surprisingly, only approximately 5\% of papers made use of statistical tests.} %

\begin{figure*}[t]
	\centering
	\vspace{0.2cm}
	\includegraphics[width=0.85\linewidth]{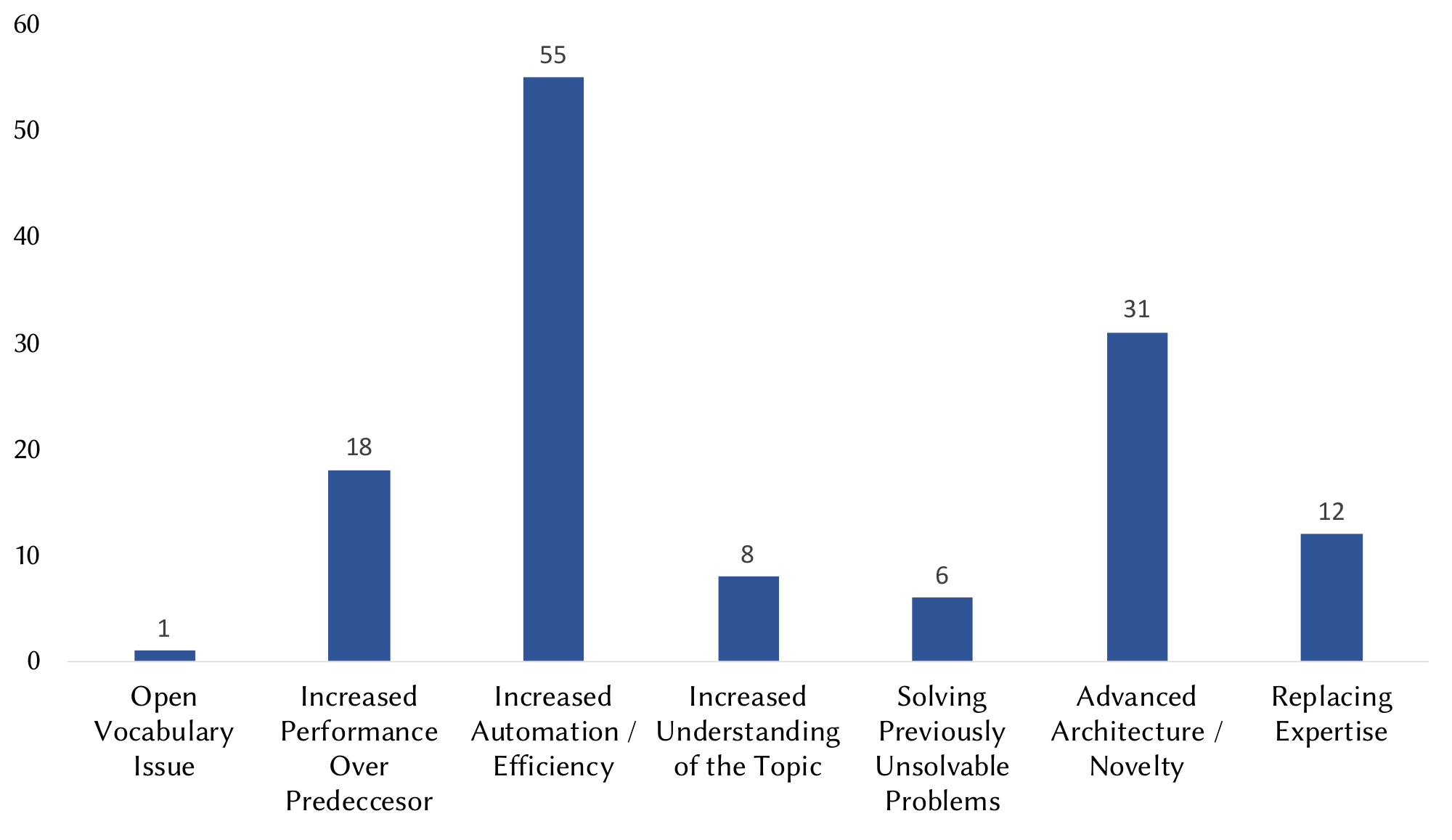}
	\vspace{-0.2cm}
	\caption{Impact of DL4SE}	
	\label{fig:impact}
	\vspace{-0.0cm}
\end{figure*}

We classified each primary study into seven categories, which represents the major contribution of the work. The result of this inquiry can be seen in Figure \ref{fig:impact}. %
\revision{We found three primary objectives that the implementation of a DL model is meant to address: (i) in $\approx43\%$ of papers observed, a DL approach was implemented with the main goal of increasing automation efficiency; (ii) in $\approx24\%$ of the papers observed, a DL approach was implemented with the main goal of advancing or introducing a novel architecture; (iii) in $\approx14\%$ of the papers observed, a DL approach was implemented with the main goal of increasing performance over a prior technique.} %

\begin{wrapfigure}{l}{2.6in}
 \centering
    \vspace{-0.50cm}
	\includegraphics[width=2.4in]{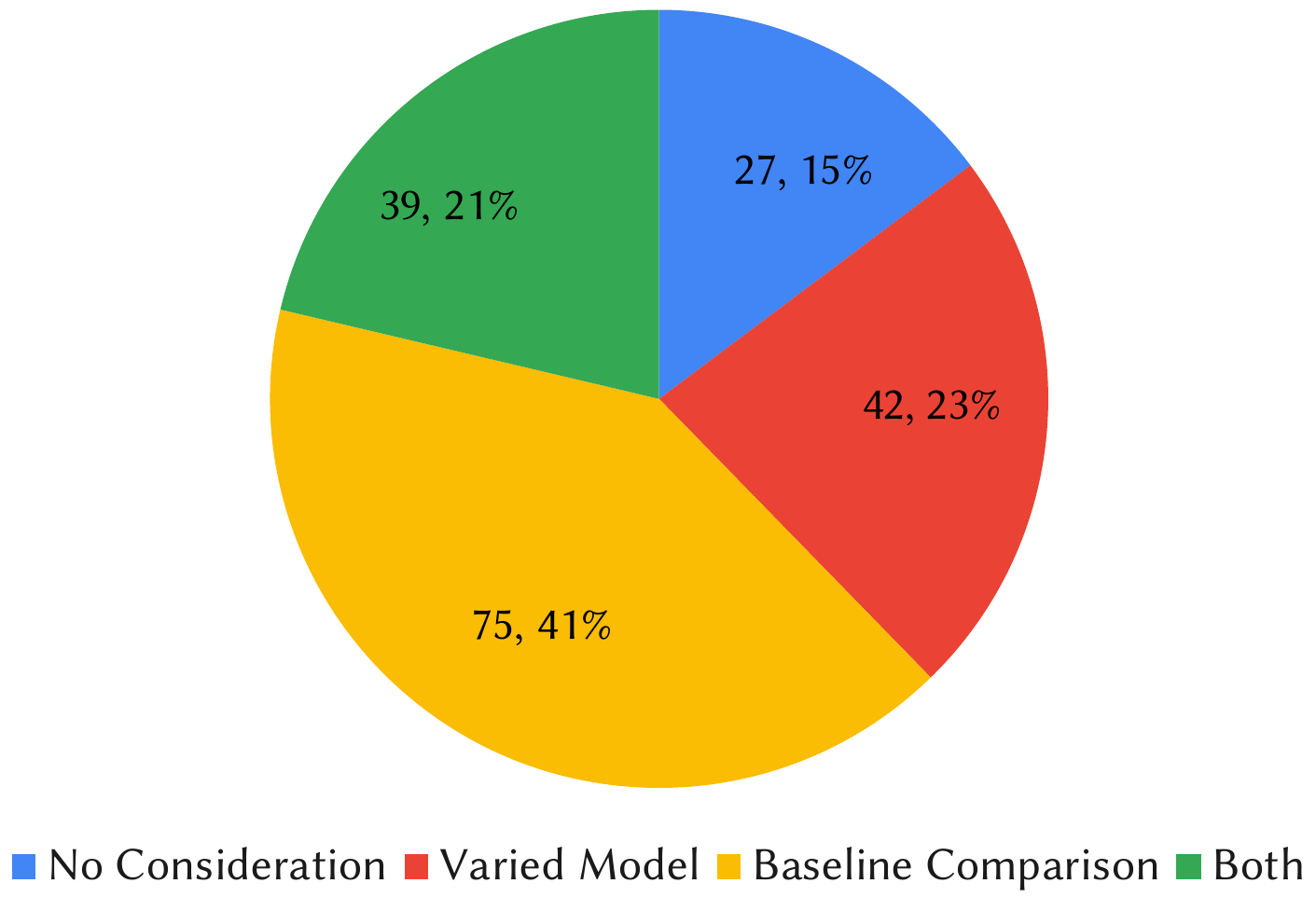}
	\vspace{-0.1cm}
	\caption{Evidence of Occam's Razor}	
	\vspace{-0.20cm}
	\label{fig:occams}
\end{wrapfigure}

In addition to the primary studies major objectives, we also observed that many papers did not analyze the complexity or generalizability of their implemented models. Thus to examine this further, we analyzed our primary studies through the lends of Occam's Razor and model efficiency. A valid question for many proposed DL techniques applied to SE tasks is whether the complexity of the model is worth the gains in effectiveness or automation for a given task, as recent research has illustrated~\cite{FuFSE16}. This concept is captured in a notion known as \textit{Occam's Razor}. Occam's Razor is defined by two major viewpoints: 1) "Given two models with the same generalization error, the simpler one should be preferred because simplicity is desirable"~\cite{10.5555/3000292.3000299}, 2) "Given two models with the same training-set error, the simpler one should be preferred because it is likely to have lower generalization error" \cite{10.5555/3000292.3000299}. %
In the context of our SLR, we aimed to investigate the concept of Occam's Razor through analyzing whether authors considered technically ``simpler'' baseline techniques in evaluating their approaches.  In Figure \ref{fig:occams} we break the primary studies into four groups: 1) those that compare against less complex models and analyze the results; 2) those that manipulate the complexity of their own model by testing a variety of layers or nodes per layer; 3) those that perform both; 4) those that did not have any Occam's Razor consideration. Note that these are overlapping groupings and so the sum of papers exceeds the number of papers in our SLR.

Although a majority of the primary studies do consider Occam's Razor, there are still $\approx16\%$ of DL4SE studies that do not consider the principle. %
Without a consideration of Occam's Razor, it is possible that a canonical machine learning model or a simple statistical based approach could yield an optimal solution. This idea coincides with the findings mentioned by Fu et al. \cite{FuFSE16}, who discovered that by applying a simpler optimizer to fine tune an SVM they were able to outperform a DL model applied to the same task. Fu~\etal warn against the blind use of DL models without a thorough evaluation regarding whether the DL technology is a necessary fit for the problem~\cite{FuFSE16}. Interestingly, in $\approx23\%$ of the primary studies, the author's considered Occam's Razor by adjusting the complexity of the model being evaluated. This is done by varying the number of layers, the number of nodes, the size of embeddings, etc. The downside to this method is that there is no way to determine if the extraction of complex hierarchical features is more complex than what is necessary to address the SE task. The only way to properly answer this question is to compare against baseline approaches that are not as complex. In our DL4SE studies, this often took the form of a comparison to a canonical ML technique. %

\subsubsection{Results of Exploratory Data Analysis}

\revision{Our exploratory data analysis revealed  papers that combat overfitting, excluding data augmentation, omit ROC or AUC evaluations with a confidence level of 
$\approx0.95$. This metric is a common means by which comparisons to baseline approaches can be performed. Our exploratory data analysis of this RQ revealed that the automation impact is correlated to the SE task deduced from a mutual information of $0.71B$. This means that there is a subtle association between the SE task and the claimed automation impact of the approach.}

\subsubsection{Opportunities for Future Research} Throughout our analysis regarding the evaluation of DL4SE studies, it became apparent that there is a troubling lack of consistency of analysis, even within a given application to an SE task. Thus, there is an opportunity to develop guidelines and supporting evaluation infrastructure for metrics and approach comparisons. Such work would allow for clearer and more concise evaluations of new approaches, solidifying claims made from the results of a given evaluation.  DL models are evaluated on their ability to be generalizable, this is normally accomplished through the use of a testing set, which the model has not been trained on. However, these testing sets can suffer from under representing certain class of data that can be found in the real world. More work is needed on evaluating the quality of testing sets and determining how representative they are when being used for evaluation. Having the limitations of DL approaches well document will create a greater opportunity for these DL solutions to be applied in the context of industry and real software development scenarios.  %
Lastly, it would be advantageous for the research community to develop a methodology that could demonstrate the \textit{need} for the complexity that DL offers when addressing a particular problem.

\mybox{\textbf{Summary of Results for RQ$_{4b}$}:}{gray!60}{gray!20}{\revision{Our analysis illustrates that a variety of metrics have been used to evaluate DL4SE techniques, with \textit{accuracy} (\accpercent), \textit{precision} (\precpercent), \textit{recall} (\recallpercent), and \textit{F1-measure} (\fpercent) being the most prominent.} In terms of claimed impact of our primary studies, the most claimed was \textit{increased automation or efficiency}, followed by \textit{advancing a DL architecture}, and \textit{replacing human expertise.} We also found that most studies did consider the concept of Occam's Razor and offered a comparison to a conceptually simpler learning model.}

\section{\textbf{RQ$_5$: What common factors contribute to the difficulty when reproducing or replicating DL4SE studies?}}
\label{sec:rq5}

DL models carry with them significant complexity, thus even small, seemingly nuanced changes can lead to drastic affects in the approach's performance. %
Such changes could encompass the model, the extraction and preprocessing of the data, the learning algorithm, the training process, or the hyperparameters. In this RQ, we synthesize important details related to reporting a \textit{complete} set of elements of computation learning in DL4SE. We examined through the lens of \textit{replicability}, or the ability to reproduce a given described approach using the same experimental setup or author-provided artifacts, and \textit{reproducibility}, or the ability to reproduce a given approach in a different experimental setup by independently developing the approach as described in the paper~\cite{reprovrepro}.

\begin{wrapfigure}{r}{2.5in}
	\centering
	\vspace{-0.4cm}
	\includegraphics[width=0.5\columnwidth]{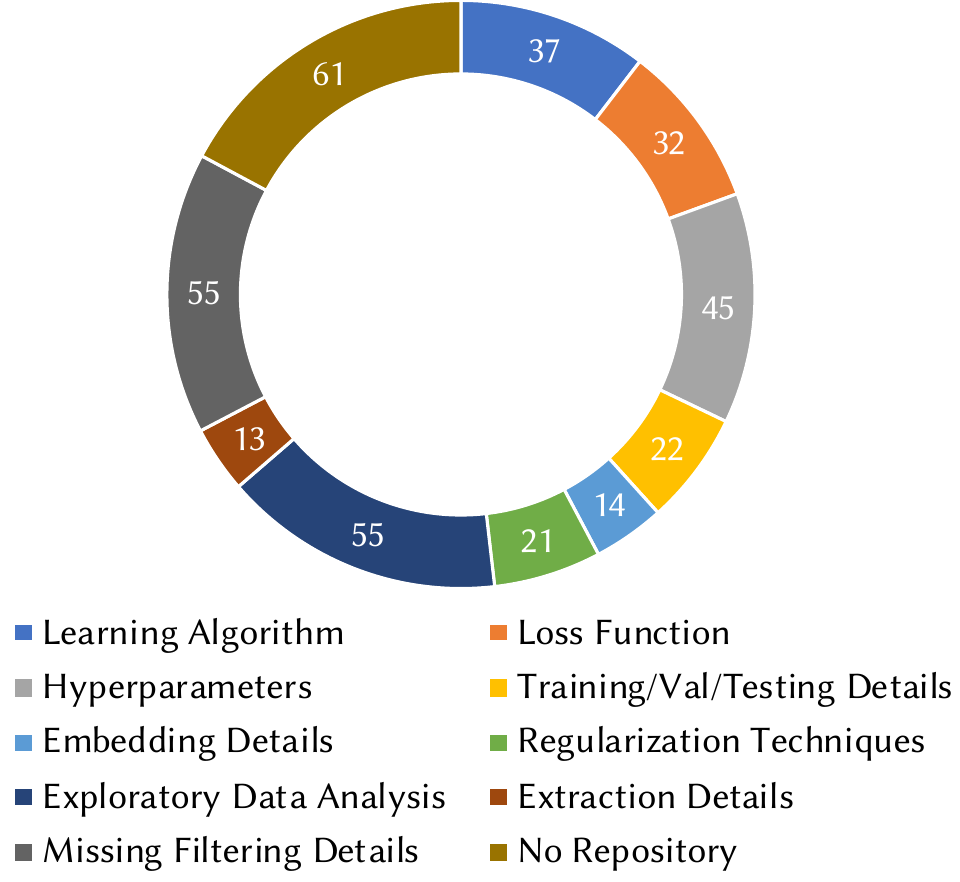}
	\vspace{-0.5cm}
	\caption{Non-Reproducibility Factors}
	\vspace{-0.4cm}
	\label{fig:reproduce}
\end{wrapfigure}

In terms of replicability, we found that \revision{\replipercent~of the primary studies provided enough information through an online repository to reasonably replicate the approach (derived via a thorough open coding with four authors).} This means that the vast majority of DL4SE studies either did not provide the implementation of the approach or did not provide the dataset to train and test the approach. \revision{In terms of reproducibility, we found that \reprodpercent~of the studies we analyzed provided enough detail in the publication regarding the various elements of computational learning such that a given technique could be reasonably expected to be reproduced from the text itself (according to a thorough open coding procedure between four authors). Out of the \reprodcount~studies that can be reproduced, only 11 of those studies were also replicable.} %
We found that there were ten major factors that contributed to the lack of reproducibility. The breakdown of this analysis for the primary studies are show in in Figure \ref{fig:reproduce}.

In Figure \ref{fig:reproduce} we show areas where DL approaches in SE may be lacking the necessary details to reproduce or reimplement a given approach. \revision{The first two areas that contribute to the difficult of reproducibility pertain to the \textit{learning algorithm} (\learnalgocount~papers) and the \textit{hyperparameters} (\hypercount~papers).} %
We found that there were missing details pertaining to either the method of tuning the weights, the manner in which the error was calculated or the optimization algorithm. All three aspects of the learning algorithm are important for an accurate reproduction of a described technique, and omission of these details can jeopardize the reproduction process. %
In addition to the learning algorithm, the hyperparameters also serve a crucial role in reproducibility. Consequently, if the hyperparameters are not reported, then it is impossible to know how many parameters contributed to the estimation of the target function, since hyperparameters control the number of layers and the number of nodes per layer. Additionally, the manner in which the learning rate is adjusted ultimately controls the parameters that estimate the target function. An incorrect learning rate can lead to incorrect parameters, which in can in turn lead to modeling of a different target function.

Additional details we often found omitted from papers pertained to \textit{the data} and the way it was extracted, filtered, and formatted. DL models are data-driven, meaning that they extract features from the data without any human intervention. In order for the study to be reproducible, three pieces of information need to be accessible. \revision{First is the \textit{extraction details} ($13$~papers) . In order for a study to be reproduced, the data must be accessible which means either the dataset must be readily available or the details about how the data was extracted need to be reported. The second piece of information that is needed is the \textit{preprocessing details} ($14$~papers) . Once the data is extracted, it needs to be formatted into a representation that the DL model can accept as input. The manner in which the data is represented within the model, at least partially, controls the features that are able to be extracted. Thus, if the data is represented differently from an original study, then the results of the reproduced study may be invalid. The last attribute of the data that is required is the \textit{filtering details} ($55$~papers) . Data can be inherently noisy and authors will frequently need to filter out noise in order for the DL model to learn an effective relationship between an input and a target. This process typically involves the removal of certain data points, or an entire section of the original dataset, based upon a criteria that the authors stipulate. We discovered that $55$  primary studies are missing crucial details about the filtering of their data. Reporting these details related to the filtering of data should not only include the filtering criteria, but should also explain the steps and methodology taken to remove this data from the dataset.}

\subsubsection{Opportunities for Future Research}

Our analysis highlights the importance of open source approach implementations and datasets. This is not only the best mechanism for replicability and reproducibility, but allows for future research to more effectively build upon past techniques. We have mentioned before that DL models are extremely data-intensive. However, there are relatively few publicly available SE-related datasets tailored for DL techniques. This not only inhibits the use of DL in SE, but it also makes comparative studies with different DL models difficult. This lack of data hurts the evaluation of DL-based approaches because there are very few baselines to compare a newly synthesized model to. This can lead to claims regarding the effectiveness of a DL implementation that can be difficult to refute or verify. Therefore, we encourage future researchers to make the datasets and DL models publicly available. We believe that this will drive a greater quality of research and allow for verifiable comparisons of DL-based approaches in SE.  

This SLR also highlights a number of factors that contribute to the difficulty in reproducing some of the DL4SE studies. Based on two of those factors, details regarding the exploratory data analysis and data filtering, there exists an opportunity to generate guidelines dedicated toward the preprocessing of data for specific DL algorithms. In particular, filtering steps can have large implications on overfitting of the model and lead to a reduced capacity to generalize to unseen data. One interesting research direction could be to analyze how impactful unique filtering steps can be within a specific SE task. This could provide an indication on the trade-off between generalizability and the model's performance.

\mybox{\textbf{Summary of Results for RQ$_{5}$}:}{gray!60}{gray!20}{\revision{Our analysis illustrates that only \replicount~ of our primary studies could be conceivably labeled as \textit{replicable}, whereas only \reprodcount~studies could be reasonably \textit{reproduced} based upon the description given in the study. In addition to a lack of published open source implementations and datasets, the major contributing factors to these issues were mainly due to the~\textit{missing data filtering details} (55 papers) and a lack of description of \textit{hyperparameters} (45 papers).}}

\section{Threats to Validity}
\label{sec:threats}

Our systematic literature review was conducted according to the guidelines set forth by Kitchenham et al.~\cite{Kitchenham2007}. However, as with any SLR our review does exhibit certain limitations primarily related to our search methodology and our data extraction process employed to build our paper taxonomy.

\subsubsection{External Validity}

Issues related to external validity typically concern the generalization of the conclusions drawn by a given study. A potential threat to the external validity to our systematic literature review is the search string and filtering process used to identify meaningful DL4SE studies. It is possible that our search string missed studies that should have been included in our review. This could be due to a missed term or combination of terms that may have returned more significant results. We mitigated this threat by testing a variety of DL and SE terms such as:

\begin{enumerate}
    \item (“Deep Learning” OR “Neural Network”)
    \item (“Learning”) AND (“Neural” OR “Automatic” OR “Autoencoder” OR “Represent”)
    \item (“Learning”) AND (“Supervised” OR “Reinforcement” OR “Unsupervised” OR “Semi-supervised”)
    \item (“Learning” OR “Deep” OR “Neural” OR “Network”)
    \item (“Learning” OR “Deep” OR “Neural”)
    \item (“Artificial Intelligence” OR “Learning” OR “Representational” OR “Neural” OR “Network”)
\end{enumerate}

We evaluated these potential search strings through an iterative process as outlined by Kitchenham et al. The utilized search string "Deep" \texttt{OR} "Learning" \texttt{OR} "Neural" returned the greatest number of DL4SE studies. This search string was also chosen to limit selection bias since it ``cast the widest net'' in order to bolster completeness and limit potential biases introduced by more restrictive search strings. However, the trade-off was that it required a much more substantial effort to remove studies that were not applicable to DL4SE.

We also face potential selection bias of the studies to be included into our SLR. We attempt to mitigate this threat through the use of inclusion and exclusion criteria, which is predefined before the filtering process begins, and which we have listed in Sec. \ref{appendix:A}. This criteria is also helpful in reducing the manual effort of filtering papers given our broad search string. We also perform snowballing as a means to mitigate selection bias. In this method, we collect all the references from the primary studies that passed our inclusion and exclusion criteria and determine if any of those references should be considered for the SLR.

Additionally, to further illustrate the generalizability of our paper sampling methodology, we perform a probability sampling to determine if we capture a significant proportion of DL4SE papers. We found that our expert sampling strategy captures a statistically significant number of studies, such that we are confident in our taxonomy's representation. Therefore, we feel that the trends highlighted in this review can be generalized to the entire body of DL4SE work. We discuss more details pertaining to our statistical sampling of studies in Sec. \ref{sec:methodology}.

Another potential threat to our systematic literature review consists of the venues chosen for consideration. For our review, we included the top SE, PL, and AI related conferences and journals. We included venues with at least a C CORE ranking~\cite{core}, which helped us to determine venue impact. Although it is possible that not considering other conferences and journals caused us to miss some pertinent studies, we wanted to capture trends as seen in top SE, PL, and AI venues. Furthermore, we provide our current taxonomy and list of venues on our website, and welcome contributions from the broader research community. We intend for our online appendix to serve as a "living" document that continues to survey and categorize DL4SE research.

\subsubsection{Internal Validity}

A major contribution of this paper lies in our derived taxonomy that characterizes the field of DL4SE. To mitigate any mistakes in our taxonomy, we followed a process inspired by open coding in constructivist grounded theory~\cite{Charmaz:groundedtheory} where each attribute classification of a primary study within our SLR was reviewed by at least three authors. However, while multiple evaluators limit the potential bias of this process, the classifications are still potentially affected by the collective views and opinions of the authors. Therefore, in effort to provide transparency into our process and bolster the integrity of our taxonomy, we have released all data extraction and classifications in our online repository \cite{watson_palacio_cooper_moran_poshyvanyk, cody_watson_2021_4768587}. In releasing this information, authors of the works included in the SLR can review our classifications.

\subsubsection{Construct Validity}

One point of construct validity is the conclusions we draw at the end of each research question. In order to draw these conclusions, we performed an exploratory data analysis using rule association mining. In this analysis, we mine associations between attributes of DL solutions to SE tasks, which provides inspiration to further research why certain attributes show a strong or weak correlation. Additional details surrounding our correlation discovery and association rule learning processes can be found in Sec. \ref{sec:methodology}.

Another threat to construct validity is our methodology for data synthesis and taxonomy derivation. To mitigate this threat we followed a systematic and reproducible process for analyzing the primary studies and creating a resulting taxonomy. To reduce the potential bias of data extraction, the authors developed and agreed upon a data extraction form to apply to each study. 
For our taxonomy, primary studies were categorized by three authors and refined by one additional authors. Through this process, we limit the number of individual mistakes in extracting the data and synthesizing the taxonomy.

\section{Guidelines for Future Work on DL4SE}
\label{sec:guidelines}

\begin{figure*}
	\includegraphics[width=\columnwidth]{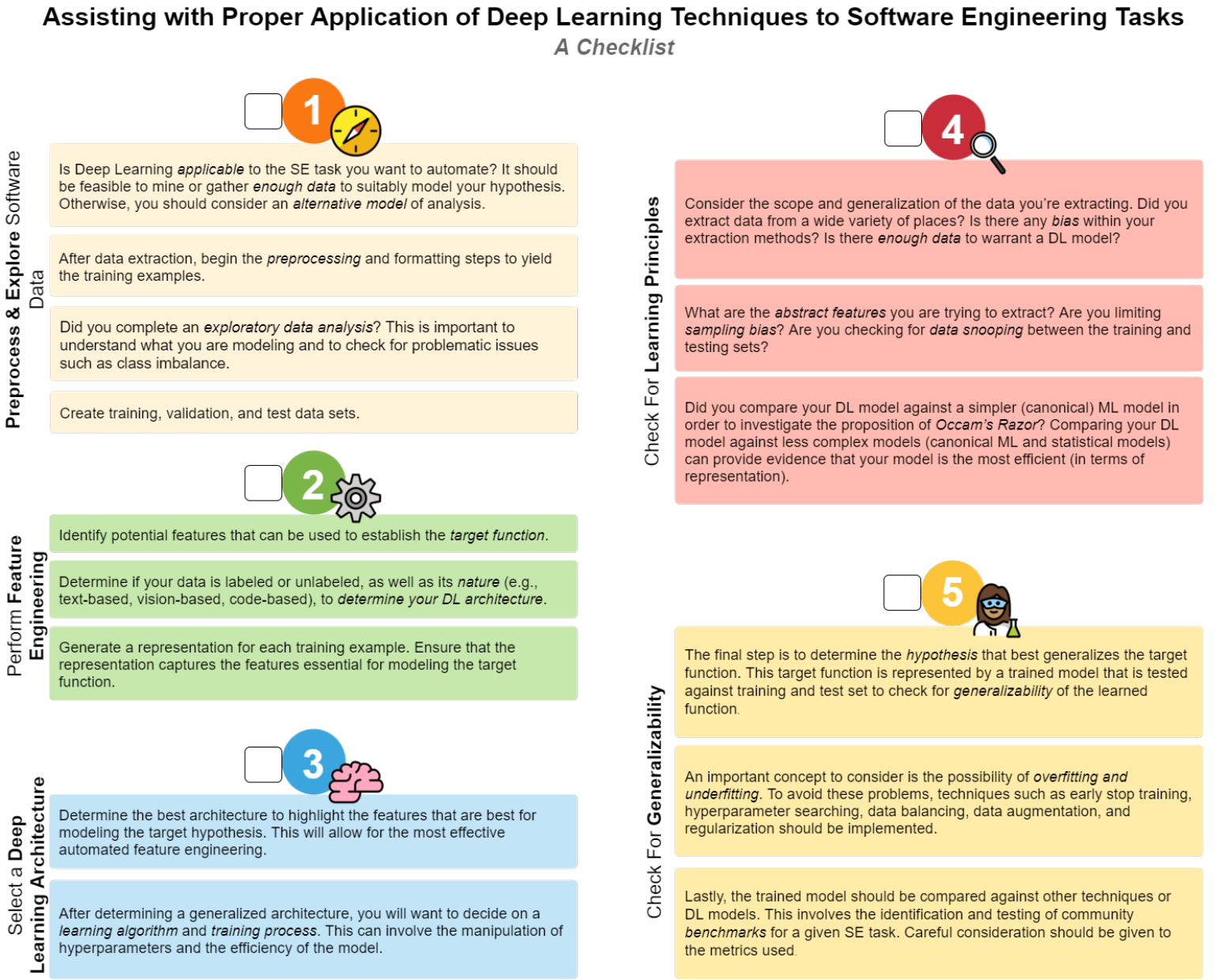}
	\vspace{0.3cm}
	\caption{Guidelines for Applying DL to SE Research}	
	\label{fig:guidelines}
	\vspace{0.5cm}
\end{figure*}

In this section, we provide guidelines for conducting future work on DL4SE based upon the findings of our SLR. As illustrated in Figure~\ref{fig:guidelines}, we synthesized a checklist with five prescribed steps that should aid in guiding researchers through the process of applying DL in SE. %
We do not intend for this to serve as a \textit{comprehensive} guide, as each implementation of these complex models come with their own nuances. However, we attempt to cover many of the \textit{essential} aspects of applying DL to SE in order to help guide future researchers.

\noindent \textbf{\textit{Step 1: Preprocessing and Exploring Software Data:}} This step focuses on building a proper data pipeline. As part of this step, one must determine whether deep learning techniques are a viable solution for representing the data associated with a given software engineering task. This involves determining if enough data is available for learning, performing data preprocessing such as filtering, implementing an exploratory data analysis (EDA) study to better understanding the data, and performing standard data splitting to avoid data snooping.

\noindent{\textit{What to report - }} Steps within the data formatting process should be explicitly explained to the point where replication is possible; this would include any abstractions, filters, vocabulary limitations or normalizations that the data is subjected to. Lastly, researchers should consider any potential limitations or biases uncovered as part of an exploratory data analysis together with the steps taken to mitigate these factors.

\noindent \textbf{\textit{Step 2: Perform Feature Engineering:}} While one of the advantages of Deep Learning models is that they provide some degree of automated feature engineering, it is still important for researchers to decide upon \textit{what information} should be used to provide a meaningful signal to their model. As such, the process of feature engineering involves identifying important features from data and accounting for different modalities (\eg textual, structured, or visual) of such data. Additionally, this step will require defining the target function that the model will attempt to learn. This requires knowing whether the model will be trained in a  supervised manner, \ie with labels, an unsupervised manner, \ie without labels, or through reinforcement learning. Lastly, the data processing to convert the preprocessed data into a format that can be fed directly into a model should be performed. Similarly to the last step, all steps should be thoroughly documented in order to support replication.

\noindent{\textit{What to report - }} For this step researchers should report their definition of their target function, the explicit features they used (e.g., code, natural language, snippets of execution traces, etc.) as well as reasoning that explains why they believe that the features and learning function chosen map well to their given problem domain. In addition, the researchers should report how they vectorize the resulting features, which includes the format of the vector or matrix subjected to the DL model.

\noindent \textbf{\textit{Step 3: Select a Deep Learning Architecture:}} This step comprises the determination of the learning algorithm, DL architecture, hyper-parameterization, optimization methods, and the training process. Researchers should carefully consider various aspects of existing architectures and determine whether a given architecture could be adapted for their task or whether the creation of a new architecture should be considered. In general these decisions can be guided by the successes and failures of past work. 

\noindent{\textit{What to report - }} Particular attention to the learning algorithm is required as this can significantly affect the results of the model. In addition to the learning algorithm, a full and thorough explanation of the optimization algorithm should be included, since the optimization algorithm describes the process of adjusting the parameters of the model to enable effective learning. When researchers report their approach, they should include a exhaustive description regarding the DL architecture used in their approach, which might include aspects such as the number of nodes, number of layers, type of layers, activation function(s), etc. Without all the aforementioned information pertaining to the architecture, future researchers will be forced to make assumptions which would likely affect reproducibility. Lastly, authors should communicate the hyper-parameters employed and training process designed. The hyper-parameters directly affect the performance of the model as they control the ability to the model to "learn". Hyper-parameters can be determined by researchers or through empirical analysis on a reserved portion of the dataset. Therefore, the process of finding the optimal set of hyper-parameters should be explicitly stated. Similarly, researchers should communicate details pertaining to the training process of the DL model, this commonly includes the number of training iterations, the time taken to train, and any techniques employed to combat overfitting.

\noindent \textbf{\textit{Step 4: Check for Learning Principles:}} This step is for assessing whether the basic learning principles are being considered to avoid biases, data snooping, or unnecessarily complex models (\ie Occam's Razor). Specifically, it is relevant to ensure your \textit{dataset} is diverse, representative, and sufficient. Moreover, it is relevant to ensure your \textit{model} considers less complex approaches than deep learning architectures. When considering Occam's Razor through the comparison of simpler (canonical) ML models, it is important to report the type of models and a brief overview of how those models were turned. This allows for additional reproducibility within the evaluation of the approach, it also substantiates the claim that the implemented DL model is appropriate for the problem being addressed.

\noindent{\textit{What to report - }} For this step in the process, researchers should explicitly state which learning principles they considered in their analysis. This might include techniques to check for bias in a dataset, comparisons against simpler techniques, or techniques for examining whether data snooping was occurring in practice. 

\noindent \textbf{\textit{Step 5: Check for Generalizability:}} The final step involves properly evaluating a DL model. This requires carefully choosing metrics for the task, testing the model against a held out test set, ensure the model has reached optimum capacity, and using standards (or creating standards) through the usage of benchmarks. This step demonstrates a meaningful attempt to accurately represent how the model would perform in a "real world" scenario. Therefore, researchers should provide evidence of generalizability that the model has achieved and describe which benchmarks were used to support those claims.

\noindent{\textit{What to report - }} Here researchers should strive both include the details of their evaluation plan, as well as provide rationale for the choices made. For example, this may include detailed descriptions of various metrics and justifications as to why those metrics are an appropriate measure of model efficacy for a given SE task.

In addition to the above steps steps, our findings corroborate many of the issues discussed in the study by Humbatova et al. \cite{humbatova2019taxonomy} This study analyzed real faults discovered in DL systems from GitHub. They found three major areas of error when implementing DL systems: errors in the input, the model, and the training process. They also interviewed developers to determine the severity and effort required to address many of these pitfalls. They found that developers thought the most severe errors related to (i) the proper implementation of the optimizer, (ii) deriving correct input data, and (iii) correctly implementing the models with the proper number of layers. However, they also learned that developers typically found that such errors require a relatively low amount of effort to fix \cite{humbatova2019taxonomy}. 

Our research demonstrates the components of learning within DL4SE work that are often not discussed or properly accounted for. For example, the selection and appropriate values of hyper-parameters can be an impactful when using a DL model. However, developers rated this issue to be the third highest in the amount of effort needed to address it. Similarly, properly preparing the training data and training process correct are ranked number two and number one, respectively, for the most amount of effort required to address the issue. %

All details mentioned in the above guidelines should be accessible in any DL4SE publication in order to facilitate reproduction and replication of the experiments done. We also encourage authors to promote transparency of their approach by making all datasets, models, and implementation scripts available via an online repository. This should lead to increased quality of future DL research in SE and allow for more meaningful comparisons of different DL approaches addressing SE tasks.

\section{Fringe Research not Included in this Review}
\label{sec:others}

In the midst of performing this SLR, we encountered studies which passed our initial inclusion criteria, but were eventually excluded based on their lack of a DL implementation. This SLR maintains the definition that deep learning models must automatically extract complex, hierarchical features from the data it is given. This implies that the data must be subjected to multiple, nonlinear transformations by passing it through multiple hidden layers within a neural network. This type of model would exclude certain algorithms that represent more canonical machine learning techniques. This hierarchy and definition of deep learning is shared by Goodfellow \etal \cite{Goodfellow2016} in a widely recognized textbook. %

There were two common types of papers we encountered when performing our SLR that we felt deserved a brief introduction and explanation as to why they were not included. The first is primary studies which use Word2Vec or some variation of Word2Vec in order to embed some type of sequential data. We frequently observed Word2Vec used as a pre-processing or embedding approach in order to draw relationships between textual software artifacts using some form of similarity measure. However, we contend that such techniques do not fall within the scope of this SLR, as
Word2Vec does not constitute a sufficiently ``deep'' architecture due to it only having a single embedding layer making it unable to model more complex representations that are normally associated with ``deep'' architecture, and are often used in conjunction with classical machine learning algorithms. Thus, including such works in our review would have significantly diluted the body of work that applies true DL techniques to SE problems.

We also identified a new field of research that has gained popularity known as SE4DL where the concepts learned from software testing and maintenance are being applied to the development of software based on DL algorithms. The process of our SLR captured some of these primary studies that have worked in applying SE ideals to the models generated by DL algorithms. These works focus on problems such as concolic testing for deep neural networks, addressing the lack of interoperability of DL systems by developing multi-granularity testing criteria, generating unique testing data through the use of generative adversarial networks, detecting bugs in the implementations of these DL models, seeding mutation of input data to create new testing data, detecting bugs when using certain ML frameworks such as TensorFlow, and detecting erroneous behaviors in autonomous vehicles powered by deep neural networks \cite{Sun2018a, Ma2018a, Zhang2018a, Falcini2017, Ma2018b, Zahavy2016, Zintgraf2017, Guo2018, Falcini2017a, Kindermans2017, Mirman2018, Zhang2018b, Tian2018a}. These pieces of work are only the beginning of a growing field that will attempt to understand and interpret the inner-workings of DL models.

\section{Related Studies and Literature Reviews}

In this literature review, we systematically collect, analyze, and report on DL approaches applied to a variety of SE tasks. While the focus of our work is specifically on Deep Learning, given its increasing popularity, there also exists a body of work which looks to analyze related concepts of the applications of ML more generally in SE. These studies vary in terms of scope and level of detail in their analysis of the implementations of the components of (deep) learning. Previous to our approach, a number of systematic literature reviews analyzed the  applications of Machine Learning and Artificial Intelligence to the field of SE \cite{WEN201241,oke08,muenchaisri_2019}. These works focus on implementations of ML models that do not contain the complexity or hierarchical feature extraction that DL possesses. In contrast, our work solely focuses on the technology of DL applications involving the extraction of complex, hierarchical features from the training data.

More recent works have analyzed the use of DL strategies within the scope of a specific SE task. For example, there have been a variety of SLRs performed for the use of DL applied to defect prediction or anomaly detection~\cite{DBLPjournalscorrabs190103407,Kwon2017ASO,akmel18,7272910}. These studies take a detailed look at the application of DL only within the scope of defect prediction or anomaly detection. However, our SLR examines the application of DL to a multitude of SE tasks and analyzes the variety of issues and correlations found generally when applying DL to any field of SE.

The most highly related works to the work presented in this study also looks at the application of DL to a variety of SE tasks. The two most closely related papers in this space are non-peer reviewed literature reviews hosted on arXiv by Li \etal~\cite{DBLPjournalscorrabs180504825} and Ferreira \etal~\cite{ferreira2019software}, which we briefly discuss here for completeness. Li \etal's study analyzes 98 DL studies for general trends and applications in DL. The primary findings of this work discusses the SE tasks addressed and the most common architectures employed. The main discussion points of this survey revolve around issues that are inherent to the implementation of DL methods. This includes problems with efficiency, understandability and testability of DL approaches. This SLR also mentions the difference between the application of DL in research and industry, which could be a result of not having a suitable way to apply DL toward SE tasks in practice.

Similarly, Ferreira \etal \cite{ferreira2019software} provide similar analysis to the paper presented by Li et al. However, Ferreira \etal provides a brief description of the works they studied as well as highlights some of their strengths in regards to the specific tasks they addressed. They also perform a general survey of the type of DL architectures implemented. This study's discussions involve highlighting the strengths and weakness of DL applications such as the lack of need for feature engineering, the ability to represent unstructured data, high computational costs, and large numbers of parameters.

Our work differs significantly as we implement a much more detailed and methodological analysis than both of these SLRs.  Our analysis is rooted in an exploration of the \textit{components of learning}, offering a much more comprehensive view of the entirety of applied DL approaches. Additionally, we carry out an exploratory data analysis to examine trends in the attributes we extracted from our primary studies. In addition to discussing common architectures and SE tasks, our SLR goes further by discussing additional factors such as data types, preprocessing techniques, exploratory data analysis methods,  learning algorithms, loss functions, hyperparamter tuning methods, techniques to prevent over/under - fitting, baseline techniques used in evaluations, benchmarks used in evaluations, consideration of the principle of Occam's Razor, metrics used in evaluations and reproducibility concerns, all of which are omitted from the previous two studies. In addition to a thorough discussion of these aspects, we also provide a correlation analysis between these different attributes of DL approaches applied to SE. We outline many meaningful relationships that can aid future researchers in their development of new DL solutions. Along with our analysis, we synthesized common pitfalls and difficulties that exist when applying DL-based solutions, and provide guidelines for avoiding these. Lastly, we include actionable next steps for future researchers to continue exploring and improving on various aspects of DL techniques applied to SE.

\section{Conclusions}
\label{sec:conclusion}

In this paper, we present a systematic literature review on the primary studies related to DL4SE from the top software engineering research venues. Our work heavily relied on the guidelines laid out by Kitchenham \etal for performing systematic literature reviews in software engineering. We began by establishing a set of research questions that we wanted to answer pertaining to applications of DL models to SE tasks. We then empirically developed a search string to extract the relevant primary studies to the research questions we wanted to answer. We supplemented our searching process with snowballing and manual additions of papers that were not captured by our systematic approach but were relevant to our study. We then classified the relevant pieces of work using a set of agreed upon inclusion and exclusion criteria. After distilling out a set of relevant papers, we extracted the necessary information from those papers to answer our research questions. Through the extraction process and the nature of our research questions, we inherently generated a taxonomy which pertains to different aspects of applying a DL-based approach to a SE task. Our hope is that this SLR provides future SE researchers with the necessary information and intuitions for applying DL in new and interesting ways within the field of SE. The concepts described in this review should aid researchers and developers in understanding where DL can be applied and necessary considerations for applying these complex models to automate SE tasks.

\section*{Acknowledgment} %

This material is supported by the NSF grants CCF-1927679, CCF-1955853, CCF-2007246,  and CCF-1815186. Any opinions, findings, conclusions, or recommendations expressed in this material are those of the authors and do not necessarily reflect the views of the sponsors.

\bibliographystyle{ACM-Reference-Format}
\bibliography{dlslr}

\section{Appendix - Supplemental Information}
\label{appendix:A}

\begin{table}[H]
\centering
\footnotesize
\caption{Software Terms used as Additional Search Parameters for DL/ML Venues}
\label{tab:se-terms}
\begin{tabular}{|l|l|}
\hline
Agile software						 		&				 Apps  \\ \hline
Autonomic systems					   & 			   Cloud computing \\ \hline
Component-based software  		 & 				 Configuration management \\ \hline
Crowd sourced software       		& 				Cyber physical systems \\ \hline
Debugging 									& 				Fault localization \\ \hline
Repair											&				Distributed software \\ \hline
Embedded software					  &				  Empirical software \\ \hline
End-user software						&				Formal methods \\ \hline
Human software							&				Green software \\ \hline
Human-computer interaction		 &				 Middleware \\ \hline
Frameworks								   &				APIs \\ \hline
Mining software							  &				   Mobile applications \\ \hline
Model-driven 								&				Parallel systems \\ \hline
Distributed systems		   			    &				Concurrent systems \\ \hline
Performance									&				Program analysis \\ \hline
Program comprehension				&				Program Synthesis \\ \hline
Programming languages				&				 Recommendation systems \\ \hline
Refactoring									 &				  Requirements engineering \\ \hline
Reverse engineering						&				 Search-based software \\ \hline
Secure software							  &				   Privacy software \\ \hline
Software architecture					&				 Software economics \\ \hline
Software metrics						  &				   Software evolution \\ \hline
Software maintenance				  &				   Software modeling \\ \hline
Software design							  &				   Software performance \\ \hline
Software process						 &				  Software product lines \\ \hline
Software reuse							   &				Software services \\ \hline
Software testing						   &				Software visualization \\ \hline
Modeling languages						&				 Software tools \\ \hline
Traceability								   &				Ubiquitous software \\ \hline
Validation 										&				Verification \\ \hline
\end{tabular}
\end{table}
\begin{table}[H]
\centering
\footnotesize
\caption{SLR Inclusion Criteria}
\begin{tabular}{|l|}
\hline
Study is published in the year range January 1, 2009 - June 1, 2019. \\ \hline
Study clearly defines a SE task. \\ \hline
Study was published in the predefined venues. \\ \hline
Study must identify and address a DL based approach. \\ \hline
Study must contain one of the terms learning, deep, or neural in the full body of the text. \\ \hline
Study was manually added by authors. \\ \hline
\end{tabular}
\end{table}

\begin{table}[H]
\centering
\footnotesize
\caption{SLR Exclusion Criteria}
\begin{tabular}{|l|}
\hline
Study was published before January 1, 2009 or after June 1, 2019.\\ \hline
Study does not address a SE task.\\ \hline
Study only applied DL in the evaluation as a baseline for comparison. \\ \hline
Study is outside the scope of software engineering or is published in an excluded venue.\\ \hline
Study does not fully implement a DL based approach. \\ \hline
Study does not evaluate a DL based approach. \\ \hline
Study is an extended abstract. \\ \hline
Study is solely based on a representational learning approach (word2vec, doc2vec, etc.).\\ \hline
\end{tabular}
\end{table}

\begin{table}[H]
\centering
\footnotesize
\caption{Top Search Strings Tested}
\begin{tabular}{|l|}
\hline
    (“Deep Learning” OR “Neural Network”)\\ \hline
    (“Learning”) AND (“Neural” OR “Automatic” OR “Autoencoder” OR “Represent”)\\ \hline
    (“Learning”) AND (“Supervised” OR “Reinforcement” OR “Unsupervised” OR “Semi-supervised”)\\ \hline
    (“Learning” OR “Deep” OR “Neural” OR “Network”)\\ \hline
    (“Learning” OR “Deep” OR “Neural”)\\ \hline
    (“Artificial Intelligence” OR “Learning” OR “Representational” OR “Neural” OR “Network”)\\ \hline
\end{tabular}
\end{table}

\begin{table}[H]
\centering
\footnotesize
\caption{Search String Terms Considered}
\begin{tabular}{|l|l|}
\hline
    Represent           &    Autoencoder\\ \hline
    Learning            &    Artificial \\ \hline
    Engineering         &    Automated \\ \hline
    Recurrent           &    Context \\ \hline
    Training            &    Layers \\ \hline
    Representational    &    Feature \\ \hline
    Neural              &    Network \\ \hline
    Recurrent           &    Convolution \\ \hline
    Machine             &    Deep \\ \hline
    Intelligence        &    Back-propagation \\ \hline
    Gradient            &    Hyper-parameters \\ \hline
\end{tabular}
\end{table}
\begin{table}[H]
\centering
\footnotesize
\caption{Column Discriptions for SE Tasks}
\label{tab:classifications}
\begin{tabular}{| p{0.35\linewidth} | p{0.65\linewidth} |}
\hline
Code Comprehension						 	&				 Research focused on the understanding of source code or program functionality.  \\ \hline
Source Code Generation					    & 			     Research focused on the creation or automatic synthesis of source code.   \\ \hline
Source Code Retrieval \& Traceability 		& 				 Research focused on the location of source code within software projects. \\ \hline
Source Code Summarization       		    & 				 Research focused on the summarization of source code or projects, also includes comment generation.\\ \hline
Bug-Fixing Process							& 				 Research focused on the location, understanding, or patching of bugs found in source code.  \\ \hline
Code Smells									&				 Research focused on locating, fixing or better understanding source code smells.  \\ \hline
Software Testing						    &				 Research focused on the synthesis, improvement or implementation of software tests. This includes both traditional software and mobile software testing.\\ \hline
Generating Non Code Artifacts				&				 Research focused on generating non code artifacts found in software repositories such as commit messages, story points, issue tracking, etc. \\ \hline
Clone Detection		                        &				 Research focused on the ability to find and classify source code clones. \\ \hline
Software Energy Metrics					    &				 Research focused on making software more energy efficient.\\ \hline
Program Synthesis						    &				 Research focused on the generation of programs or improving the synthesis process. \\ \hline
Image To Structured Representation 			&				 Research focused on taking images or sketches and translating them to structured source code.   \\ \hline
Software Security	   			            &				 Research focused on the privacy or security of software.   \\ \hline
Program Repair								&				 Research focused on program repair which focus exclusively on automatic patch generation.    \\ \hline
Software Reliability / Defect Prediction    &				 Research focused on predicting the reliability of software or the potential for defects within software.   \\ \hline
Feature Location				            &				 Research focused on the activity of identifying an initial location in the source code that implements functionality in a software system.   \\ \hline
Developer Forum Analysis				    &				 Research focused on mining developer forums to better understand the software engineering life cycle or the development process.  \\ \hline
Program Translation						    &				 Research focused on translating programs from one language to another.   \\ \hline
Software Categorization					    &				 Research focused on classifying different software projects into discrete categories, specifically for software within an app store.   \\ \hline
Code Location Within Media			        &				 Research focused on identifying and extracting source code snippets in videos or images.  \\ \hline
Developer Intention Mining				    &				 Research focused on developer communication and extracting developers intentions for a software project based on these communications.   \\ \hline
Software Resource Control				    &				 Research focused on improving or optimizing the number of resources needed for a particular software project at run time.    \\ \hline
\end{tabular}
\end{table}

\end{document}